\documentclass[preprintnumbers,prd,superscriptaddress,preprint]{revtex4-1}

\usepackage{CJK}
\usepackage{amsmath}
\usepackage{subfigure}
\usepackage{enumerate}
\usepackage{amssymb}
\usepackage{amsfonts}
\usepackage{mathrsfs}
\usepackage{latexsym}
\usepackage{bm}
\usepackage{amscd}
\usepackage{physics} 
\usepackage{graphicx}
\usepackage{epsfig}
\usepackage{hhline,multirow}
\usepackage{dcolumn}
\usepackage{url}
\usepackage[colorlinks=true,linkcolor=blue,citecolor=blue,urlcolor=blue]{hyperref}
\usepackage{color}
\usepackage{xcolor}
\usepackage{indentfirst}
\usepackage{subfigure}
\usepackage{psfrag}
\usepackage{slashed}
\usepackage{float}
\usepackage{setspace}
\usepackage{mathtools}
\allowdisplaybreaks[2]

\newcommand{\MBbar}{\overline{M}_{\!B}}
\newcommand{\MTbar}{\overline{M}_T}
\newcommand{\MTBbar}{\overline{M}_{TB}}

\begin{document}

\preprint{JLAB-THY-22-3553, \, ADP-22-1/T1172}

\title{Generalized parton distributions of sea quarks in the proton \\ from nonlocal chiral effective theory}

\author{Fangcheng He}
% \affiliation{Institute of High Energy Physics, CAS,
% 	Beijing 100049, China}
\affiliation{CAS Key Laboratory of Theoretical Physics, Institute of Theoretical Physics, Chinese Academy of Sciences, Beijing 100190, China}
\author{Chueng-Ryong Ji}
\affiliation{Department of Physics, North Carolina State University,
	Raleigh, North Carolina 27695, USA}
\author{W. Melnitchouk}
\affiliation{Jefferson Lab, Newport News,
	Virginia 23606, USA}
\author{A. W. Thomas}
\affiliation{CDMPP and CSSM, Department of Physics, University of Adelaide, Adelaide, SA 5005, Australia}
\author{P. Wang}
\affiliation{Institute of High Energy Physics, Chinese Academy of Sciences,
	Beijing 100049, China}
\affiliation{College of Physics Sciences, University of Chinese Academy of Sciences, Beijing 100049, China}

\begin{abstract}
We calculate the spin-averaged generalized parton distributions (GPDs) of sea quarks in the proton at zero skewness % ($\xi=0$) 
from nonlocal covariant chiral effective theory, including one-loop contributions from intermediate states with pseudoscalar mesons and octet and decuplet baryons.
A~relativistic regulator is generated from the nonlocal Lagrangian where a gauge link is introduced to guarantee local gauge invariance, with additional diagrams from the expansion of the gauge link ensuring conservation of electric charge and strangeness.
Flavor asymmetries for sea quarks at zero and finite momentum transfer, as well as strange form factors, are obtained from the calculated GPDs, and results compared with phenomenological extractions and lattice QCD.
% Without fine tuning the parameters 
\end{abstract}


\maketitle

%%%%%%%%%%%%%%%%%%%%%%%%%%%%%%%%%%%%%%%%%%%%%%%%%%%%%%%%%%%%%%%%%%%%%%%%%%%%
\section{Introduction}

Reconstructing the three-dimensional structure of the nucleon and other hadrons in terms of their fundamental quark and gluon (or parton) constituents is one of the defining problems in modern nuclear physics, and one which is a major driver of experimental programs at facilities such as Jefferson Lab and the future Electron-Ion Collider (EIC)~\cite{Accardi:2012qut, Avakian:2019csr}.
A central element of this endeavor is the extraction of generalized parton distributions (GPDs), which, as Fourier transforms of nonforward (and nondiagonal) matrix elements of nonlocal operators, contain rich information on the partonic structure of the nucleon.
GPDs interpolate between exclusive form factors, when integrated over parton momentum fraction $x$, and parton distribution functions (PDFs) in the forward limit, and contain considerably more information about the nucleon's internal structure than do PDFs or form factors alone (for reviews of GPDs see, {\it e.g.}, Refs.~\cite{Goeke:2001tz, Belitsky:2005qn}).
% By integrating GPDs with different powers of the momentum fraction $x$, GPDs can be transformed into Mellin moments. 

The mapping of nucleon GPDs requires a comprehensive program of experimental studies of hard exclusive processes, such as deeply-virtual Compton scattering (DVCS) and hard exclusive meson production (HEMP), over a broad kinematic range.
While theoretical tools have been developed to formally factorize GPDs from the process-dependent, hard scattering amplitudes~\cite{Ji:1998pc, Ji:1996nm, Collins:1998be}, the reconstruction of the full functional dependence of the GPDs, including their flavor and spin dependence, from limited experimental data is a formidable challenge~\cite{Kumericki:2016ehc}.
Experimental data were obtained at the HERA collider by the H1~\cite{Adloff:2001cn, Aktas:2005ty} and ZEUS~\cite{Breitweg:1998nh, Chekanov:2003ya} collaborations and by the HERMES~\cite{Airapetian:2001yk, Airapetian:2011uq, Airapetian:2012mq} fixed target experiment, as well as by COMPASS at CERN~\cite{dHose:2004usi, Fuchey:2015frv}.
A rich program of DVCS and HEMP measurements is also underway at Jefferson Lab with the 12~GeV energy upgraded, high-luminosity CEBAF accelerator~\cite{Stepanyan:2001sm, Defurne:2015kxq, Jo:2015ema, Seder:2014cdc}.

In addition to the experimental efforts, considerable progress has also been made on the theoretical front.
Because of the complex, nonperturbative properties of QCD, it is extremely challenging to calculate GPDs from first principles.
Since parton distributions and other light-cone correlation functions are defined in Minkowski space, it has also been very difficult to simulate GPDs on the Euclidean lattice.
Recent breakthroughs, however, have enabled the $x$ dependence of PDFs to be inferred from matrix elements of nonlocal operators on the lattice, in the form of quasi-parton distributions using the large momentum effective theory~\cite{Ji:2013dva}, pseudo-PDFs~\cite{Orginos:2017kos}, and lattice good cross sections~\cite{Ma:2014jla, Ma:2017pxb}.

As with PDFs, the simulation of GPDs on the lattice is still at a relatively early stage of development.
Much of the work on GPDs has focused on finding effective ways to parametrize their dependence on kinematic variables~\cite{Guidal:2004nd}.
From more phenomenological perspectives, characteristics of GPDs have been studied within nonperturbative approaches, such as the MIT and cloudy bag models~\cite{Ji:1997gm, Pasquini:2006ib}, the constituent quark model~\cite{Boffi:2002yy, Scopetta:2003et}, the NJL model~\cite{Mineo:2005vs}, the light-front quark model~\cite{Choi:2001fc, Choi:2002ic}, the color glass condensate model~\cite{Goeke:2008jz}, the chiral quark-soliton model~\cite{Petrov:1998kf, Penttinen:1999th, Goeke:2001tz, Schweitzer:2002nm, Ossmann:2004bp, Wakamatsu:2005vk}, and the Bethe-Salpeter approach~\cite{Tiburzi:2001je, Theussl:2002xp}.

In addition to the phenomenological models, more systematic approaches using heavy baryon and relativistic chiral effective field theory (EFT) have been widely used to study hadron structure at small momentum transfer \cite{Fuchs:2003ir, Kubis:2000zd}.
Perturbative calculations in chiral EFT expand observables as series in the pseudoscalar meson mass ${\cal O}(m_\phi/\Lambda_{\chi})$ or small external momentum ${\cal O}(q/\Lambda_{\chi})$, where $\Lambda_{\chi} \sim 1$~GeV is the scale associated with the chiral EFT.
%The perturbative calculation in chiral EFT can be expanded as the series of pesudoscalar meson mass $\cal {O}$$(m_\phi/\Lambda_{\chi})$ or small external momentum $\cal {O}$$(q/\Lambda_{\chi})$, where $\Lambda_{\chi}$ is scale of chiral EFT and is about 1 GeV.

Historically, most formulations of EFT have been based on dimensional or infrared regularization.
Recently, a {\it nonlocal} chiral effective Lagrangian was proposed~\cite{Wang:2010rib, He:2017viu, He:2018eyz}, which makes it possible to extend the range of momentum transfers over which hadron properties can be described. 
The method is a relativistic extension of finite range regularization, which has been applied extensively to extrapolate lattice QCD calculations of quantities such as the vector meson mass, magnetic moments, magnetic and strange form factors, charge radii, and moments of PDFs and GPDs~\cite{Young:2002ib, Leinweber:2003dg, Wang:2007iw, Allton:2005fb, Wang:1900ta, Wang:2012hj, Hall:2013dva, Shanahan:2012wh, Shanahan:2014uka} from unphysically large quark masses to the physical region.
The nonlocal interaction generates both the regulator which makes the loop integral convergent and the momentum dependence of the form factors at tree level. 
The electromagnetic and strange form factors of the nucleon obtained in this approach have been found to be in excellent agreement with experiment up to values of the four-momentum transfer squared of $\approx 1$~GeV$^2$ \cite{He:2017viu, He:2018eyz}.
Recently, the method has also been applied to calculate the $\bar{d}-\bar{u}$ flavor asymmetry in the proton~\cite{Salamu:2014pka}, the strange--antistrange PDF asymmetry $s-\bar{s}$~\cite{Wang:2016eoq, Wang:2016ndh, Salamu:2018cny, Salamu:2019dok}, and the sea quark Sivers function~\cite{He:2019fzn} in the proton.

In this paper we apply the nonlocal chiral effective theory for the first time to GPDs of sea quarks in the proton.
The study is timely, given the ongoing experimental program of DVCS and HEMP measurements at Jefferson Lab, and plans for future studies of high-$Q^2$ exclusive reactions at the EIC. 
We begin in Sec.~\ref{sec.framework} by introducing the local and nonlocal chiral Lagrangian, including a derivation of the currents which couple to the external vector field. 
The GPDs may be written as convolutions of splitting functions, describing the nucleon to meson plus baryon process, with the GPDs of the bare hadrons.
The one-loop nucleon $\to$ meson plus octet and decuplet baryon splitting functions are derived in Sec.~\ref{sec.splitting} from the full set of  rainbow, Kroll-Ruderman~\cite{Kroll:1953vq}, tadpole, and bubble diagrams.
Unlike the one-dimensional splitting functions relevant for PDFs, the splitting functions for nonforward GPDs also depend on the momentum transfer squared, in addition to the dependence on the longitudinal momentum fraction variable.
Taking moments of the nonforward splitting functions and expanding in powers of the pseudoscalar meson mass, in Sec.~\ref{sec.lna} we derive their nonanalytic behavior, which serves as a model-independent constraint on phenomenological models.
The convolution formalism is discussed in Sec.~\ref{sec.convolution}, where we present explicit expressions for the unpolarized electric ($H$) and magnetic ($E$) GPDs in terms of the splitting functions and GPDs of the pseudoscalar mesons and intermediate state baryons.
Numerical results are presented in Sec.~\ref{sec.numerical} for the nonperturbative sea quark contributions to the $H$ and $E$ GPDs for light quark and strange flavors, interpolating the corresponding constraints from the sea quark contributions to PDFs and form factors.
Finally, Sec.~\ref{sec.summary} summarises our results and anticipates future extensions of this analysis.
Explicit formulas for splitting function integrals are compiled in Appendix~\ref{sec.appendix}.

%%%%%%%%%%%%%%%%%%%%%%%%%%%%%%%%%%%%%%%%%%%%%%%%%%%%%%%%%%%%%%%%%%%%%%%%%%%%%
\section{Theoretical framework}
\label{sec.framework}

In this section we introduce the basic chiral Lagrangian which defines the theoretical basis of our calculations, as well as its nonlocal generalization which generates the ultraviolet regulator for loop integrals in a natural way, respecting Lorentz and gauge invariance.
The nonlocal formulation relevant for PDFs was presented in Refs.~\cite{Salamu:2018cny, Salamu:2019dok}; here we generalize the formalism to the case of nonforward matrix elements needed to compute GPDs.

% ...........................................................................
\subsection{Local chiral effective Lagrangian}
\label{ssec.chirallag}

We begin by introducing the lowest-order local Lagrangian of chiral SU(3)$_L\times$SU(3)$_R$ effective theory that describes the interaction of pseudoscalar mesons ($\phi$) with octet ($B$) and decuplet ($T_\mu$) baryons~\cite{Jenkins:1991ts, Ledwig:2014rfa}, \\
\begin{eqnarray}
\label{eq:ch8}
{\cal L}
&=& {\rm Tr} \big[ \bar B (i\slashed{D} - M_B) B \big]
 -  \frac{D}{2} \, {\rm Tr} \big[ \bar B \gamma^\mu \gamma_5 \{u_\mu, B\} \big]
 -  \frac{F}{2}\, {\rm Tr} \big[ \bar B \gamma^\mu \gamma_5 [ u_\mu ,B ] \big]
\nonumber\\
&+& \overline{T}_\mu^{ijk}
   (i\gamma^{\mu \nu \alpha} D_\alpha - M_T \gamma^{\mu\nu}) T_\nu^{ijk}
 -\ \frac{{\cal H}}{2}\,
(\overline{T}_\mu)^{ijk} \gamma^{\alpha} \gamma_5 (u_\alpha)^{kl}\,
	      (T^\mu)^{ijl}
\nonumber\\
&-& \frac{{\cal C}}{2}
   \big[ \epsilon^{ijk}\, \overline{T}_\mu^{ilm}
	  \Theta^{\mu\nu} (u_\nu)^{lj} B^{mk} + {\rm H.c.}
   \big]
 +\ \frac{f^2}{4}
    {\rm Tr} \big[ D_\mu U (D^\mu U)^\dag \big],
\end{eqnarray}
where $M_B$ and $M_T$ are the octet and decuplet masses, $D$, $F$, $\cal C$ and $\cal H$ are the baryon-meson coupling constants, and $f=93$~MeV is the pseudoscalar decay constant.
The octet–decuplet transition operator $\Theta^{\mu\nu}$ is given by
\begin{equation}
\Theta^{\mu\nu}
= g^{\mu\nu} - \Big(Z+\frac12\Big) \gamma^\mu \gamma^\nu,
\label{eq:Theta}
\end{equation}
where $Z$ is the decuplet off-shell parameter, chosen here to be 1/2~\cite{Nath:1971wp}.
We define the tensors
  $\gamma^{\mu\nu}
   = \frac{1}{2} [\gamma^\mu,\gamma^\nu] = -i \sigma^{\mu\nu}$
and
  $\gamma^{\mu\nu\alpha}
   = \frac{1}{2} \{\gamma ^{\mu\nu}, \gamma^\alpha\}$,
and $\epsilon^{ijk}$ is the antisymmetric tensor in
flavor space.
The SU(3) baryon octet fields $B^{ij}$ and decuplet fields $T^{ijk}_\mu$ are represented by the matrix
\begin{eqnarray}
\label{e.B}
B =
\left(
\begin{array}{ccc}
  \frac{1}{\sqrt 2} \Sigma^0 + \frac{1}{\sqrt 6} \Lambda
& \Sigma^+
& p					\\
  \Sigma^-
&-\frac{1}{\sqrt 2} \Sigma^0 + \frac{1}{\sqrt 6} \Lambda
& n					\\
  \Xi^-
& \Xi^0
&-\frac{2}{\sqrt6} \Lambda
\end{array}
\right)
\end{eqnarray}
and by symmetric tensors with components
\begin{eqnarray}
\label{e.T}
%\begin{array}{c}
&&T^{111} = \Delta^{++},\ \
T^{112} = \frac{1}{\sqrt 3} \Delta^+,\ \
T^{122} = \frac{1}{\sqrt 3} \Delta^0,\ \
T^{222} = \Delta^-,				\notag\\
&&T^{113} = \frac{1}{\sqrt 3} \Sigma^{*+},\ \
T^{123} = \frac{1}{\sqrt 6} \Sigma^{*0},\ \
T^{223} = \frac{1}{\sqrt 3} \Sigma^{*-},	\\
&&T^{133} = \frac{1}{\sqrt 3} \Xi^{*0},\ \
T^{233} = \frac{1}{\sqrt 3} \Xi^{*-},		\notag\\
&&T^{333} = \Omega^-,           \notag
%\end{array}
%
% T^{111} = \Delta^{++},\ \
% T^{112} = \frac{1}{\sqrt 3} \Delta^+,\ \
% T^{122} = \frac{1}{\sqrt 3} \Delta^0,\ \
% T^{222} = \Delta^-,				\notag\\
%
% T^{113} = \frac{1}{\sqrt 3} \Sigma^{*+},\ \
% T^{123} = \frac{1}{\sqrt 6} \Sigma^{*0},\ \
% T^{223} = \frac{1}{\sqrt 3} \Sigma^{*-},\ \	\notag\\
%
% T^{133} = \frac{1}{\sqrt 3} \Xi^{*0},\ \
% T^{233} = \frac{1}{\sqrt 3} \Xi^{*-},\ \	\\
%
% T^{333} = \Omega^-.				\notag
\end{eqnarray}
respectively.
The operator $U$ is defined in terms of the matrix of pseudoscalar meson fields~$\phi$,
\begin{equation}
U \equiv u^2 = \exp\bigg(i \frac{\sqrt2\phi}{f}\bigg),
\end{equation}
where the matrix
\begin{eqnarray}
\label{e.phi}
\phi =
\left(
{\begin{array}{*{20}{c}}
  \frac{1}{\sqrt 2} \pi^0 + \frac{1}{\sqrt 6} \eta
& \pi^+
& K^+						\\
  \pi^-
& -\frac{1}{\sqrt 2} \pi^0 + \frac{1}{\sqrt 6} \eta
& K^0						\\
  K^-
& \overline K^0
& -\frac{2}{\sqrt 6} \eta
\end{array}}
\right)
\end{eqnarray}
represents the $\pi$, $K$ and $\eta$ mesons.
The covariant derivatives of the octet and decuplet baryon fields in Eq.~(\ref{eq:ch8}) are given by~\cite{Hemmert:1998pi, Hemmert:1999mr}
\begin{eqnarray}
D_\mu B
&=& \partial_\mu B
 + [\Gamma_\mu, B]
 - i \langle \lambda^0 \rangle \upsilon_\mu^0\, B,	\\
D_\mu T_\nu^{ijk}
&=& \partial_\mu T_\nu^{ijk}
 + (\Gamma_\mu, T_\nu )^{ijk}
 - i \langle \lambda^0 \rangle \upsilon_\mu^0\, T_\nu^{ijk},
\label{eq:11}
\end{eqnarray}
respectively, where $\upsilon_\mu^0$ denotes an external singlet vector field, $\lambda^0$ is the unit matrix, and $\langle\, \cdots \rangle$ represents a trace in flavor space.
For the covariant derivative of the decuplet field, we employ the shorthand notation
\begin{equation}
(\Gamma_\mu, T_\nu)^{ijk}
= (\Gamma_\mu)_l^i\, T_\nu^{ljk}
+ (\Gamma_\mu)_l^j\, T_\nu^{ilk}
+ (\Gamma_\mu)_l^k\, T_\nu^{ijl}.
\end{equation}
For the meson fields, the covariant derivative is given by
\begin{eqnarray}
D_\mu U
&=& \partial_\mu U
 + (iU \lambda^a - i \lambda^a U)\, \upsilon_\mu^a.
\end{eqnarray}
The mesons couple to the baryon fields through the vector and axial vector combinations
\begin{eqnarray}
\Gamma_\mu
&=& \frac12
    \left( u \partial_\mu u^\dagger + u^\dagger \partial_\mu u \right)
 -  \frac{i}{2}
    \left( u \lambda^a u^\dagger + u^\dagger \lambda^a u \right) \upsilon_\mu^a,
\\
  u_\mu
&=& i
    \left( u^\dagger \partial_\mu u - u \partial_\mu u^\dagger \right)
 +  \left( u^\dagger \lambda^a u - u \lambda^a u^\dagger \right) \upsilon_\mu^a,
\label{eq:22}
\end{eqnarray}
where $\upsilon_\mu^a$ corresponds to an external octet vector field,
and $\lambda^a$ ($a=1, \ldots, 8$) are the Gell-Mann matrices.

While the unpolarized twist-two GPD $H$ receives contributions from each quark flavor from the lowest-order Lagrangian in Eq.~(\ref{eq:ch8}), to compute the effects of meson loops on the magnetic GPD $E$ requires an additional contribution to the Lagrangian for the magnetic interaction, which enters at a higher order.
The magnetic Lagrangian at ${\cal O}$$(q^2)$ for the octet, decuplet and octet-decuplet transition interaction is given by~\cite{He:2017viu, He:2018eyz, Yang:2020rpi, Jones:1972ky, Geng:2009ys}
\begin{eqnarray}
\label{lomag}
{\cal L}_{\rm mag}
&=& \frac{1}{4 M_B} 
\Big( c_1{\rm Tr}\left[\bar{B} \sigma^{\mu\nu}
  \left\{F^+_{\mu\nu},B\right\}\right]+c_2{\rm Tr}\left[\bar{B}
  \sigma^{\mu\nu} \left[F^+_{\mu\nu},B \right]\right]+c_3{\rm Tr}\left[\bar{B}
  \sigma^{\mu\nu}B \right]{\rm Tr}\left[F^+_{\mu\nu}\right]
\Big) \nonumber \\
&+&
\frac{i}{4 M_B} c_4 F_{\mu\nu}\Big(\epsilon_{ijk}(\lambda_q)^i_l\bar{B}^j_m\gamma^\mu\gamma_5(T^\nu)^{klm} 
+\epsilon^{ijk}(\lambda_q)^l_i(\overline{T}^\mu)_{klm}\gamma^\nu\gamma_5 B^m_j\Big)
  \nonumber \\
&+&
\frac{F_2^T}{2 M_T}
(\overline{T}_\mu)^{abc}\sigma^{\rho\sigma}\partial_{\sigma} \upsilon_\rho^q(\lambda_q)^a_e (T^\mu)^{ebc},
\end{eqnarray}
where we adopt the notation $c_1$, $c_2$ and $c_3$ for the octet baryon interaction from Ref.~\cite{Yang:2020rpi} and $c_4$ for the octet-decuplet transition, which corresponds to the constant $\mu_T$ in Refs.~\cite{He:2017viu, He:2018eyz}, and following Refs.~\cite{He:2017viu,He:2018eyz} denote by $F_2^T$ the coupling for the decuplet interaction.
In Eq.~(\ref{lomag}) the electromagnetic interaction with the individual quark flavors is introduced by the field strength tensor
\begin{equation}
F^+_{\mu\nu} = \frac12
\left(u^\dag F^q_{\mu\nu}\lambda_q u + u F^q_{\mu\nu}\lambda_q u^\dag\right),
\end{equation}
where $F^q_{\mu\nu} = \partial_\mu \upsilon^q_\nu-\partial_\nu \upsilon^q_\mu$ for the external field $\upsilon^q_\mu$ interacting with the quark flavor $q=u,d,s$ with unit charge, and the matrix $\lambda_q$ is the diagonal quark flavor matrix defined as
\mbox{$\lambda_q = {\rm diag}\{\delta_{qu},\delta_{qd},\delta_{qs}\}$}.
At this order, the magnetic Lagrangian ${\cal L}_{\rm mag}$ generates the following quark flavor decomposition for the proton anomalous magnetic moment, given by the proton's Pauli form factor $F_2^p(t)$ at $t=0$,
\begin{subequations}
\begin{eqnarray}
\label{treemag}
F^{p(u)}_2(0) &=& c_1+c_2+c_3, \\
F^{p(d)}_2(0) &=& c_3,         \\
F^{p(s)}_2(0) &=& c_1-c_2+c_3.
\end{eqnarray}
\end{subequations}
Since at tree level there is no strange quark contribution to the proton, we take $c_3=c_2-c_1$.
Furthermore, from SU(3) symmetry one also obtains relationships between the octet and decuplet constants~\cite{He:2017viu,He:2018eyz},
\begin{subequations}
\label{eq:F2}
\begin{eqnarray}
c_4   &=& 4 c_1,  \\
F_2^T &=& c_1+3c_2.
\end{eqnarray}
\end{subequations}
Within the flavor SU(3) framework, the magnetic moments of the octet and decuplet baryons, and the transition moments between the octet and decuplet baryons, can be expressed in terms of quark magnetic moments, $\mu_q$.
For example, for the proton and neutron one would have
    $\mu_p = \frac43 \mu_u - \frac13 \mu_d$
and $\mu_n = \frac43 \mu_d - \frac13 \mu_u$,
respectively, while for the $\Delta^{++}$ baryon $\mu_{\Delta^{++}} = 3 \mu_u$.

If we include the higher order magnetic Lagrangian ${\cal L}_{\rm mag}$ in Eq.~(\ref{lomag}), for consistency in the power counting we also need to consider the next-to-leading order Lagrangian for the baryon-meson interaction.
Generalizing Eq.~(\ref{eq:ch8}), and using the notation from Ref.~\cite{Kubis:2000aa}, we therefore include the additional baryon contribution involving two derivatives~\cite{Kubis:2000aa}
\begin{eqnarray}
\label{adm}
\!\!{\cal L}_{B\phi}' &=& \frac{i}{2}\sigma^{\mu\nu} 
\Big( 
  b_9\, {\rm Tr} \left[\bar{B}u_\mu\right]{\rm Tr}\left[u_\nu B\right]
+ b_{10}\, {\rm Tr} \left[\bar{B}\{[u_\mu,u_\nu],B\}\right]
+ b_{11}\, {\rm Tr} \left[\bar{B}[[u_\mu,u_\nu],B]\right]\!
\Big),~~~
\end{eqnarray}
where the values of the coefficients have been determined to be
    $b_9=1.36$~GeV$^{-1}$,
    $b_{10}=1.24$~GeV$^{-1}$,
and
    $b_{11}=0.46$~GeV$^{-1}$~\cite{Kubis:2000aa}.
Expanding the Lagrangians $\cal L$ in Eq.~(\ref{eq:ch8}) and ${\cal L}_{B\phi}'$ in Eq.~(\ref{adm}), the lowest order baryon-meson interaction involving the proton can then be written as
\begin{equation}
\begin{split}
{\cal L}_{\rm int}
&
= \frac{(D+F)}{2f}
  \left( \bar p\, \gamma^\mu \gamma_5 p\, \partial_\mu \pi^0
       + \sqrt2\, \bar p\, \gamma^\mu \gamma_5 n\, \partial_\mu \pi^+
  \right)
- \frac{(D-3F)}{\sqrt{12} f}
  \bar p\, \gamma^\mu \gamma_5 p\, \partial_\mu \eta        \\
&
+ \frac{(D-F)}{2 f}
  \left( \sqrt2\, \bar p\, \gamma^\mu \gamma_5 \Sigma^+\, \partial_\mu K^0
       + \bar p\, \gamma^\mu \gamma_5 \Sigma^0\, \partial_\mu K^+
  \right)                                                   
- \frac{(D+3F)}{\sqrt{12} f}
  \bar p\, \gamma^\mu \gamma_5 \Lambda\, \partial_\mu K^0   \\
&
+ \frac{\cal C}{\sqrt{12} f}
  \left(
  - 2\, \bar p\, \Theta^{\nu\mu} \Delta_\mu^+\, \partial_\nu \pi^0
  - \sqrt2\, \bar p\, \Theta^{\nu\mu} \Delta_\mu^0\, \partial_\nu \pi^+
  + \sqrt6\, \bar p\, \Theta^{\nu\mu} \Delta_\mu^{++}\, \partial_\nu \pi^-
  \right.							\\
&
  \hspace*{1.8cm}
  \left.
  - \bar p\, \Theta^{\nu\mu} \Sigma_\mu^{*0}\, \partial_\nu K^+
  + \sqrt2\, \bar p\, \Theta^{\nu\mu} \Sigma_\mu^{*+}\, \partial_\nu K^0
  + {\rm H.c.}
  \right)							\\
&
+ \frac{i}{4f^2} \bar p\, \gamma^\mu p
  \Big[
    (\pi^+ \partial_\mu \pi^-  -  \pi^- \partial_\mu \pi^+)
    + 2 (K^+ \partial_\mu K^-  -  K^- \partial_\mu K^+)
    + (K^0 \partial_\mu \bar K^0  -  \bar K^0 \partial_\mu K^0)
  \Big]   \\
&
+ \frac{i}{f^2} \bar p\, \sigma^{\mu\nu} p
  \Big( 2(b_{10}+b_{11})
    \partial_\mu\pi^+ \partial_\nu \pi^-
    + (4b_{11}+b_9) \partial_\mu K^+ \partial_\nu K^-  
    + 2(b_{10}-{b_{11}}) \partial_\mu K^0 \partial_\nu \bar K^0 
  \Big).
\end{split}
\label{eq:j1}
\end{equation}
For the interactions with the external field $\upsilon_\mu^a$, from the Lagrangian $\cal L$ in Eq.~(\ref{eq:ch8}) one can obtain the vector current 
\begin{eqnarray}
J^\mu_a
&=& \frac12 {\rm Tr}
   \big[
   \bar B \gamma^\mu
   \left[ u \lambda^a u^\dagger + u^\dagger \lambda^a u, B
   \right]	
      + \frac{D}{2}{\rm Tr}
   \big[
   \bar B \gamma^\mu \gamma_5
   \left\{ u \lambda^a u^\dagger - u^\dagger \lambda^a u, B
   \right\}
   \big]						\notag\\
&+& \frac{F}{2}{\rm Tr}
   \big[
   \bar B \gamma^\mu \gamma_5
   \left[ u \lambda^a u^\dagger - u^\dagger \lambda^a u, B
   \right]
   \big]						\notag\\
&+& \frac12\,
   \overline{T}_\nu \gamma^{\nu\alpha\mu}
   \left( u \lambda^a u^\dagger + u^\dagger \lambda^a u, T_\alpha
   \right)
   + \frac{\cal C}{2}
   \left[
   \overline{T}_\nu \Theta^{\nu\mu}
   (u \lambda^a u^\dagger - u^\dagger \lambda^a u) B
   + {\rm H.c.}
   \right]						\notag\\
&+& \frac{f^2}{4}{\rm Tr}
   \big[
   \partial^\mu U
   (U^\dagger i \lambda^a
    - i \lambda^a U^\dagger)
+  (U i \lambda^a
    - i \lambda^a U)
   \partial^\mu U^\dagger
   \big].
\label{eq:ch1}
\end{eqnarray}
For the SU(3) flavor singlet case, the current coupling to the external field $\upsilon_\mu^0$ can be written
\begin{eqnarray}
J^\mu_0
&=& \langle \lambda^0 \rangle\,
    {\rm Tr}[\bar B \gamma^\mu B]
 +  \langle \lambda^0 \rangle\,
    \overline{T}_\nu \gamma^{\nu\alpha\mu}\, T_\alpha.
\label{eq:ch2}
\end{eqnarray}
The magnetic current coupling to the external field $\upsilon_\mu^q$ can be obtained from the magnetic Lagrangian in Eq.~(\ref{lomag}), % $J_q^{\mu,\rm mag}$
\begin{eqnarray}
J_{q, \rm mag}^\mu
&=&
\frac{\partial_\nu}{4 M_B}
\Big(
  c_1 {\rm Tr}\bar{B} \sigma^{\mu\nu}
    \left\{u^\dag \lambda_q u + u \lambda_q u^\dag,B\right\}
+ c_2 {\rm Tr}\bar{B} \sigma^{\mu\nu}
    \left[u^\dag \lambda_q u + u \lambda_q u^\dag,B \right]
\nonumber \\
& & \qquad
+\, c_3 {\rm Tr}\bar{B} \sigma^{\mu\nu} B\, 
        {\rm Tr}(u^\dag \lambda_q u + u \lambda_q u^\dag)
\Big)
-\
\frac{F_2^T}{2 M_T}
\partial_\sigma
\Big( (\overline{T}_\nu)^{abc}\sigma^{\mu\sigma}(\lambda_q)^a_e (T^\nu)^{ebc}
\Big)
\nonumber \\
&-&
\frac{ic_4}{4 M_B} 
(g^{\mu\nu} \partial^\sigma - g^{\mu\sigma} \partial^\nu)
\Big( \epsilon_{ijk} 
      (\lambda_q)^i_l\, \bar{B}^j_m \gamma_\sigma \gamma_5 T_\nu^{klm}
    + \epsilon^{ijk}
      (\lambda_q)^l_i\, \overline{T}_{\sigma,klm} \gamma_\nu \gamma_5 B^m_j
\Big),
\end{eqnarray}
which satisfies current conservation,
    $\partial_\mu J_{q, \rm mag}^\mu = 0$.
The quark flavor currents can be written in terms of the SU(3) singlet ($a=0$) and octet ($a=3, 8$), and quark magnetic currents,
\begin{subequations}
\label{eq:ch3}
\begin{eqnarray}
J^\mu_u
&=& \frac13 J^\mu_0
 +  \frac12 J^\mu_3
 +  \frac{1}{2\sqrt3} J^\mu_8 
 + J_{u,\rm mag}^\mu,			\\
J^\mu_d
&=& \frac13 J^\mu_0
 -  \frac12 J^\mu_3
 +  \frac{1}{2\sqrt3} J^\mu_8 
 + J_{d,\rm mag}^\mu,			\\
J^\mu_s
&=& \frac13 J^\mu_0
 -  \frac{1}{\sqrt3} J^\mu_8 
 + J_{s,\rm mag}^\mu.
\end{eqnarray}
\end{subequations}
Using Eqs.~(\ref{eq:ch1}), (\ref{eq:ch2}) and (\ref{eq:ch3}), the quark flavor currents can be written more explicitly in the form
% "Please rewrite Eq.(~\ref{eq:jq}) expressing  % \partial^\mu$ and $q^\mu$ in terms of $p^\mu$ and $p'^\mu$"
%
\begin{subequations}
\label{eq:jq}
\begin{eqnarray}
J_u^\mu
&=& 2 \bar p \gamma^\mu p + \bar n \gamma^\mu n
+ \bar\Lambda \gamma^\mu \Lambda
+ 2 \overline{\Sigma}^+ \gamma^\mu \Sigma^+
+ {\overline\Sigma}^0 \gamma^\mu \Sigma^0
- \frac{1}{2f^2}\, \bar p \gamma^\mu p\, 
  \big( \pi^+ \pi^- + 2 K^+ K^- \big)
\notag\\
&+&
  3 \overline{\Delta}_\alpha^{++} \gamma^{\alpha\beta\mu} \Delta_\beta^{++}
+ 2 \overline{\Delta}_\alpha^+    \gamma^{\alpha\beta\mu} \Delta_\beta^+
+   \overline{\Delta}_\alpha^0    \gamma^{\alpha\beta\mu} \Delta_\beta^0
+ 2 \overline{\Sigma}_\alpha^{*+} \gamma^{\alpha\beta\mu} \Sigma_\beta^{*+}
+   \overline{\Sigma}_\alpha^{*0} \gamma^{\alpha\beta\mu} \Sigma_\beta^{*0}
\notag\\
&+&
  i \left( \pi^- \partial^\mu\pi^+ - \pi^+ \partial^\mu\pi^- \right)
+ i \left( K^-   \partial^\mu K^+  - K^+   \partial^\mu K^-   \right)
\notag\\
&-&
  \frac{i(D+F)}{\sqrt 2f} \bar p \gamma^\mu \gamma_5 n\, \pi^+
+ \frac{i(D+3F)}{\sqrt{12}f} \bar p \gamma^\mu \gamma_5 \Lambda\, K^+
- \frac{i(D-F)}{2f} \bar p \gamma^\mu \gamma_5 \Sigma^0\, K^+
\notag\\
&+&
  \frac{i\, \cal C}{\sqrt{12}f}
  \left(
    \sqrt{6}\, \bar p\, \Theta^{\mu\nu} \Delta_\nu^{++}\, \pi^-
  + \sqrt{2}\, \bar p\, \Theta^{\mu\nu} \Delta_\nu^0\, \pi^+
  + \,\bar p\, \Theta^{\mu\nu} \Sigma_\nu^{*0}\, K^+
  + {\rm H.c.}
  \right)
\notag\\
&+& \frac{1}{4M_B} \partial_\nu (\bar{p}\sigma^{\mu\nu}p)
    \bigg[
      4 c_2 \Big( 1 - \frac{1}{2 f^2} K^+K^- \Big) - \frac{(c_1+c_2)}{f^2} \pi^+\pi^- 
    \bigg]
 +  \frac{c_2-c_1}{2M_B} \partial_\nu(\bar{n}\sigma^{\mu\nu}n)
\notag\\
&+& \frac{3c_2-2c_1}{6M_B} \partial_\nu (\bar{\Lambda}\sigma^{\mu\nu}\Lambda) 
 +  \frac{c_1}{2\sqrt3 M_B} \partial_\nu (\overline{\Lambda}\sigma^{\mu\nu}\Sigma^0) 
 +  \frac{c_2}{M_B} \partial_\nu (\overline{\Sigma}^+\sigma^{\mu\nu}\Sigma^+)
 +  \frac{c_2}{2M_B} \partial_\nu (\overline{\Sigma}^0\sigma^{\mu\nu}\Sigma^0) 
\notag\\
&+&
\frac{i c_4}{4\sqrt3 M_B} \partial^\nu
\bigg[
  \bar{p}(\gamma_\nu \gamma_5 \Delta^{+\mu} - \gamma^\mu\gamma_5 \Delta^+_\nu)
+ \bar{n}(\gamma_\nu \gamma_5 \Delta^{0\mu} - \gamma^\mu\gamma_5 \Delta^0_\nu)
- \overline{\Sigma}^+ (\gamma_\nu \gamma_5 \Sigma^{*+\mu} - \gamma^\mu \gamma_5 \Sigma^{*+}_\nu)
\notag\\
& & \hspace{2cm}
- \frac{\sqrt3}{2} 
  \bar{\Lambda} (\gamma_\nu \gamma_5 \Sigma^{*0\mu} - \gamma^\mu \gamma_5 \Sigma^{*0}_\nu)
+ \frac12 
  \overline{\Sigma}^0 (\gamma_\nu \gamma_5 \Sigma^{*0\mu} - \gamma^\mu \gamma_5 \Sigma^{*0}_\nu)   
\bigg]
\notag\\
&-& \frac{F_2^T}{6 M_T}\partial_\nu
\Big[
  3\overline{\Delta}^{++}_\alpha \sigma^{\mu\nu} \Delta^{++\alpha}
+ 2\overline{\Delta}^{+}_\alpha  \sigma^{\mu\nu} \Delta^{+\alpha}   
+  \overline{\Delta}^{0}_\alpha  \sigma^{\mu\nu} \Delta^{0\alpha}
+ 2\overline{\Sigma}^{*+}_\alpha \sigma^{\mu\nu} \Sigma^{*+\alpha}
+  \overline{\Sigma}^{*0}_\alpha \sigma^{\mu\nu} \Sigma^{*0\alpha}
\Big],
\notag\\
& &
\label{eq:ju}
\end{eqnarray}
\begin{eqnarray}
J^\mu_d
&=& \bar p \gamma^\mu p
+ 2 \bar n \gamma^\mu n
+ 2 \overline{\Sigma}^- \gamma^\mu \Sigma^-
+ \overline{\Sigma}^0 \gamma^\mu \Sigma^0
+ \bar\Lambda \gamma^\mu \Lambda
+ \frac{1}{2f^2}\, \bar p \gamma^\mu p\, 
  \big( \pi^+ \pi^- - \overline{K}^0 K^0 \big)
\notag\\
&+&
    \overline{\Delta}_\alpha^+ \gamma^{\alpha\beta\mu} \Delta_\beta^+
+ 2 \overline{\Delta}_\alpha^0 \gamma^{\alpha\beta\mu} \Delta_\beta^0
+ 3 \overline{\Delta}_\alpha^- \gamma^{\alpha\beta\mu} \Delta_\beta^-
+   \overline{\Sigma}_\alpha^{*0}  \gamma^{\alpha\beta\mu} \Sigma_\beta^{*0}
+ 2 \overline{\Sigma}_\alpha^{*0-} \gamma^{\alpha\beta\mu} \Sigma_\beta^{*-}
\notag\\
&-&
  i (\pi^- \partial^\mu \pi^+  -  \pi^+ \partial^\mu \pi^-)
+ i (\overline{K}^0 \partial^\mu K^0  -  K^0 \partial^\mu \overline{K}^0)
\notag\\
&+&
  \frac{i(D+F)}{\sqrt2 f}
  \bar p \gamma^\mu \gamma_5 n\, \pi^+
- \frac{i(D-F)}{\sqrt2 f}
  \bar p \gamma^\mu \gamma_5 \Sigma^+\, K^0
\notag\\
&-&
\frac{i\, \cal C}{\sqrt6 f}
  \left(
    \sqrt3\, \bar p\, \Theta^{\mu\nu} \Delta_\nu^{++}\, \pi^-
  + \bar p\, \Theta^{\mu\nu} \Delta_\nu^0\, \pi^+
  + \bar p\, \Theta^{\mu\nu} \Sigma_\nu^{*+}\, K^0
  + {\rm H.c.}
  \right)
\notag\\
&+& \frac{1}{4M_B} \partial_\nu (\bar{p}\sigma^{\mu\nu} p)
    \bigg[ (c_2-c_1) \Big( 2 - \frac{1}{f^2} \overline{K}^0 K^0 \Big)
        + \frac{(c_1+c_2)}{f^2} \pi^+\pi^-
    \bigg]
 +  \frac{c_2}{M_B} \partial_\nu (\bar{n}\sigma^{\mu\nu}n)
\notag\\
&+& \frac{3c_2-2c_1}{6M_B} \partial_\nu (\bar{\Lambda}\sigma^{\mu\nu}\Lambda) 
 +  \frac{c_2}{M_B} \partial_\nu (\overline{\Sigma}^-\sigma^{\mu\nu}\Sigma^-)    
 +  \frac{c_2}{2M_B} \partial_\nu (\overline{\Sigma}^0\sigma^{\mu\nu}\Sigma^0)
 -  \frac{c_1}{2\sqrt3 M_B} \partial_\nu (\bar{\Lambda}\sigma^{\mu\nu}\Sigma^0) 
\notag\\
&-& \frac{i c_4}{4\sqrt3 M_B} \partial^\nu
\bigg[
  \bar{p}(\gamma_\nu \gamma_5 \Delta^{+\mu} - \gamma^\mu \gamma_5 \Delta^+_\nu)
+ \bar{n}(\gamma_\nu \gamma_5 \Delta^{0\mu} - \gamma^\mu \gamma_5 \Delta^0_\nu)
- \overline{\Sigma}^-(\gamma_\nu \gamma_5 \Sigma^{*-\mu} - \gamma^\mu \gamma_5 \Sigma^{*-}_\nu)
\notag\\  
& & \hspace{2cm}
- \frac{\sqrt3}{2} \bar{\Lambda}(\gamma_\nu \gamma_5 \Sigma^{*0\mu} - \gamma^\mu \gamma_5 \Sigma^{*0}_\nu)
- \frac12 \overline{\Sigma}^0(\gamma_\nu \gamma_5 \Sigma^{*0\mu} - \gamma^\mu \gamma_5 \Sigma^{*0}_\nu)
\bigg]
\notag\\
&-& \frac{F_2^T}{6 M_T}\partial_\nu
\Big[
  3\overline{\Delta}^-_\alpha \sigma^{\mu\nu} \Delta^{-\alpha}
+ 2\overline{\Delta}^0_\alpha \sigma^{\mu\nu} \Delta^{0\alpha}
+  \overline{\Delta}^+_\alpha \sigma^{\mu\nu} \Delta^{+\alpha}
+ 2\overline{\Sigma}^{*-}_\alpha \sigma^{\mu\nu} \Sigma^{*-\alpha}
+  \overline{\Sigma}^{*0}_\alpha \sigma^{\mu\nu} \Sigma^{*0\alpha}
\Big],
\notag\\
& & \\
& & \notag
\label{eq:jd}
\end{eqnarray}
\begin{eqnarray}
\label{eq:js}
J^\mu_s
&=&\overline{\Sigma}^+ \gamma^\mu \Sigma^+
 + \overline{\Sigma}^0 \gamma^\mu \Sigma^0
 + \bar\Lambda \gamma^\mu \Lambda
 + \frac{1}{2f^2}\, \bar p \gamma^\mu p\, 
   \big( 2 K^+ K^- + \overline{K}^0 K^0 \big)
\notag\\
&+&
   \overline{\Sigma}_\alpha^{*+} \gamma^{\alpha\beta\mu} \Sigma_\beta^{*+}
 + \overline{\Sigma}_\alpha^{*0} \gamma^{\alpha\beta\mu} \Sigma_\beta^{*0}
 - i (K^- \partial^\mu K^+  -  K^+ \partial^\mu K^-)
 - i (\overline{K}^0 \partial^\mu K^0 - K^0 \partial^\mu \overline{K}^0)
\notag\\
&+&
   \frac{i(D-F)}{\sqrt2 f}
   \bar p \gamma^\mu \gamma_5 \Sigma^+\, K^0
 + \frac{i(D-F)}{2f}
   \bar p \gamma^\mu \gamma_5 \Sigma^0\, K^+
 - \frac{i(D+3F)}{\sqrt{12} f}
   \bar p \gamma^\mu \gamma_5 \Lambda\, K^+		\notag\\
&-&
   \frac{i\, \cal C}{\sqrt{12} f}
   \left(
     \bar p\, \Theta^{\mu\nu} \Sigma_\nu^{*0}\, K^+
   - \sqrt2\, \bar p\, \Theta^{\mu\nu} \Sigma_\nu^{*+}\, K^0
   + {\rm H.c.}
   \right)                                              \notag\\
&+& \frac{1}{4 M_B f^2} \partial_\nu (\bar{p}\sigma^{\mu\nu} p) 
\bigg[ 2 c_2 K^+K^- + (c_2-c_1) \overline{K}^0 K^0 \bigg]
 +  \frac{(c_1+3c_2)}{6M_B} \partial_\nu (\bar{\Lambda}\sigma^{\mu\nu}\Lambda)
\notag\\
&+& \frac{(c_2-c_1)}{2M_B} \partial_\nu (\overline{\Sigma}^+ \sigma^{\mu\nu}  \Sigma^+)
 +  \frac{(c_2-c_1)}{2M_B} \partial_\nu (\overline{\Sigma}^- \sigma^{\mu\nu} \Sigma^-) 
 +  \frac{(c_2-c_1)}{2M_B} \partial_\nu (\overline{\Sigma}^0 \sigma^{\mu\nu} \Sigma^0)
\notag\\
&-& \frac{\, ic_4}{4\sqrt3 M_B} \partial^\nu
\bigg[
  \overline{\Sigma}^0 (\gamma_\nu \gamma_5 \Sigma^{*0\mu} - \gamma^\mu \gamma_5 \Sigma^{*0}_\nu)
+ \overline{\Sigma}^- (\gamma_\nu \gamma_5 \Sigma^{*-\mu} - \gamma^\mu \gamma_5 \Sigma^{*-}_\nu)
\notag\\
& & \hspace*{2cm}
- \overline{\Sigma}^+ (\gamma_\nu \gamma_5 \Sigma^{*+\mu} - \gamma^\mu \gamma_5 \Sigma^{*+}_\nu) 
\bigg],
\notag\\
&-& \frac{F_2^T}{6 M_T} \partial_\nu 
\Big[
  \overline{\Sigma}^{*-}_\alpha \sigma^{\mu\nu} \Sigma^{*-\alpha}   
+ \overline{\Sigma}^{*0}_\alpha \sigma^{\mu\nu} \Sigma^{*0\alpha}
+ \overline{\Sigma}^{*+}_\alpha \sigma^{\mu\nu} \Sigma^{*+\alpha}
\Big],
\end{eqnarray}
\end{subequations}
for the $u$, $d$ and $s$ quark flavors, respectively.
As in Ref.~\cite{Salamu:2018cny}, terms involving the doubly-strange $\Xi^{0,-}$ and $\Xi^{*0,-}$ hyperons and the triply-strange $\Omega^-$ baryon do not couple directly 
to proton states and not included here.

% ..................................................................................
\subsection{Nonlocal chiral Lagrangian}
\label{ssec:nonlocal}

In this section we outline the generalization of the effective local chiral Lagrangian to the case of nonlocal interactions.
Taking the traces in Eqs.~(\ref{eq:ch8}), (\ref{lomag}) and (\ref{adm}) in Sec.~\ref{ssec.chirallag}, we can write the local Lagrangian density in the form
\begin{eqnarray}
{\cal L}^{\rm (local)}(x)
&=&\bar B(x)(i \gamma^\mu \mathscr{D}_{\mu} - M_B) B(x)
 + \frac{C_{B\phi}}{f}
   \big[ \bar{p}(x) \gamma^\mu \gamma_5 B(x)\,
	  \mathscr{D}_{\mu} \phi(x) + {\rm H.c.}
   \big]						\notag\\
&+&\overline{T}_\mu(x)
   (i \gamma^{\mu\nu\alpha} \mathscr{D}_{\alpha} - M_T \gamma^{\mu\nu})\,
   T_\nu(x)
 + \frac{C_{T\phi}}{f}
   \left[\, \bar{p}(x) \Theta^{\mu\nu} T_\nu(x)\,
	 \mathscr{D}_{\mu} \phi(x) + {\rm H.c.}
   \right]						\notag\\
&+&\mathscr{D}_{\mu} \phi(x) (\mathscr{D}_{\mu} \phi)^\dag(x)\,
 + \frac{i C_{\phi\phi^\dag}}{2 f^2}
   \bar p(x) \gamma^\mu p(x)
   \left[ \phi(x) (\mathscr{D}_\mu \phi)^\dag(x)
	- \mathscr{D}_{\mu} \phi(x) \phi^\dag(x)
   \right]
\notag\\
&+&\frac{i C_{\phi\phi^\dag}'}{2 f^2}
   \bar p(x) \sigma^{\mu\nu} p(x)\mathscr{D}_\mu \phi(x)
    (\mathscr{D}_{\nu} \phi)^\dag(x)
\notag\\
&+&\frac{C_{\phi\phi^\dag}^{\rm mag}}{4 M_B f^2}
   \bar p(x) \sigma^{\mu\nu} p(x) F_{\mu\nu}(x) \phi(x)\phi^\dag(x)
 + \frac{C_{B}^{\rm mag}}{4 M_B}
   \bar B(x) \sigma^{\mu\nu} B(x) F_{\mu\nu}(x) 
\notag\\
&+&\frac{i C_{BT}^{\rm mag}}{4 M_B}
   \bar B(x) \gamma^\mu\gamma_5 T^\nu (x) F_{\mu\nu}(x)		
 - \frac{C_{T}^{\rm mag}}{4 M_T}
   \overline T_\alpha(x) \sigma^{\mu\nu} T^\alpha(x) F_{\mu\nu}(x)	
 + \cdots ,
\label{eq:Llocal}
\end{eqnarray}
where the dependence on the space-time coordinate $x$ is shown explicitly, and for the interaction part we show only those terms that contribute to the proton GPDs.
The covariant derivatives in Eq.~(\ref{eq:Llocal}) are given by
\begin{subequations}
\begin{eqnarray}
\mathscr{D}_{\mu} B(x)
&=& \left[ \partial_\mu - i e^q_B\, \mathscr{A_\mu}(x)
    \right] B(x),
\\
\mathscr{D}_{\mu} T^\nu(x)
&=& \left[ \partial_\mu - i e^q_T\, \mathscr{A_\mu}(x)
    \right] T^\nu(x),
\\
\mathscr{D}_{\mu} \phi(x)
&=& \left[ \partial_\mu - i e^q_\phi\, \mathscr{A_\mu}(x)
    \right] \phi(x),
\end{eqnarray}
\end{subequations}
where $\mathscr{A_\mu}$ is the electromagnetic gauge field, and $e^q_B$, $e^q_T$ and $e^q_\phi$ denote, respectively, the quark flavor charges of the octet and decuplet baryons, $B$ and $T$, and meson $\phi$.
In the case of the proton, for instance, one has the flavor charges
$e^u_p = 2 e^d_p = 2$, $e^s_p = 0$,
while for the $\Sigma^+$ hyperon
$e^u_{\Sigma^+} = 2 e^s_{\Sigma^+} = 2$, $e^d_{\Sigma^+} = 0$,
and similarly for the other baryons.
For the pion and kaon, the flavor charges are
$e^u_{\pi^+} = -e^d_{\pi^+} = 1$, $e^q_{\pi^0} = 0$ for all $q$,
and 
$e^u_{K^+} = -e^s_{K^+} = 1$, $e^d_{K^+} = 0$,
with the values for other mesons obtained by charge conjugation.
The coefficients $C_{B\phi}$, $C_{T\phi}$, $C_{\phi\phi^\dag}$, $C_{\phi\phi^\dag}'$, $C_B^{\rm mag}$, $C_{BT}^{\rm mag}$, $C_T^{\rm mag}$ and $C_{\phi\phi^\dag}^{\rm mag}$ in Eq.~(\ref{eq:Llocal}) are given explicitly in Table~\ref{tab:C} for the various processes discussed in this work.

Following Ref.~\cite{Salamu:2018cny}, we sketch here the derivation of the nonlocal Lagrangian from Eq.~(\ref{eq:Llocal}).
Details of the methodology used here can be found in Refs.~\cite{Terning:1991yt, Holdom:1992fn, Faessler:2003yf, Wang:1996zu, He:2017viu, He:2018eyz}.
The nonlocal analog of the local Lagrangian (\ref{eq:Llocal}) can be written as
\begin{eqnarray}
{\cal L}^{\rm (nonloc)}(x)
&=& \bar B(x) (i\gamma^\mu \mathscr{D}_\mu - M_B) B(x)
+ \overline{T}_\mu(x)
  (i\gamma^{\mu\nu\alpha} \mathscr{D}_\alpha - M_T \gamma^{\mu\nu})
  T_\nu(x)
\notag\\
& & \hspace*{-2.2cm}
+\ \bar{p}(x)
  \left[
    \frac{C_{B\phi}}{f} \gamma^\mu \gamma_5 B(x)\,
  + \frac{C_{T\phi}}{f} \Theta^{\mu\nu} T_\nu(x)
  \right]
  \mathscr{D}_\mu\!
  \int\!\dd[4]{a}\, {\cal G}_\phi^q(x,x+a) F(a) \phi(x+a)
  + {\rm H.c.}
\notag\\
& & \hspace*{-2.2cm}
+\ \frac{i C_{\phi\phi^\dag}}{2 f^2}\,
  \bar{p}(x) \gamma^\mu p(x)\!                           
  \int\!\dd[4]{a}\, {\cal G}_\phi^q(x,x+a) F(a) \phi(x+a)\,
  \mathscr{D}_\mu\!\!
  \int\!\dd[4]{b}\, {\cal G}_\phi^q(x+b,x) F(b) \phi^\dag(x+b)
%  + {\rm H.c.}
\notag\\
& & \hspace*{-2.2cm}
+\ \frac{i C_{\phi\phi^\dag}'}{2 f^2}\,
  \bar{p}(x) \sigma^{\mu\nu} p(x)\,				
  \mathscr{D}_\mu\! \int\!\dd[4]{a}\, {\cal G}_\phi^q(x,x+a) F(a) \phi(x+a)\,
  \mathscr{D}_\nu\! \int\!\dd[4]{b}\, {\cal G}_\phi^q(x+b,x) F(b) \phi^\dag(x+b)
\notag\\
& & \hspace*{-2.2cm}
+\ \frac{C_B^{\rm mag}}{4 M_B}\,
   \bar B(x) \sigma^{\mu\nu} B(x) F_{\mu\nu}(x)
+  \frac{i C_{BT}^{\rm mag}}{4 M_B}
   \bar B(x) \gamma^\mu\gamma_5 T^\nu(x) F_{\mu\nu}(x)
-  \frac{C_T^{\rm mag}}{4 M_T}
   \overline T_\alpha(x) \sigma^{\mu\nu} T^\alpha(x) F_{\mu\nu}(x)	
\notag\\
& & \hspace*{-2.2cm}
+\ \frac{C_{\phi\phi^\dag}^{\rm mag}}{4 M_B f^2}\,
  \bar{p}(x) \sigma^{\mu\nu} p(x) \int\!\dd[4]{a}\!\int\!\dd[4]{b}\, F_{\mu\nu}(x)\,
  {\cal G}_\phi^q(x+b,x+a) F(a) F(b)\, \phi(x+a) \phi^\dag(x+b)
\notag\\
& & \hspace*{-2.2cm}
+\ \mathscr{D}_\mu \phi(x) (\mathscr{D}_\mu \phi)^\dag(x)
+\ \cdots,
\label{eq:j4}
\end{eqnarray}
where
%$\mathscr{D}_{\mu} \Psi= \left[\partial_\mu -ie_\Psi^q \mathscr{A}_\mu(x)\right]\Psi(x)$ 
the gauge link ${\cal G}_\phi^q$ is introduced for local gauge invariance,
\begin{equation}
{\cal G}_\phi^q(x,y)
= \exp\left[ -i e^q_\phi \int_x^y \dd{z}^\mu \mathscr{A}_\mu(z)  \right],
\label{eq:link}
\end{equation}
and $F(a)$ is the meson--baryon vertex form factor in coordinate space.
Note that both the nonlocal Lagrangian in Eq.~(\ref{eq:j4}) and the local Lagrangian in Eq.~(\ref{eq:Llocal}) are invariant under the gauge transformations,
\begin{subequations}
\begin{eqnarray}
B(x) &\to& 
B'(x) = B(x) \exp\big[i e^q_B\, \theta(x)\big],
\\
T_\mu(x) &\to&
T_\mu'(x) = T_\mu(x) \exp\big[i e^q_T\, \theta(x)\big],
\\
\phi(x) &\to&
\phi'(x) = \phi(x) \exp\big[i e^q_\phi\, \theta(x)\big],
\end{eqnarray}
for the baryon and meson fields, and
\begin{equation}
\mathscr{A}^\mu(x) \to
\mathscr{A}'^\mu(x) = \mathscr{A}^\mu(x) + \partial^\mu \theta(x)
\end{equation}
\end{subequations}
for the electromagnetic field, where $\theta(x)$ is an auxiliary function.

\begin{table}[b]
\begin{center}
\caption{Coupling constants $C_{B\phi}$ and $C_{T\phi}$ for the $p B \phi$ and $p T \phi$ interactions, respectively, and $C_{\phi\phi^\dag}$ and $C'_{\phi\phi^\dag}$ for the $p p \phi \phi^\dag$ coupling, and the tree level magnetic moments $C_B^{\rm mag}$, $C_T^{\rm mag}$, $C_{BT}^{\rm mag}$ and $C_{\phi\phi^\dag}^{\rm mag}$, respectively, for all the allowed flavor channels.}
{\normalsize
\begin{tabular}{c|ccccccc} \hline
&&&&&&& \\
\hspace*{0.1cm}$\bm B$ \hspace*{0.1cm}
& \hspace*{0.1cm}$\bm p$ \hspace*{0.1cm}
& \hspace*{0.1cm}$\bm n$ \hspace*{0.1cm}
& \hspace*{0.1cm}$\bm \Sigma^+$\hspace*{0.1cm}
& \hspace*{0.1cm}$\bm \Sigma^0$\hspace*{0.1cm}
& \hspace*{0.1cm}$\bm \Sigma^-$\hspace*{0.1cm}
& \hspace*{0.1cm}$\bm \Lambda$\hspace*{0.1cm}	
& \hspace*{0.1cm}$\bm{\Lambda\Sigma^0}$\hspace*{0.1cm}
\\
\hspace*{0.1cm}$C_{B}^{\text {mag}}$\hspace*{0.1cm}
& \hspace*{0.1cm}$\frac13c_1 +c_2$\hspace*{0.1cm}
& \hspace*{0.1cm}$-\frac23 c_1$\hspace*{0.1cm}
& \hspace*{0.1cm}$\frac13c_1 +c_2$\hspace*{0.1cm}
& \hspace*{0.1cm}$\frac13 c_1$\hspace*{0.1cm}
& \hspace*{0.1cm}$\frac13 c_1-c_2$\hspace*{0.1cm}
& \hspace*{0.1cm}$-\frac13 c_1$\hspace*{0.1cm}	
& \hspace*{0.1cm}$\frac{1}{\sqrt{3}} c_1$\hspace*{0.1cm}	 
\\ 
&&&&&&& \\
\hline
&&&&&&& \\
\hspace*{0.1cm}$\bm T$\hspace*{0.1cm}
& \hspace*{0.1cm}$\bm \Delta^{++}$\hspace*{0.1cm}
& \hspace*{0.1cm}$\bm \Delta^+$\hspace*{0.1cm}
& \hspace*{0.1cm}$\bm \Delta^0$\hspace*{0.1cm}
& \hspace*{0.1cm}$\bm \Delta^-$ \hspace*{0.1cm}
& \hspace*{0.1cm}$\bm \Sigma^{*+}$\hspace*{0.1cm}
& \hspace*{0.1cm}$\bm \Sigma^{*0}$\hspace*{0.1cm}
& \hspace*{0.1cm}$\bm \Sigma^{*-}$\hspace*{0.1cm}
\\
\hspace*{0.1cm}$C_T^{\text {mag}}$\hspace*{0.1cm}
& \hspace*{0.1cm}$\frac23 F_2^T$\hspace*{0.1cm}
& \hspace*{0.1cm}$\frac13 F_2^T$\hspace*{0.1cm}
& \hspace*{0.1cm}$0$\hspace*{0.1cm}
& \hspace*{0.1cm}$-\frac13 F_2^T$\hspace*{0.1cm}
& \hspace*{0.1cm}$\frac13 F_2^T$\hspace*{0.1cm}
& \hspace*{0.1cm}$0$\hspace*{0.1cm}
& \hspace*{0.1cm}$-\frac13 F_2^T$\hspace*{0.1cm}	
\\ 
&&&&&&& \\
\hline
&&&&&&& \\
\hspace*{0.1cm}$\bm{BT}$\hspace*{0.1cm}
& \hspace*{0.1cm}$\bm{p\Delta^+}$\hspace*{0.1cm}
& \hspace*{0.1cm}$\bm{\Delta^0}$\hspace*{0.1cm}
& \hspace*{0.1cm}$\bm{\Sigma^+\Sigma^{*+}}$ \hspace*{0.1cm}
& \hspace*{0.1cm}$\bm{\Sigma^0\Sigma^{*0}}$\hspace*{0.1cm}
& \hspace*{0.1cm}$\bm{\Lambda\Sigma^{*0}}$\hspace*{0.1cm}
& \hspace*{0.1cm}$\bm{\Sigma^-\Sigma^{*-}}$\hspace*{0.1cm}
\\
\hspace*{0.1cm}$C_{BT}^{\text {mag}}$\hspace*{0.1cm}
& \hspace*{0.1cm}$-\frac{1}{\sqrt3}c_4$\hspace*{0.1cm}
& \hspace*{0.1cm}$-\frac{1}{\sqrt3}c_4$\hspace*{0.1cm}
& \hspace*{0.1cm}$\frac{1}{\sqrt3}c_4$\hspace*{0.1cm}
& \hspace*{0.1cm}$\frac{1}{2\sqrt3}c_4$\hspace*{0.1cm}
& \hspace*{0.1cm}$\frac12 c_4$\hspace*{0.1cm}
& \hspace*{0.1cm}$0$\hspace*{0.1cm}	
\\ 
&&&&&&& \\
\hline
&&&&&&& \\
\hspace*{0.1cm}$\bm{B\phi}$\hspace*{0.1cm}
& \hspace*{0.1cm}$\bm{p \pi^0}$\hspace*{0.1cm}
& \hspace*{0.1cm}$\bm{n \pi^+}$\hspace*{0.1cm}
& \hspace*{0.1cm}$\bm{\Sigma^+ K^0}$\hspace*{0.1cm}
& \hspace*{0.1cm}$\bm{\Sigma^0 K^+}$\hspace*{0.1cm}
& \hspace*{0.1cm}$\bm{\Lambda K^+}$\hspace*{0.1cm}	
\\
\hspace*{0.1cm}$C_{B\phi}$\hspace*{0.1cm}
& \hspace*{0.1cm}$\frac12 (D+F)$\hspace*{0.1cm}
& \hspace*{0.1cm}$\frac{1}{\sqrt2} (D+F)$\hspace*{0.1cm}
& \hspace*{0.1cm}$\frac{1}{\sqrt2} (D-F)$\hspace*{0.1cm}
& \hspace*{0.1cm}$\frac12 (D-F)$\hspace*{0.1cm}
& \hspace*{0.1cm}$-\frac{1}{\sqrt{12}} (D+3F)$\hspace*{0.1cm} 
\\ 
&&&&&&& \\
\hline
&&&&&&& \\
$\bm{T\phi}$
& $\bm{\Delta^0 \pi^+}$
& $\bm{\Delta^+ \pi^0}$
& $\bm{\Delta^{++} \pi^-}$
& $\bm{\Sigma^{*+} K^0}$
& $\bm{\Sigma^{*0} K^+}$				
\\
$C_{T\phi}$
& $-\frac{1}{\sqrt 6} {\cal C}$
& $-\frac{1}{\sqrt 3} {\cal C}$
& $\frac{1}{\sqrt 2} {\cal C}$
& $\frac{1}{\sqrt 6} {\cal C}$
& $-\frac{1}{\sqrt {12}} {\cal C}$
\\ 
&&&&&&& \\
\hline
&&&&&&& \\
$\bm{\phi\phi^\dag}$
& $\bm{\pi^+\pi^-}$
& $\bm{K^0 \overline{K}^0}$
& $\bm{K^+ K^-}$
&
&						\\
$C_{\phi\phi^\dag}$
& $\frac{1}{2}$
& $\frac{1}{2}$
& 1
&
&						\\
$C_{\phi\phi^\dag}'$
& $4(b_{10}+b_{11})$
& $4(b_{11}-b_{10})$
& $8b_{11}+2b_{9}$
&
&						\\
$C_{\phi\phi^\dag}^{\rm mag}$
& $-\frac{1}{2}(c_1+c_2)$
& $0$
& $-c_2$
&
&						\\
&&&&&&& \\
\hline
\end{tabular}
}
\label{tab:C}
\end{center}
\end{table}

The gauge link (\ref{eq:link}) can next be expanded in powers of the charge $e_\phi^q$,
\begin{eqnarray}
{\cal G}^q_\phi(x+b,x+a)
&=& \exp \Big[ -i e^q_\phi\, (a-b)^\mu
	      \int_0^1 \dd{t}\, \mathscr{A}_\mu\big(x+at+b(1-t)\big)
	 \Big]
\notag\\
&=& 1\ +\ \delta {\cal G}^q_\phi\
       +\ \cdots,
\label{eq:linkexpand}
\end{eqnarray}
using a change of variables	$z^\mu \to x^\mu + a^\mu\, t + b^\mu\, (1-t)$, % where the term of ${\cal O}(e^q_\phi)$ can be written as
\begin{eqnarray}
\delta {\cal G}^q_\phi
&=& -\ i e^q_\phi\, (a-b)^\mu
       \int_0^1 \dd{t}\, \mathscr{A}_\mu\big(x+at+b(1-t)\big).
\label{eq:deltaG}
\end{eqnarray}
This allows the nonlocal Lagrangian ${\cal L}^{\rm (nonloc)}$ in Eq.~(\ref{eq:j4}) to be further decomposed into free and interacting parts, with the latter consisting of
purely hadronic (${\cal L}^{\rm (nonloc)}_{\rm had}$), electromagnetic (${\cal L}^{\rm (nonloc)}_{\rm em}$), and gauge link (${\cal L}^{\rm (nonloc)}_{\rm link}$) contributions.
The hadronic and electromagnetic interaction parts of ${\cal L}^{\rm (nonloc)}$ are obtained from the first term in Eq.~(\ref{eq:linkexpand}), and given by
\begin{eqnarray}
{\cal L}^{\rm (nonloc)}_{\rm had}(x)
&=& \bar{p}(x)
    \left[ \frac{C_{B\phi}}{f}\, \gamma^\mu \gamma_5 B(x)
	 + \frac{C_{T\phi}}{f}\, \Theta^{\mu\nu} T_\nu(x)
    \right]
    \!\int\!\dd[4]{a}\, F(a)\, \partial_\mu \phi(x+a)
	+ {\rm H.c.}
\notag\\
& & \hspace*{-2.5cm}
 +\ \frac{iC_{\phi\phi^\dag}}{2f^2}
    \bar{p}(x) \gamma^\mu p(x)
    \int\!\dd[4]{a}\!\int\!\dd[4]{b}\ F(a) F(b)
    \left[ \phi(x+a) \partial_\mu \phi^\dag(x+b)
	 - \partial_\mu \phi(x+a) \phi^\dag(x+b)
    \right]
\notag \\
& & \hspace*{-2.5cm}
 +\ \frac{iC_{\phi\phi^\dag}'}{2f^2}
    \bar{p}(x) \sigma^{\mu\nu} p(x)
    \int\!\dd[4]{a}\!\int\!\dd[4]{b}\ F(a) F(b)
    \left[ \partial_\mu\phi(x+a) \partial_\nu \phi^\dag(x+b)
	 - \partial_\mu \phi(x+a) \partial_\nu \phi^\dag(x+b)
    \right],
\notag\\
& &
\label{eq:Lnonloc_had}
\end{eqnarray}
and
\begin{eqnarray}
{\cal L}^{\rm (nonloc)}_{\rm em}(x)
&=& e^q_B\, \bar{B}(x) \gamma^\mu B(x)\, \mathscr{A}_\mu(x)\
 +\ e^q_T\, \overline{T}_\mu(x) \gamma^{\mu\nu\alpha} T_\nu(x)\,
	    \mathscr{A}_\alpha(x)
\notag\\
& & \hspace*{-2.3cm}
 + i e^q_\phi \left[ \partial^\mu \phi(x) \phi^\dag(x)
		    - \phi(x) \partial^\mu \phi^\dag(x)
	       \right] \mathscr{A}_\mu(x)
\notag\\
& & \hspace*{-2.3cm}
 - i e^q_\phi\, \bar{p}(x)
    \left[
      \frac{C_{B\phi}}{f}\, \gamma^\mu \gamma_5 B(x)
    + \frac{C_{T\phi}}{f}\, \Theta^{\mu\nu} T_\nu(x)
    \right]
    \int\!\dd[4]{a}\, F(a)\, 
    \phi(x+a) \mathscr{A}_\mu(x) + {\rm H.c.}
\notag\\
& & \hspace*{-2.3cm}
 -  \frac{e^q_\phi C_{\phi\phi^\dag}}{2f^2}\,
    \bar{p}(x) \gamma^\mu p(x)
    \int\!\dd[4]{a} \int\!\dd[4]{b}\ 
    F(a) F(b)\, \phi(x+a) \phi^\dag(x+b)\mathscr{A}_\mu(x)
\notag\\
& & \hspace*{-2.3cm}
 -  \frac{C_{\phi\phi^\dag}'}{f^2}\,
    \bar{p}(x) \sigma^{\mu\nu} p(x) 
    \int\!\dd[4]{a} \int\!\dd[4]{b}\ 
    F(a) F(b)\, \phi(x+a) \partial_\nu\phi^\dag(x+b) \mathscr{A}_\mu(x)
\notag\\
& & \hspace*{-2.3cm}
 +  \frac{C_B^{\rm mag}}{4 M_B}
    \bar B(x) \sigma^{\mu\nu} B(x) F_{\mu\nu}(x)
 +  \frac{i C_{BT}^{\rm mag}}{4 M_B}
    \bar B(x) \gamma^\mu\gamma_5 T^\nu (x) F_{\mu\nu}(x)	
 -  \frac{C_T^{\rm mag}}{4 M_T}
    \overline T_\alpha(x) \sigma^{\mu\nu} T^\alpha(x) F_{\mu\nu}(x)	
\notag\\
& & \hspace*{-2.3cm}
 +  \frac{C_{\phi\phi^\dag}^{\rm mag}}{4 M_B f^2}\,
    \bar{p}(x) \sigma^{\mu\nu} p(x) \int\!\dd[4]{a} \int\!\dd[4]{b}\,
    F(a) F(b)\, \phi(x+a) \phi^\dag(x+b) F_{\mu\nu}(x),
\label{eq:Lnonloc_em}
\end{eqnarray}
respectively.
The second term in Eq.~(\ref{eq:linkexpand}), which explicitly depends on the gauge link, gives rise to an additional contribution to the Lagrangian density that can be expanded as
\begin{eqnarray}
\hspace*{-1cm}{\cal L}^{\rm (nonloc)}_{\rm link}(x)
&=& -i e^q_\phi\, \bar{p}(x)
    \left[ \frac{C_{B\phi}}{f}\, \gamma^\rho \gamma_5 B(x)
	     + \frac{C_{T\phi}}{f}\, \Theta^{\rho\nu} T_\nu(x)
    \right]
\notag\\
& & \hspace*{0.3cm} \times
    \int_0^1 \dd{t} \int\!\dd[4]{a}\,  %\int\!d^4b \, 
    F(a) a^\mu\, \partial_\rho (\phi(x+a) \mathscr{A}_\mu\big(x+at)\big)
    + {\rm H.c.}
\notag\\
&+& \frac{e^q_\phi C_{\phi\phi^\dag}}{2f^2}\,
    \bar{p}(x) \gamma^\rho p(x)
    \int_0^1 \dd{t} \int\!\dd[4]{a}\! \int\!\dd[4]{b}\, %\int\!d^4c\,
	F(a)\, F(b)\, (a-b)^\mu\,
\notag\\
& & \hspace*{0.3cm} \times
    \left[ \phi(x+a) \partial_\rho \phi^\dag(x+b)
	     - \partial_\rho \phi(x+a) \phi^\dag(x+b)
    \right] \mathscr{A}_\mu\big(x+at+b(1-t)\big).
\label{eq:Lnonloc_link}
\end{eqnarray}
Finally, the quark current for the nonlocal theory can be written as a sum of two terms, from the usual electromagnetic current obtained from Eq.~(\ref{eq:Lnonloc_em}) with minimal substitution, $J_{q, \rm em}^\mu$, and from the additional term associated with the gauge link, $\delta J_q^\mu$,
\begin{equation}
J_{q, \rm em}^\mu(x)
\equiv \frac{\delta \int\!\dd[4]{y}\,
		{\cal L}_{\rm em}^{\rm (nonloc)}(y)}
	      {\delta \mathscr{A_\mu}(x)}, 
\qquad
\delta J_q^\mu(x)
\equiv \frac{\delta \int\!\dd[4]{y}\, {\cal L}^{\rm (nonloc)}_{\rm link}(y)}
              {\delta \mathscr{A_\mu}(x)},
\end{equation}
where, explicitly,
\begin{eqnarray}
\label{eq:Jem}
J_{q, \rm em}^\mu(x)
&=& e^q_B\,\bar{B}(x) \gamma^\mu B(x)
 +  e^q_T\, \overline{T}_\alpha(x) \gamma^{\alpha\nu\mu} T_\nu(x)
 +  ie^q_\phi \big[ \partial^\mu \phi(x) \phi^\dag(x)
                  - \phi(x) \partial^\mu \phi^\dag(x)
              \big]
\notag\\
&-& i e^q_\phi\,
\bigg(    \int\!\dd[4]{a} F(a)\, \bar{p}(x)
    \left[ \frac{C_{B\phi}}{f}\, \gamma^\mu \gamma_5 B(x)
	     + \frac{C_{T\phi}}{f}\, \Theta^{\mu\nu} T_\nu(x)
    \right] \phi(x+a)\ +\ {\rm H.c.}
\bigg)
\notag\\
&-& \frac{e^q_\phi C_{\phi\phi^\dag}}{2f^2}
    \int\!\dd[4]{a} \int\!\dd[4]{b} F(a) F(b)\,
    \bar{p}(x) \gamma^\mu p(x)\, \phi(x+a) \phi^\dag(x+b)
\notag\\
&+& \frac{C_B^{\rm mag}}{2 M_B}\, \int\!\dd[4]{a} F(a)\,
    \partial_\nu\big(\bar{p}(x) \sigma^{\mu\nu} p(x) \big)
 -  \frac{C_T^{\rm mag}}{2 M_T}\, \int\!\dd[4]{a} F(a)\,
    \partial_\nu\big(\overline {T}_\alpha(x) \sigma^{\mu\nu} T^\alpha(x) \big )
\notag\\
&+& \frac{i C_{BT}^{\rm mag}}{4 M_B}\, 
    ( \partial_\nu\big(\bar{p}(x) \gamma^\nu \gamma_5 T^\mu(x) \big)
    - \partial_\nu\big(\bar{p}(x) \gamma^\mu \gamma_5 T^\nu(x) \big)
\notag\\
&+& \frac{C_{\phi\phi^\dag}^{\rm mag}}{2 M_B f^2}\,
    \int\!\dd[4]{a} \int\!\dd[4]{b} F(a) F(b)\,
   \partial_\nu \big(\bar{p}(x) \sigma^{\mu\nu} p(x) \phi(x+a) \phi^\dag(x+b) \big),
\\
& &
\notag\\
\delta J_q^\mu(x)
&=& i e^q_\phi
    \int_0^1 \dd{t} \int\!\dd[4]{a} F(a)\, a^\mu\,
\notag\\
& & \times
    \partial_\rho 
    \left( \bar{p}(x-at)
    \left[ \frac{C_{B\phi}}{f}\, \gamma^\rho \gamma_5 B(x-at)\,
       +\, \frac{C_{T\phi}}{f}\, \Theta^{\rho\nu} T_\nu(x-at)
    \right]
    \right) \phi(x+a\bar{t}\big) + {\rm H.c.}
\notag\\
&-& \frac{e^q_\phi C_{\phi\phi^\dag}}{2f^2}
    \int_0^1 \dd{t}\!\int\!\dd[4]{a}\!\int\!\dd[4]{b} F(a) F(b)\, (a-b)^\mu\,
\notag\\
& & \times
\bigg[ 
    \partial_\rho
    \Big(
        \bar{p}\big(x-at-b\bar{t}\big) \gamma^\rho p\big(x-at-b\bar{t}\big)
    \phi\,\big(x+(a-b)\bar{t}\big)
    \Big)
    \phi^\dag\big(x-(a-b)t)
\notag\\
& & \hspace*{0.2cm}
 -  \partial_\rho
    \Big( 
        \bar{p}\big(x-at-b\bar{t}\big) \gamma^\rho p\big(x-at-b\bar{t}\big)
    \phi^\dag\big(x-(a-b)t\big)
	\Big)
    \phi\big(x+(a-b)\bar{t}\big)
\bigg],
\label{eq:Jlink}
\end{eqnarray}
with $\bar t \equiv 1-t$.
Compared with the local theory, Eqs.~(\ref{eq:j1}) and (\ref{eq:jq}), the nonlocal  formulation in Eqs.~(\ref{eq:Lnonloc_had})--(\ref{eq:Jlink}) includes the regulator function $F(a)$.
In the limit where $F(a) \to \delta^{(4)}(a)$, which corresponds to taking the momentum space form factor to unity, the local limit can be obtained from the nonlocal result.
In the next section we will apply the nonlocal interaction derived here to compute the hadronic splitting functions for protons transitioning to baryons and pseudoscalar mesons.

%%%%%%%%%%%%%%%%%%%%%%%%%%%%%%%%%%%%%%%%%%%%%%%%%%%%%%%%%%%%%%%%%%%%%%%%%%%%%%%%%%%%
\section{Nonforward splitting functions}
\label{sec.splitting}

The GPDs for a quark flavor $q$ in a proton with initial momentum $p$ and final momentum $p'$ are defined by the Fourier transform of the matrix elements of the quark bilocal field operators $\psi_q$ as~\cite{Ji:1996nm}
\begin{eqnarray}
\label{eq:GPD}
\int_{-\infty}^{\infty}\frac{\dd{\lambda}}{2\pi} e^{-ix\lambda}
\langle p'| 
    \bar\psi_q (\tfrac12\lambda n) \slashed{n}\, \psi_q(-\tfrac12\lambda n)
|p \rangle 
&=& \bar u(p') 
  \Big[ \slashed n H^q(x,\xi,t) 
      + \frac{i\sigma^{\mu\nu}n_\mu \Delta_\nu}{2M}\, E^q(x,\xi,t)
  \Big] u(p),
\notag\\
& &
\end{eqnarray}
where $n_\mu$ is the light-cone vector which projects the ``plus" component of momenta and $\lambda$ is a dimensionless parameter.
From Lorentz invariance, the Dirac ($H^q$) and Pauli ($E^q$) GPDs are typically written as functions of the light-cone momentum fraction $x$ of the proton carried by the initial quark with momentum $k_q$ and the skewness parameter $\xi$, which are defined as
\begin{equation}
x\,   \equiv\, \frac{k_q^+}{P^+}, \qquad
\xi\, \equiv\, -\frac{\Delta^+}{2P^+},
\end{equation}
where
\begin{equation}
P\, =\, \frac12 \big(p + p'\big), \qquad
\Delta\, =\, p' - p,
\end{equation}
are the average of the initial and final proton momenta and the momentum difference, respectively. The light-cone components $k^+$ and $k^-$ of any four-vector $k^\mu$ are defined 
as $k^+ = \frac{1}{\sqrt{2}}(k^0+k^3)$ and $k^- = \frac{1}{\sqrt{2}}(k^0-k^3)$.
The GPDs are also functions of the hadronic four-momentum transfer squared, $t \equiv \Delta^2$. % = (p'-p)^2$.
The dependence of the GPDs on the fourth variable, typically taken to be the four-momentum transfer squared from the incident lepton, $Q^2$, is suppressed.
%
%In addition, $x = \frac{k^+}{P^+}$ is the quark longitudinal momentum fraction and $P=\frac{p'+p}2$, while $\xi$ is the skewness parameter, defined as $\xi = -\frac{p'^+-p^+}{2P^+} = -\frac{\Delta^+}{2P^+}$.

Integrating the $H^q$ and $E^q$ GPDs over $x$, one obtains the Dirac and Pauli form factors for a given quark flavor $q$, respectively,
\begin{eqnarray}
F_1^q(t) = \int_{-1}^1\,\dd{x} H^q(x,\xi,t), \qquad
F_2^q(t) = \int_{-1}^1\,\dd{x} E^q(x,\xi,t),
\label{eq.F12q}
\end{eqnarray}
and summing over the quark flavors gives the nucleon Dirac, $F_1^N$, and Pauli, $F_1^N$, form factors,
\begin{eqnarray}
F_{1,2}^N(t) = \sum_q e_q\, F_{1,2}^{\, q}(t).
\end{eqnarray}
The combination of these form factors can generate the usual Sachs electric and magnetic form factors as 
\begin{eqnarray}\label{EM}
G_E^N(t)=F_1^N(t)+{t\over{4M^2}}F_2^N(t), \qquad
G_M^N(t)=F_1^N(t)+F_2^N(t).
\label{eq.GEMN}
\end{eqnarray}
In our calculations we consider only non-skewed GPDs, and henceforth set $\xi=0$, in which case the hadron momenta are parametrized as~\cite{Brodsky:1997de}
\begin{eqnarray}
% \Delta^\mu &=& (0,\,0,\,\bm \Delta_\perp),\, \nonumber\\
p^\mu   &=& \Big(P^+,P^-,-\frac12 {\bm \Delta_\perp}\Big), \qquad
p'^\mu\, =\, \Big(P^+,P^-,\frac12 {\bm \Delta_\perp}\Big),
\end{eqnarray}
where the momentum transfer $\Delta^\mu$ is purely in the transverse direction.

In the application of our nonlocal EFT framework to GPDs, we need to compute the nonforward splitting functions defined by the matrix elements of the hadronic level currents derived in the previous section.
The electromagnetic vertex is given by
\begin{equation}
\label{eq.emvertex}
\langle N(p')|J^{\mu}|N(p) \rangle\,
=\, \bar{u}(p')
    \Big[ \gamma^\mu F_1^N(t) + \frac{i\sigma^{\mu\nu}\Delta_\nu}{2M}F_2^N(t)
    \Big]
    u(p)\,
\equiv
\int\!\dd[4]{k} \widetilde{\Gamma}^\mu(k),
\end{equation}
where the integrand $\widetilde{\Gamma}^\mu (k)$ depends on the internal meson momentum, $k$.
Defining the light-cone momentum fraction of the target nucleon carried by the interacting hadron, $y = k^+/P^+$, the Dirac-like splitting function $f(y,t)$ and Pauli-like splitting function $g(y,t)$ are related to the vertex by
\begin{equation}
\bar{u}(p')
\Big[ \gamma^+ f(y,t) + \frac{i\sigma^{+\nu}\Delta_\nu}{2M}\, g(y,t)
\Big] 
u(p)
= \int \dd[4]{k} \widetilde{\Gamma}^+(k)\, \delta\Big(y-\frac{k^+}{P^+}\Big)\,
\equiv\, \Gamma^+.
\label{eq.Gamma+def}
\end{equation}
One can easily verify that the integral of the splitting functions over $y$ leads to the corresponding form factors in (\ref{eq.emvertex}).

\begin{figure}[tp]
\begin{center}
\includegraphics[scale=0.85]{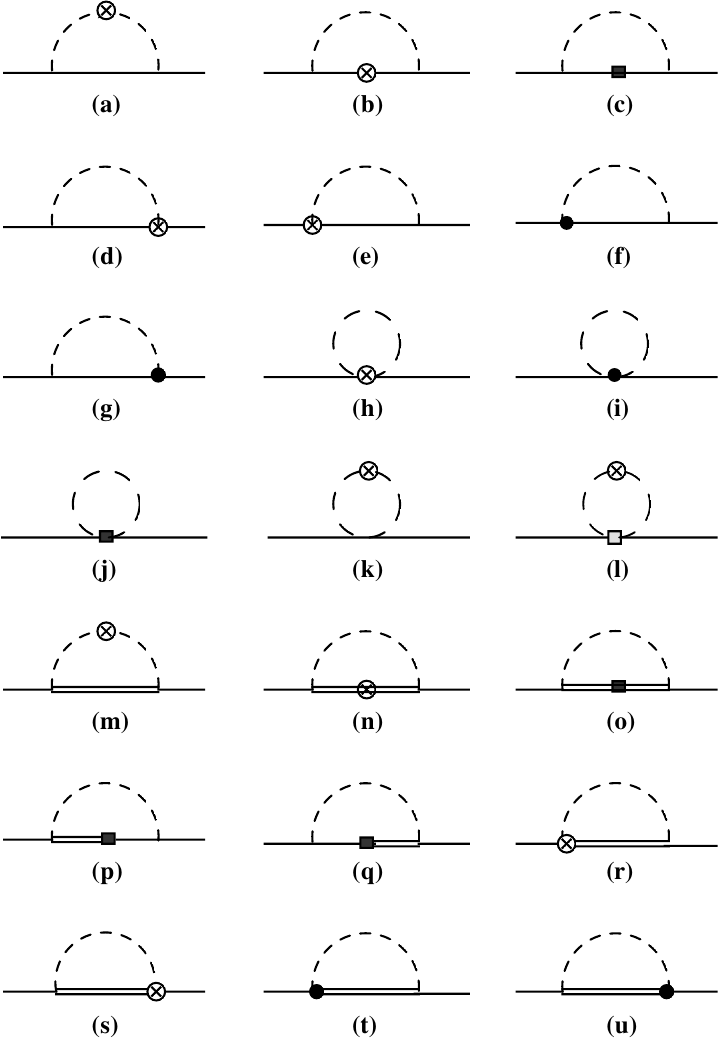}
\caption{One-loop diagrams for the proton to pseudoscalar meson (dashed lines) and octet baryon (solid lines) or decuplet baryon (double solid lines) splitting functions up to the fourth chiral order:
{\bf (a)--(c)} octet baryon rainbow diagrams,
{\bf (d)--(g)} octet baryon Kroll-Ruderman diagrams,
{\bf (h)--(j)} tadpole diagrams,
\mbox{{\bf (k)--(l)} bubble} diagrams,
{\bf (m)--(o)} decuplet baryon rainbow diagrams,
{\bf (p)--(q)} octet-decuplet transition rainbow diagrams,
{\bf (r)--(u)} decuplet baryon Kroll-Ruderman diagrams.
The crossed circles ($\otimes$) represent the interaction with external vector field from the minimal substitution, 
filled circles ({\Large$\bullet$}) denote additional gauge link interaction with the external field,
black squares ($\blacksquare$) represent the magnetic interaction in Eq.~(\ref{lomag}), and
gray squares (${\textcolor{gray}{\blacksquare}}$) denote the interaction in Eq.~(\ref{adm}).} 
\label{diagrams}
\end{center}
\end{figure}

The diagrams that are relevant for the calculation of the one-meson loop contributions to GPDs up to the fourth chiral order are shown in Fig.~\ref{diagrams}.
In the following we outline the calculation of the corresponding splitting functions, beginning with the diagrams involving only octet baryons [Fig.~\ref{diagrams}(a)--\ref{diagrams}(l)], and then presenting results for contributions with intermediate decuplet baryons.
Since the final results, after integration, for the latter are rather lengthy, we collect the complete expressions for these in Appendix~\ref{sec.appendix}.

\clearpage
% ............................................................................
\subsection{Octet baryon intermediate states}
\label{ssec.fy_octet}

Starting with the octet baryon rainbow diagram in Fig.~\ref{diagrams}(a), in which the external field couples to the meson, the contribution of this diagram to the matrix element $\Gamma^+$ in Eq.~(\ref{eq.Gamma+def}) is given by
\begin{eqnarray} 
\Gamma_{(\rm a)}^+
&=& \bar{u}(p') 
\frac{C^2_{B\phi}}{f^2}
\int\!\frac{\dd[4]{k}}{(2\pi)^4}
(\slashed{k}+\slashed{\Delta})\gamma_5\,
\widetilde{F}(k+\Delta) 
\frac{i}{D_\phi(k+\Delta)}\, 2k^+                 \nonumber\\
&&\hspace*{2.3cm}
\times \frac{i}{D_\phi(k)} \frac{i(\slashed{p}-\slashed{k}+M_B)}{D_B(p-k)}
\gamma_5 \slashed{k}\, 
\widetilde{F}(k)\,
\delta\Big(y-\frac{k^+}{p^+}\Big)\,
u(p)                                            \label{eq.Gamma_a}\\
&\equiv& \bar{u}(p')
\bigg[
  \gamma^+ f_{\phi B}^{({\rm rbw})}(y,t)
  + \frac{i\sigma^{+\nu}\Delta_\nu}{2M} g_{\phi B}^{({\rm rbw})}(y,t)
\bigg] u(p),
\label{eq.splitfnproj}
\end{eqnarray}
where the propagator factors $D_\phi(k)$ and $D_B(p)$ are defined as
\begin{equation} 
D_\phi(k) = k^2-m_\phi^2+i\epsilon, \qquad
D_B(p) = p^2-M_B^2+i\epsilon,
\end{equation}
and $m_\phi$ is the meson mass.
The function $\widetilde{F}$ regulates the ultraviolet divergence in the loop integration, and for simplicity is chosen to be a function of the meson momentum only (see Sec.~\ref{sec.numerical} below).
After simplifying the combinations of Dirac $\gamma$ matrices in Eq.~(\ref{eq.Gamma_a}), the ``Dirac'' $f_{\phi B}^{({\rm rbw})}(y,t)$ and ``Pauli'' $g_{\phi B}^{({\rm rbw})}(y,t)$ splitting functions can be written as %
\begin{subequations}
\label{eq:split_rbw_phiB}
\begin{eqnarray}
\label{eq:split_rbw_fphiB}
f_{\phi B}^{({\rm rbw})}(y,t)
&=&\frac{C_{B\phi}^2}{f^2}
\int\!\frac{\dd[4]{k}}{(2\pi)^4}
\frac{-iF_{\phi B}^{({\rm rbw})}}{D_B(p-k) D_\phi(k+\Delta) D_\phi(k)}
\widetilde{F}(k) \widetilde{F}(k+\Delta)\,
\delta\Big(y-\frac{k^+}{p^+}\Big),
\notag\\
&& \\
g_{\phi B}^{({\rm rbw})}(y,t)
\label{eq:split_rbw_gphiB}
&=&\frac{C_{B\phi}^2}{f^2}
\int\!\frac{\dd[4]{k}}{(2\pi)^4}
\frac{-iG_{\phi B}^{({\rm rbw})}}{D_B(p-k)D_\phi(k+\Delta) D_\phi(k)}
\widetilde{F}(k) \widetilde{F}(k+\Delta)\,
\delta\Big(y-\frac{k^+}{p^+}\Big),
\notag\\
&&
\end{eqnarray}
\end{subequations}
where the trace factors in the numerator of the integrand are given by
\begin{subequations}
\begin{eqnarray}
F_{\phi B}^{({\rm rbw})}
&=& \frac12\, y k^2\, \big( 4 k\cdot P + 4 M \MBbar + y\, t \big)
-\, y\, \big( 2 (k \cdot p)^2 - M\MBbar\, (2 k \cdot \Delta + y\, t) \big)
\nonumber\\
&& 
+\, k\cdot p\, \big( 2k \cdot p' + y\, t \big),
\\
G_{\phi B}^{({\rm rbw})}
&=& -2 y M\,
%\big( \MBbar\, (2 y M^2 - k \cdot p) - (\MBbar + 2 y M)\, k\cdot p 
% \big( 2 y M^2 \MBbar - 2 (\MBbar + 2 y M)\, k\cdot p + k^2\, (\MBbar + y M)
\big( 2 y M^2 \MBbar + (\MBbar + 2 y M) (k^2 - 2k\cdot p) \big),
\end{eqnarray}
\end{subequations}
with $\MBbar \equiv M + M_B$. %\\ \\
%
%\begin{eqnarray}
%\label{eq:MBbar}
%\MBbar = M_B + M. %, \qquad \Delta_B=M_B-M.
%\end{eqnarray}

%\begin{eqnarray}
%\label{eq:deltaB}
%\Delta_B = M_B - M.
%\end{eqnarray}

For the baryon rainbow diagram in Fig.~\ref{diagrams}(b), the photon couples to the intermediate octet baryon through an electric vertex. 
The contribution of this diagram to the nonforward matrix element $\Gamma^+$ is given by
\begin{eqnarray}  
\Gamma_{\rm (b)}^+
&=&\bar{u}(p')\, \frac{C^2_{B\phi}}{f^2} 
\int\!\frac{\dd[4]{k}}{(2\pi)^4}\, \slashed{k}\gamma_5\,\widetilde{F}(k)\frac{i}{D_\phi(k)}
\frac{i(\slashed{p'}-\slashed{k}+M_B)}{D_B(p'-k)} \gamma^+
\frac{i(\slashed{p}-\slashed{k}+M_B)}{D_B(p-k)}
\nonumber\\
&&\qquad\qquad\qquad\quad
\times\,
\gamma_5\slashed{k}\, \widetilde{F}(k)\,
\delta\Big(y-\frac{k^+}{p^+}\Big)\,
u(p).
\end{eqnarray} 
Using a similar projection as in Eq.~(\ref{eq.splitfnproj}), the splitting functions $f_{B\phi}^{({\rm rbw})}(y,t)$ and $g_{B\phi}^{({\rm rbw})}(y,t)$ are obtained as 
\begin{subequations}
\begin{eqnarray}
f_{B\phi}^{({\rm rbw})}(y,t)
&=& \frac{C^2_{B\phi}}{f^2} 
\int\!\frac{\dd[4]{k}}{(2\pi)^4}
\frac{-i F_{B\phi}^{({\rm rbw})}}{D_B(p'-k)D_B(p-k)D_\phi(k)}\,
\widetilde{F}^2(k)\,
\delta\Big(y-\frac{k^+}{p^+}\Big),
\\
g_{B\phi}^{({\rm rbw})}(y,t)
&=& \frac{C^2_{B\phi}}{f^2} 
\int\!\frac{\dd[4]{k}}{(2\pi)^4}
\frac{-iG_{B\phi}^{({\rm rbw})}}{D_B(p'-k)D_B(p-k)D_\phi(k)}\,
\widetilde{F}^2(k)\,
\delta\Big(y-\frac{k^+}{p^+}\Big),
\end{eqnarray}
\end{subequations}
where the trace factors in the integrand are given by
\begin{subequations}
\begin{eqnarray}
F_{B\phi}^{({\rm rbw})}
&=& -k\cdot p\, \big( 4k \cdot p' + y\MBbar \Delta_B \big)
+ k^2 \big( 4 k\cdot P + \MBbar (\MBbar - 2y M)\big) 
\nonumber\\
&&
-\, \frac12 y\MBbar 
    \big(y \MBbar\, t + 2\Delta_B\, k\cdot p' \big) - k^4,
\\
G_{B\phi}^{({\rm rbw})}
&=&-\frac{2M \MBbar}{t}
\Big( 2(k\cdot \Delta)^2 
    + y\, t\, \big( 4 k\cdot p + k\cdot p' - k^2 - y M \MBbar \big)
\Big),
\end{eqnarray}
\end{subequations}
with $\Delta_B \equiv M_B - M$.

The diagram in Fig.~\ref{diagrams}(c) involves the magnetic photon-baryon interaction. 
The contribution of this diagram to $\Gamma^+$ can be written as 
\begin{eqnarray}
\Gamma_{\rm (c)}^+
&=& \bar{u}(p')\, \frac{C^2_{B\phi}}{f^2}
\int\!\frac{\dd[4]{k}}{(2\pi)^4}\,
\slashed{k}\gamma_5 \widetilde{F}(k)
\frac{i(\slashed{p'}-\slashed{k}+M_B)}{D_B(p'-k)}
\frac{i\sigma^{+\nu}\Delta_\nu}{2 M_B}
\frac{i(\slashed{p}-\slashed{k}+M_B)}{D_B(p-k)}
\frac{i}{D_\phi(k)}
\nonumber\\
&& \qquad\qquad\qquad\quad
\times\,
\gamma_5\slashed{k}\,\widetilde{F}(k)\,
\delta\Big(y-\frac{k^+}{p^+}\Big)\,
u(p).
\end{eqnarray}
The splitting functions $f_{B\phi}^{({\rm rbw\, mag})}(y,t)$ and $g_{B\phi}^{({\rm rbw\, mag})}(y,t)$ in this case are given by
\begin{subequations}
\begin{eqnarray}
f_{B\phi}^{({\rm rbw\, mag})}(y,t)
&=& \frac{C^2_{B\phi}}{2 f^2}
\int\!\frac{\dd[4]{k}}{(2\pi)^4}
\frac{-iF_{B\phi}^{({\rm rbw\, mag})}}{D_B(p'-k)D_B(p-k)D_\phi(k)}
\widetilde{F}^2(k)\,
\delta\Big(y-\frac{k^+}{p^+}\Big),
\\
g_{B\phi}^{({\rm rbw\, mag})}(y,t)
&=& \frac{C^2_{B\phi}}{2 f^2}
\int\!\frac{\dd[4]{k}}{(2\pi)^4}
\frac{-iG_{B\phi}^{({\rm rbw\, mag})}}{D_B(p'-k)D_B(p-k)D_\phi(k)}
\widetilde{F}^2(k)\,
\delta\Big(y-\frac{k^+}{p^+}\Big),
\end{eqnarray}
\end{subequations}
where 
\begin{subequations}
\begin{eqnarray}
F_{B\phi}^{({\rm rbw\, mag})}
&=& \frac{\MBbar}{M_B}
\Big( 2 (k\cdot \Delta)^2 
    + y\, t \big( 2 k\cdot P - k^2 - y\, M \MBbar \big)
\Big),
\\
G_{B\phi}^{({\rm rbw\, mag})}
&=& -\frac{2M}{M_B\, t}
\Big( 
2 \MBbar^2\, (k\cdot \Delta)^2
+ 2 y \MBbar (2M+\MBbar)\, t\, k \cdot P
- 2 y^2 M^2 \MBbar^2\, t
\nonumber\\
&& \qquad\qquad
-\, 4\, t\, k\cdot p~k\cdot p'
+\, t k^2 \big( 4k\cdot P - \MBbar (\MBbar + 2yM) + k^2 \big) 
\Big).
\end{eqnarray}
\end{subequations}
The magnetic coupling constant $C_B^{\rm mag}$ does not appear in the splitting functions, but rather is included in the input GPDs.

The photon couples to the pseudoscalar meson and nucleon in Figs.~\ref{diagrams}(d) and 1(e), as first pointed out by Kroll and Ruderman in their study photo-meson production near threshold~\cite{Kroll:1953vq}.
The contribution of these two diagrams to the matrix element $\Gamma^+$ is given by
%
%\begin{eqnarray} 
%\Gamma_{d+e}^{\mu(p)}&=&-(D+F)^2I_{(d+e)\pi}^{NN}-\frac{(3F+D)^2}{6}I_{(d+e)K}^{N\Lambda}-\frac{(D-F)^2}{2}I_{(d+e)K}^{N\Sigma},\\
%\Gamma_{d+e}^{\mu(n)}&=&(D+F)^2I_{(d+e)\pi}^{NN}-(D-F)^2I_{(d+e)K}^{N\Sigma},
%\end{eqnarray} 
%
\begin{eqnarray}
\Gamma_{\rm (d)+(e)}^+
&=& \bar{u}(p')\,
\frac{C^2_{B\phi}}{f^2}
\int\!\frac{\dd[4]{k}}{(2\pi)^4}
\bigg[
  \slashed{k}\gamma_5
  \frac{i(\slashed{p'}-\slashed{k}+M_B)}{D_B(p'-k)}
  i\gamma_5\gamma^+
- i\gamma^+\gamma_5
  \frac{i(\slashed{p}-\slashed{k}+M_B)}{D_B(p-k)}
  \gamma_5\slashed{k}
\bigg]
\nonumber\\
& & \hspace*{3.2cm} \times
\frac{i}{D_\phi(k)}\, 
\widetilde{F}^2(k)\,
\delta\Big(y-\frac{k^+}{p^+}\Big)\,
u(p).
\end{eqnarray} 
The corresponding Kroll-Ruderman splitting functions $f^{({\rm KR})}_{B\phi}(y,t)$ and $g^{({\rm KR})}_{B\phi}(y,t)$ are then defined as
\begin{subequations}
\begin{eqnarray}
f^{({\rm KR})}_{B\phi}(y,t)
&=&\frac{C^2_{B\phi}}{f^2}
\int\!\frac{\dd[4]{k}}{(2\pi)^4}
\bigg[
  \frac{-iF^{({\rm KR\, 1})}_{B\phi}}{D_B(p'-k)}
+ \frac{-iF^{({\rm KR\, 2})}_{B\phi}}{D_B(p-k)}
\bigg]
\frac{1}{D_\phi(k)}
\widetilde{F}^2(k)\,
\delta\Big(y-\frac{k^+}{p^+}\Big),
\\
g^{({\rm KR})}_{B\phi}(y,t)
&=&\frac{C^2_{B\phi}}{f^2}
\int\!\frac{\dd[4]{k}}{(2\pi)^4}
\bigg[
  \frac{-iG^{({\rm KR\, 1})}_{B\phi}}{D_B(p'-k)}
+ \frac{-iG^{({\rm KR\, 2})}_{B\phi}}{D_B(p-k)}
\bigg]
\frac{1}{D_\phi(k)}
\widetilde{F}^2(k)\,
\delta\Big(y-\frac{k^+}{p^+}\Big),
\end{eqnarray}
\end{subequations}
where the trace factors in the integrands are given by
\begin{subequations}
\begin{eqnarray}
F^{({\rm KR\, 1})}_{B\phi}
&=& k^2 - 2 k\cdot p' + y M \MBbar,
\\
F^{({\rm KR\, 2})}_{B\phi}
&=& k^2 - 2 k\cdot p\, + y M \MBbar,
\\
G^{({\rm KR\, 1})}_{B\phi}
&=& +\frac{M \MBbar}{t} \big( 2k\cdot \Delta - y\, t \big),
\\
G^{({\rm KR\, 2})}_{B\phi}
&=& -\frac{M \MBbar}{t} \big( 2k\cdot \Delta + y\, t \big).
\end{eqnarray}
\end{subequations}

The additional Kroll-Ruderman diagrams generated from the expansion of the gauge link terms are shown in Fig.~\ref{diagrams}(f) and \ref{diagrams}(g), and are important to ensure that the renormalized charge of the proton (neutron) is 1 (0).
The contribution of these two additional diagrams with intermediate octet baryons is expressed as
%\begin{eqnarray} 
%\Gamma_{f+g}^{\mu(p)}&=&-(D+F)^2I_{(f+g)\pi}^{NN}-\frac{(3F+D)^2}{6}I_{(f+g)K}^{N\Lambda}-\frac{(D-F)^2}{2}I_{(f+g)K}^{N\Sigma},\\
%\Gamma_{f+g}^{\mu(n)}&=&(D+F)^2I_{(f+g)\pi}^{NN}-(D-F)^2I_{(f+g)K}^{N\Sigma},
%\end{eqnarray} 
%where
%
\begin{eqnarray}
\Gamma_{\rm (f)+(g)}^+
&=&\bar{u}(p')
\frac{C^2_{B\phi}}{f^2}
\int\!\frac{\dd[4]{k}}{(2\pi)^4}\,
\widetilde{F}(k)
\nonumber\\
&& \hspace*{-1.8cm}
\times
\bigg[
\slashed{k} \gamma_5
\frac{i(\slashed{p'} - \slashed{k}+M_B)}{D_B(p'-k)}
\frac{i}{D_\phi(k)}
(\slashed{k}-\slashed{\Delta})
\gamma_5
\frac{2ik^+}{2k\cdot \Delta-t}\,
\big( \widetilde{F}(k-\Delta)-\widetilde{F}(k) \big)
\\
& & \hspace*{-1.5cm}
+\, (\slashed{k}+\slashed{\Delta})
\gamma_5 
\frac{-2ik^+}{2k\cdot \Delta+t}
\frac{i(\slashed{p}-\slashed{k}+M_B)}{D_B(p-k)}
\frac{i}{D_\phi(k)}
\slashed{k}\gamma_5\,
\big( \widetilde{F}(k+\Delta)-\widetilde{F}(k) \big)
\bigg]
\delta\Big(y-\frac{k^+}{p^+}\Big)\,
u(p).
\nonumber
\end{eqnarray} 
The respective splitting functions for these gauge link diagrams, $\delta f^{({\rm KR})}_{B\phi}(y,t)$ and $\delta g^{({\rm KR})}_{B\phi}(y,t)$, can be written as
\begin{subequations}
\begin{eqnarray}
\delta f^{({\rm KR})}_{B\phi}(y,t)
&=& \frac{C^2_{B\phi}}{f^2}
\int\!\frac{\dd[4]{k}}{(2\pi)^4}\,
\widetilde{F}(k)
\frac{1}{D_\phi(k)}
\delta\Big(y-\frac{k^+}{p^+}\Big)
\nonumber\\
&& \hspace*{-0.0cm} \times
\bigg[
  \frac{i\delta F^{({\rm KR\, 1})}_{B\phi}}{D_B(p'-k)}
  \frac{\widetilde{F}(k-\Delta)-\widetilde{F}(k)}{-2k\cdot \Delta+t}\,
+\,
  \frac{i\delta F^{({\rm KR\, 2})}_{B\phi}}{D_B(p-k)}
  \frac{\widetilde{F}(k)-\widetilde{F}(k+\Delta)}{2k\cdot \Delta+t}
\bigg],\hspace*{1.0cm}
\\
\delta g^{({\rm KR})}_{B\phi}(y,t)
&=& \frac{C^2_{B\phi}}{f^2}
\int\!\frac{\dd[4]{k}}{(2\pi)^4}\,
\widetilde{F}(k)
\frac{1}{D_\phi(k)}
\delta\Big(y-\frac{k^+}{p^+}\Big)
\nonumber\\
&& \hspace*{-0.0cm} \times
\bigg[
  \frac{i\delta G^{({\rm KR\, 1})}_{B\phi}}{D_B(p'-k)}
  \frac{\widetilde{F}(k-\Delta)-\widetilde{F}(k)}{-2k\cdot \Delta+t}\,
+\,
  \frac{i\delta G^{({\rm KR\, 2})}_{B\phi}}{D_B(p-k)}
  \frac{\widetilde{F}(k)-\widetilde{F}(k+\Delta)}{2k\cdot \Delta+t}
\bigg],\hspace*{1.0cm}
\end{eqnarray}
\end{subequations}
where the numerator factors are given by
\begin{subequations}
\begin{eqnarray}
\delta F^{({\rm KR\, 1})}_{B\phi}
&=& \hspace*{0.33cm} \frac{y}{2}
\Big( 
  4(k\cdot p)^2 
- k^2 \big( 4 k\cdot P + 4M\MBbar + y\, t \big)
\nonumber\\
&& \hspace*{0.8cm}
+\, 2M\MBbar \big( 2k\cdot \Delta - y\, t\big) 
+ 2k\cdot p'\, \big( 2k\cdot p + y\, t \big)
\Big),
\\
\delta F^{({\rm KR\, 2})}_{B\phi}
&=&-\frac{y}{2}
\Big( 
  4(k\cdot p)^2 
- k^2 \big( 4 k\cdot P + 4M\MBbar + y\, t \big)
\nonumber
\\
&& \hspace*{0.8cm}
-\, 2M\MBbar \big( 2k\cdot \Delta + y\, t \big)
+ 2k\cdot p\, \big( 2k\cdot p' + y\, t \big)
\Big),
\\
\delta G^{({\rm KR\, 1})}_{B\phi}
&=&-2My\,
\Big(
  \MBbar \big( 2k\cdot p' - 2y M^2 \big)
- k^2 \big( \MBbar+yM \big) + 2y M k\cdot p' 
\Big),
\\
\delta G^{({\rm KR\, 2})}_{B\phi}
&=& \hspace*{0.33cm} 2My\,
\Big( 
  \MBbar \big( 2k\cdot p\, - 2y M^2 \big)
- k^2 \big( \MBbar+yM \big) + 2y M k\cdot p
\Big).
\end{eqnarray}
\end{subequations}

The contribution to $\Gamma^+$ from the tadpole diagram in Fig.~\ref{diagrams}(h) is given by a relatively simple expression,
\begin{equation} 
\Gamma_{\rm (h)}^+
= \bar{u}(p')\,
\frac{C_{\phi\phi}}{f^2}
\int\!\frac{\dd[4]{k}}{(2\pi)^4}
\frac{i}{D_\phi(k)}\gamma^+
\widetilde{F}^2(k)\,
\delta\Big(y-\frac{k^+}{p^+}\Big)u(p).
\label{eq.Gamma+h}
\end{equation}
The splitting function $f^{({\rm tad})}_\phi(y,t)$ in this case can be easily read off from Eq.~(\ref{eq.Gamma+h}), and is given by
\begin{equation} 
f^{({\rm tad})}_\phi(y,t)
= \frac{C_{\phi\phi}}{f^2}
\int\!\frac{\dd[4]{k}}{(2\pi)^4}
\frac{i}{D_\phi(k)}
\widetilde{F}^2(k)\,
\delta\Big(y-\frac{k^+}{p^+}\Big).
\end{equation}
There is no contribution to a Pauli-like splitting function $g^{({\rm tad})}_\phi(y,t)$ from this diagram.
The related tadpole diagram that is associated with the gauge link in Fig.~\ref{diagrams}(i) makes a contribution to the $\Gamma^+$ matrix element that can be written as 
\begin{eqnarray} 
\Gamma_{\rm (i)}^+
&=& \bar{u}(p')
\frac{C_{\phi\phi}}{f^2}
\int\!\frac{\dd[4]{k}}{(2\pi)^4}\,
2\slashed{k}\,
\widetilde{F}(k)
\frac{i}{D_\phi(k)}
\frac{2k^+}{-2k\cdot \Delta+t}\,
\big( \widetilde{F}(k-\Delta)-\widetilde{F}(k) \big)\,
\delta\Big( y-\frac{k^+}{p^+}\Big)
u(p),
\notag\\
&&
\end{eqnarray}
with the corresponding splitting function $\delta f^{({\rm tad})}_\phi(y,t)$ given by
\begin{eqnarray}
\delta f^{({\rm tad})}_\phi(y,t)
&=& \frac{iC_{\phi\phi}}{f^2}
\int\!\frac{\dd[4]{k}}{(2\pi)^4}\,
\widetilde{F}(k)
\frac{y\, (4k\cdot P + y\, t)}{D_\phi(k)(-2k\cdot \Delta + t)}
\big( \widetilde{F}(k-\Delta)-\widetilde{F}(k) \big)\,
\delta\Big( y-\frac{k^+}{p^+}\Big).~~
\end{eqnarray}
The tadpole diagram in Fig.~\ref{diagrams}(j) associated with the magnetic interaction makes a contribution
\begin{equation}
\Gamma_{\rm (j)}^+
= \bar{u}(p')\,
\frac{C^{{\rm mag}}_{\phi\phi}}{f^2}
\int\!\frac{\dd[4]{k}}{(2\pi)^4}\frac{i}{D_\phi(k)}
\frac{i\sigma^{+\nu}\Delta_\nu}{2M}
\widetilde{F}^2(k)\,
\delta\Big(y-\frac{k^+}{p^+}\Big)\,
u(p),
\end{equation}
which gives rise to a Pauli-like magnetic tadpole splitting function given by,
\begin{eqnarray} 
g^{({\rm tad\, mag})}_\phi(y,t)
= \frac{C^{{\rm mag}}_{\phi\phi}}{f^2}
\int\!\frac{\dd[4]{k}}{(2\pi)^4}
\frac{i}{D_\phi(k)}
\widetilde{F}^2(k)\,
\delta\Big(y-\frac{k^+}{p^+}\Big)\,
u(p).
\end{eqnarray}

The contributions of the bubble diagrams are illustrated in Figs.~\ref{diagrams}(k) and \ref{diagrams}(l).
For the regular bubble diagram in Fig.~\ref{diagrams}(k), one has
\begin{eqnarray} 
\Gamma_{\rm (k)}^+
&=&-\bar{u}(p')\,
\frac{C_{\phi\phi}}{2f^2}
\int\!\frac{\dd[4]{k}}{(2\pi)^4}\,
i(2\slashed{k}+\slashed{\Delta})\,
\widetilde{F}(k+\Delta)
\widetilde{F}(k)\,
\frac{i}{D_\phi(k+\Delta)}
2k^+
\frac{i}{D_\phi(k)}\,
\delta\Big(y-\frac{k^+}{p^+}\Big)\,
u(p).
\notag\\
&&
\end{eqnarray}
As with the tadpole diagrams, this also generates only a Dirac-like splitting function, $f^{({\rm bub})}_\phi(y,t)$, which is expressed as
\begin{eqnarray} 
\label{eq:f_phi_bub}
f^{({\rm bub})}_\phi(y,t)
= \frac{iC_{\phi\phi}}{4 f^2}
\int\!\frac{\dd[4]{k}}{(2\pi)^4}
\frac{y\, (4k\cdot P+y\, t)}{D_\phi(k+\Delta) D_\phi(k)}\,
\widetilde{F}(k+\Delta)\,
\widetilde{F}(k)\,
\delta\Big(y-\frac{k^+}{p^+}\Big).
\end{eqnarray}
Finally, for the bubble diagram derived from the Lagrangian ${\cal L}_{B\phi}'$ in Eq.~(\ref{adm}) and illustrated in Fig.~\ref{diagrams}(l), the contribution can be written as
\begin{eqnarray} 
\Gamma_{\rm (l)}^+
&=& \bar{u}(p')\,
\frac{C'_{\phi\phi}}{2f^2}
\int\!\frac{\dd[4]{k}}{(2\pi)^4}\,
i\sigma^{\lambda\nu}\Delta_\lambda k_\nu\, \widetilde{F}(k+\Delta)
\widetilde{F}(k)
\frac{i}{D_\phi(k+\Delta)}
2k^+
\frac{i}{D_\phi(k)}\,
\delta\Big(y-\frac{k^+}{p^+}\Big)\,
u(p),
\notag\\
&&
\end{eqnarray}
and produces a Pauli-like magnetic bubble splitting function given by
\begin{eqnarray}
\label{eq:gpr_phi_bub}
g'^{({\rm bub})}_\phi(y,t)
=\frac{iC'_{\phi\phi}}{2f^2}
\int\!\frac{\dd[4]{k}}{(2\pi)^4}
\frac{y M\, (4k\cdot P + y\, t )}{D_\phi(k+\Delta)D_\phi(k)}\,
\widetilde{F}(k+\Delta)\, \widetilde{F}(k)\,
\delta\Big(y-\frac{k^+}{p^+}\Big).
\end{eqnarray}

% ............................................................................
\subsection{Decuplet baryon intermediate states}
\label{ssec.fy_decuplet}

The splitting functions for the diagrams involving decuplet baryons in the intermediate state in Figs.~\ref{diagrams}(m)--\ref{diagrams}(u) are computed in a similar way, although because of the higher spin of the decuplet baryons the expressions are typically somewhat more involved.
Here we give the basic expressions for the contributions from each diagram to the matrix element $\Gamma^+$, with the full results for the numerator trace factors in the decuplet splitting functions given in Appendix~\ref{sec.appendix}.

Beginning with the decuplet baryon rainbow diagram in Fig.~\ref{diagrams}(m), the contribution to $\Gamma^+$ is expressed in a similar form to that for the octet baryon rainbow diagram Fig.~\ref{diagrams}(a) in Eqs.~(\ref{eq.Gamma_a})--(\ref{eq.splitfnproj}),
\begin{eqnarray} 
\Gamma_{\rm (m)}^+
&=&-\bar{u}(p')\,
\frac{C_{T\phi}^2}{f^2}
\int\!\frac{\dd[4]{k}}{(2\pi)^4}\,
(k+\Delta)_\lambda\,
\Theta^{\lambda\alpha}
\widetilde{F}(k+\Delta)\frac{i}{D_\phi(k+\Delta)}\, 2k^+
\notag\\
&& \hspace*{2.2cm} \times
\frac{i}{D_\phi(k)}
\frac{i}{\slashed{p}-\slashed{k}-M_T}
S_{\alpha\beta}(p-k)\, \Theta^{\beta\rho} k_\rho\,
\widetilde{F}(k)\,
\delta\Big(y-\frac{k^+}{p^+}\Big)\,
u(p)
\\
&\equiv&
\bar{u}(p')
\bigg[
  \gamma^+ f_{\phi T}^{({\rm rbw})}(y,t) 
+ \frac{i\sigma^{+\nu}\Delta_\nu}{2M}g_{\phi T}^{({\rm rbw})}(y,t)
\bigg]
u(p),
\end{eqnarray}
where the octet-decuplet transition operator $\Theta^{\mu\nu}$ is defined in Eq.~(\ref{eq:Theta}), and the spin-3/2 projection operator $S_{\alpha\beta}$ for a particle of momentum $k$ is given by
\begin{equation} 
S_{\alpha\beta}(k)
= -g_{\alpha\beta} + \frac{\gamma_\alpha\gamma_\beta}{3}
+ \frac{2 k_\alpha\, k_\beta}{3 M_T^2}
+ \frac{\gamma_\alpha\, k_\beta-\gamma_\beta\, k_\alpha}{3 M_T}.
\end{equation}
The corresponding Dirac and Pauli decuplet rainbow splitting functions, $f^{({\rm rbw})}_{\phi T}(y,t)$ and $g^{({\rm rbw})}_{\phi T}(y,t)$, can be written as
\begin{subequations}
\label{eq.fg_phiTrbw}
\begin{eqnarray}
\label{eq.f_phiTrbw}
f^{({\rm rbw})}_{\phi T}(y,t) 
&=& \frac{C_{T\phi}^2}{f^2}
\int\!\frac{\dd[4]{k}}{(2\pi)^4}
\frac{iF_{\phi T}^{({\rm rbw})}}{D_T(p-k)D_\phi(k+\Delta)D_\phi(k)}\,
\widetilde{F}(k+\Delta)\,
\widetilde{F}(k)\,
\delta\Big(y-\frac{k^+}{p^+}\Big),~~~~~~
\\
\label{eq.g_phiTrbw}
g^{({\rm rbw})}_{\phi T}(y,t)
&=& \frac{C_{T\phi}^2}{f^2}
\int\!\frac{\dd[4]{k}}{(2\pi)^4}
\frac{iG_{\phi T}^{({\rm rbw})}}{D_T(p-k)D_\phi(k+\Delta)D_\phi(k)}\,
\widetilde{F}(k+\Delta)\,
\widetilde{F}(k)\,
\delta\Big(y-\frac{k^+}{p^+}\Big),
\end{eqnarray}
\end{subequations}
respectively.
The explicit expressions for the numerator factors $F_{\phi T}^{({\rm rbw})}$ and $G_{\phi T}^{({\rm rbw})}$ are given in Eqs.~(\ref{eq.FG_phiTrbw}) of Appendix~\ref{sec.appendix}.

In Fig.~\ref{diagrams}(n), the photon couples to the decuplet baryon in the intermediate state with an electric vertex.
The contribution of this diagram is given by 
\begin{eqnarray} 
\Gamma_{\rm (n)}^+
&=& -\bar{u}(p')\,
\frac{C^2_{T\phi}}{f^2}
\int\!\frac{\dd[4]{k}}{(2\pi)^4}\,
k_\lambda \Theta^{\lambda\sigma}\,
\widetilde{F}(k)\,
\frac{i}{D_K(k)}\,
\frac{i}{\slashed{p'}-\slashed{k}-M_T}
S_{\sigma\alpha}(p^\prime-k)\,
\gamma^{\alpha\beta+}
\nonumber\\
&& \hspace*{3.2cm} \times
\frac{i}{\slashed{p}-\slashed{k}-M_T}\,
S_{\beta\rho}(p-k)\,
\Theta^{\rho\nu}k_\nu\,
\widetilde{F}(k)\,
\delta\Big(y-\frac{k^+}{p^+}\Big)\,
u(p),~~~
\end{eqnarray} 
where the corresponding splitting functions $f^{({\rm rbw})}_{T\phi}(y,t)$ and $g^{({\rm rbw})}_{T\phi}(y,t)$ are
\begin{subequations}
\label{eq.fg_Tphirbw}
\begin{eqnarray}
f^{({\rm rbw})}_{T\phi}(y,t) 
&=& \frac{C_{T\phi}^2}{f^2}
\int\!\frac{\dd[4]{k}}{(2\pi)^4}
\frac{iF_{T\phi}^{({\rm rbw})}}{D_T(p'-k)D_T(p-k)D_\phi(k)}\,
\widetilde{F}^2(k)\,
\delta\Big(y-\frac{k^+}{p^+}\Big),
\\
g^{({\rm rbw})}_{T\phi}(y,t) 
&=& \frac{C_{T\phi}^2}{f^2}
\int\!\frac{\dd[4]{k}}{(2\pi)^4}
\frac{iG_{T\phi}^{({\rm rbw})}}{D_T(p'-k)D_T(p-k)D_\phi(k)}\,
\widetilde{F}^2(k)\,
\delta\Big(y-\frac{k^+}{p^+}\Big).
\end{eqnarray}
\end{subequations}
The factors $F_{T\phi}^{({\rm rbw})}$ and $G_{T\phi}^{({\rm rbw})}$ in the numerators here are given in Eqs.~(\ref{eq.FG_Tphirbw}).
For the magnetic photon-decuplet baryon interaction in Fig.~\ref{diagrams}(o), the contribution to $\Gamma^+$ is expressed as
%
%\begin{eqnarray} 
%\Gamma_j^{\mu(p)}&=& \frac{4{\cal C}^2}{3}I_{j\pi}^{N\Delta}+\frac{{\cal C}^2}{6}I_{jK}^{N\Sigma^*},\\
%\Gamma_j^{\mu(n)}&=& -\frac{{\cal C}^2}{3}I_{j\pi}^{N\Delta}-\frac{{\cal C}^2}{6}I_{jK}^{N\Sigma^*},
%\end{eqnarray} 
%
\begin{eqnarray} 
\Gamma_{\rm (o)}^+
&=& -\frac{C^2_{T\phi}}{f^2}
\bar{u}(p')\int\!\frac{\dd[4]{k}}{(2\pi)^4}\,
k_\lambda\Theta^{\lambda\sigma} 
\widetilde{F}(k)
\frac{i}{D_K(k)}\frac{i}{\slashed{p'}-\slashed{k}-M_T}\,
S_\sigma^{\ \, \alpha}(p^\prime-k)
\frac{-i}{2M_T}\sigma^{+\beta} \Delta_\beta,
\nonumber\\
&& \hspace*{3.2cm} \times 
\frac{i}{\slashed{p}-\slashed{k}-M_T}\,
S_{\alpha\rho}(p-k)\, 
\Theta^{\rho\nu}k_\nu\,
\widetilde{F}(k)\,
\delta\Big(y-\frac{k^+}{p^+}\Big)\,
u(p),
\end{eqnarray}
where the corresponding splitting functions $f^{\rm (rbw\, mag)}_{T\phi}(y,t)$ and $g^{\rm (rbw\, mag)}_{T\phi}(y,t)$ are given by 
\begin{subequations}
\label{eq.fg_Tphirbwmag}
\begin{eqnarray} 
f^{\rm (rbw\, mag)}_{T\phi}(y,t)
&=& \frac{C_{T\phi}^2}{f^2}
\int\!\frac{\dd[4]{k}}{(2\pi)^4}
\frac{i F_{T\phi}^{\rm (rbw\, mag)}}{D_T(p'-k) D_T(p-k) D_\phi(k)}
\widetilde{F}^2(k)\,
\delta\Big(y-\frac{k^+}{p^+}\Big),
\\
g^{\rm (rbw\, mag)}_{T\phi}(y,t) 
&=& \frac{C_{T\phi}^2}{f^2}
\int\!\frac{\dd[4]{k}}{(2\pi)^4}
\frac{i G_{T\phi}^{\rm (rbw\, mag)}}{D_T(p'-k)D_T(p-k)D_\phi(k)}\,
\widetilde{F}^2(k)\,
\delta\Big(y-\frac{k^+}{p^+}\Big),
\end{eqnarray}
\end{subequations}
and the factors $F_{T\phi}^{\rm (rbw\, mag)}$ and $G_{T\phi}^{\rm (rbw\, mag)}$ are written out in Eqs.~(\ref{eq.FG_Tphirbwmag}).

In Figs.~\ref{diagrams}(p) and \ref{diagrams}(q), the photon couples to the octet-decuplet transition vertex, whose contribution to the matrix element $\Gamma^+$ can be written as
%
%\begin{eqnarray} 
%\Gamma_{k+l}^{\mu(p)}&=& 2(D+F){\cal C}I_{(k+l)\pi}^{N\Delta}+ \frac{5}{4}(D-F){\cal C}I_{(k+l)K}^{\Sigma\Sigma^*}
%+\frac{1}{4}(3F+D){\cal C}I_{(k+l)K}^{\Lambda\Sigma^*},\\
%\Gamma_{k+l}^{\mu(n)}&=&-2(D+F){\cal C}I_{(k+l)\pi}^{N\Delta}+\frac{1}{4}(D+F){\cal C}I_{(k+l)K}^{\Sigma\Sigma^*}
%-\frac{1}{4}(3F+D){\cal C}I_{(k+l)K}^{\Lambda\Sigma^*},
%\end{eqnarray} 
%
\begin{eqnarray} 
\label{eq.FG_BTphi}
\Gamma_{\rm (p)+(q)}^+
&=& \frac{C_{B\phi}C_{T\phi}}{4 M_B f^2}\,
\bar{u}(p')
\int\!\frac{\dd[4]{k}}{(2\pi)^4}\,
\widetilde{F}^2(k)
\nonumber\\
&& \times
\bigg[
-\!\slashed{k}\gamma_5
\frac{i}{\slashed{p'}-\slashed{k}-M_B}
\slashed{\Delta}\gamma_5
\frac{i}{\slashed{p}-\slashed{k}-M_T}
S^+_{\ \, \rho}(p-k)\,
\Theta^{\rho\nu} k_\nu
\nonumber\\
&& \hspace*{0.6cm}
+\, \slashed{k}\gamma_5
\frac{i}{\slashed{p'}-\slashed{k}-M_B}
\gamma^+\gamma_5\Delta^\lambda\frac{i}{\slashed{p}-\slashed{k}-M_T}
S_{\lambda\rho}(p-k)\,
\Theta^{\rho\nu}k_\nu
\nonumber\\
&& \hspace*{0.6cm}
-\, k_\lambda\Theta^{\lambda\nu}
\frac{i}{\slashed{p'}-\slashed{k}-M_T}
S_{\nu\rho}(p^\prime-k) 
\Delta^\rho \gamma^+ \gamma_5
\frac{i}{\slashed{p}-\slashed{k}-M_B}
\slashed{k}\gamma_5
\nonumber\\
&& \hspace*{0.6cm}
+\, k_\lambda\Theta^{\lambda\nu}
\frac{i}{\slashed{p'}-\slashed{k}-M_T}
S_\nu^{\ \, +}(p^\prime-k)
\slashed{\Delta} \gamma_5\frac{1}{\slashed{p}-\slashed{k}-M_B}
\slashed{k}\gamma_5
\bigg]
\nonumber\\
&& \times\,
\frac{i}{D_\phi(k)}\,
\delta\Big(y-\frac{k^+}{p^+}\Big)\,
u(p).
\end{eqnarray}
The corresponding Dirac and Pauli splitting functions
$f^{\rm (rbw\, mag)}_{BT}(y,t)$ and $g^{\rm (rbw\, mag)}_{BT}(y,t)$ 
are in this case given by
\begin{subequations}
\label{eq.fg_BTphi}
\begin{eqnarray}
f^{\rm (rbw\, mag)}_{BT}(y,t)
&=& \frac{C_{B\phi} C_{T\phi}}{4f^2}
\int\!\frac{\dd[4]{k}}{(2\pi)^4}\,
\widetilde{F}^2(k)\,
\frac{1}{D_\phi(k)}\,
\delta\Big(y-\frac{k^+}{p^+}\Big)\,
\nonumber \\
&& \hspace*{1.6cm} \times\,
\bigg[
  \frac{i F^{\rm (rbw\, mag\, 1)}_{BT}}{D_B(p'-k) D_T(p-k)}
+ \frac{i F^{\rm (rbw\, mag\, 2)}_{BT}}{D_T(p'-k) D_B(p-k)}
\bigg],
\\
g^{\rm (rbw\, mag)}_{BT}(y,t)
&=& \frac{C_{B\phi} C_{T\phi}}{4f^2}
\int\!\frac{\dd[4]{k}}{(2\pi)^4}\,
\widetilde{F}^2(k)\,
\frac{1}{D_\phi(k)}\,
\delta\Big(y-\frac{k^+}{p^+}\Big)\,
\nonumber \\
&& \hspace*{1.6cm} \times\,
\bigg[
  \frac{i G^{\rm (rbw\, mag\, 1)}_{BT}}{D_B(p'-k) D_T(p-k)}
+ \frac{i G^{\rm (rbw\, mag\, 2)}_{BT}}{D_T(p'-k) D_B(p-k)}
\bigg],~~~~
\end{eqnarray}
\end{subequations}
with the four numerator factors $F^{\rm (rbw\, mag\, 1,2)}_{BT}$ and $G^{\rm (rbw\, mag\, 1,2)}_{BT}$ given in Eqs.~(\ref{eq.FG_BTphi}).
% of Appendix~\ref{sec.appendix}.

For the KR diagrams with decuplet baryon intermediate states in Fig.~\ref{diagrams}(r) and \ref{diagrams}(s), the contribution to $\Gamma^+$ is given by the expression
%
%\begin{eqnarray} 
%\Gamma_{m+n}^{\mu(p)}&=& \frac{2{\cal C}^2}{3}I_{(m+n)\pi}^{N\Delta}-\frac{{\cal C}^2}{6}I_{(m+n)K}^{N\Sigma^*},\\
%\Gamma_{m+n}^{\mu(n)}&=&-\frac{2{\cal C}^2}{3}I_{(m+n)\pi}^{N\Delta}-\frac{{\cal C}^2}{3}I_{(m+n)K}^{N\Sigma^*},
%\end{eqnarray}
%
\begin{eqnarray}
\Gamma_{\rm (r)+(s)}^+
&=& -\frac{C^2_{T\phi }}{f^2}\, \bar{u}(p')\, 
\int\!\frac{\dd[4]{k}}{(2\pi)^4}\,
\widetilde{F}^2(k) 
\nonumber\\
&& \times
\bigg[
   k_\nu\Theta^{\nu\sigma}
   \frac{i}{\slashed{p'}-\slashed{k}-M_T}\, 
   S_{\sigma\rho}(p^\prime-k)\,
   i\Theta^{\rho+}\,
-\ i\Theta^{+\sigma}
   \frac{i}{\slashed{p}-\slashed{k}-M_T}\,
   S_{\sigma\rho}(p-k)\,
   \Theta^{\rho\nu} k_\nu
\bigg]~~
\nonumber\\
&& \times
\frac{i}{D_\phi(k)}\,
\delta\Big(y-\frac{k^+}{p^+}\Big)\,
u(p),
\end{eqnarray}
where the corresponding KR splitting functions $f^{(\rm KR)}_{T\phi}(y,t)$ and $g_{T\phi}^{(\rm KR)}(y,t)$ are given by
\begin{subequations}
\label{eq.fg_KRTphi}
\begin{eqnarray}
f^{\rm (KR)}_{T\phi}(y,t)
&=& \frac{C^2_{T\phi}}{f^2}
\int\!\frac{\dd[4]{k}}{(2\pi)^4}
\bigg[
   \frac{iF^{\rm (KR\, 1)}_{T\phi}}{D_T(p'-k)}
+  \frac{iF^{\rm (KR\, 2)}_{T\phi}}{D_T(p-k)}
\bigg]
\widetilde{F}^2(k)\,
\frac{1}{D_\phi(k)}\,
\delta\Big(y-\frac{k^+}{p^+}\Big),
\\
g^{({\rm KR})}_{T\phi}(y,t)
&=& \frac{C^2_{T\phi}}{f^2}
\int\!\frac{\dd[4]{k}}{(2\pi)^4}
\bigg[ 
   \frac{iG^{\rm (KR\, 1)}_{T\phi}}{D_T(p'-k)}
+  \frac{iG^{\rm (KR\, 2)}_{T\phi}}{D_T(p-k)}
\bigg]
\widetilde{F}^2(k)\,
\frac{1}{D_\phi(k)}\,
\delta\Big(y-\frac{k^+}{p^+}\Big),
\end{eqnarray}
\end{subequations}
and the numerator functions $F^{\rm (rbw\, mag\, 1,2)}_{BT}$ and $G^{\rm (rbw\, mag\, 1,2)}_{BT}$ are in Eqs.~(\ref{eq.FG_KRTphi}).
Similarly, the additional KR diagrams generated from the gauge link terms with decuplet intermediate states are shown in Fig.~\ref{diagrams}(t) and \ref{diagrams}(u), and their contribution can be written as
%
%\begin{eqnarray} 
%\Gamma_{o+p}^{\mu(p)}&=& \frac{2{\cal C}^2}{3}I_{(o+p)\pi}^{N\Delta}-\frac{{\cal C}^2}{6}I_{(o+p)K}^{N\Sigma^*},\\
%\Gamma_{o+p}^{\mu(n)}&=&-\frac{2{\cal C}^2}{3}I_{(o+p)\pi}^{N\Delta}-\frac{{\cal C}^2}{3}I_{(o+p)K}^{N\Sigma^*},
%\end{eqnarray} 
%
\begin{eqnarray}
\Gamma_{\rm (t)+(u)}
&=& \frac{C^2_{\phi T}}{f^2}\,
\bar{u}(p')
\int\!\frac{\dd[4]{k}}{(2\pi)^4}
\nonumber\\
&\times&
\bigg[
k_\nu \Theta^{\nu\sigma}
\frac{i}{\slashed{p'}-\slashed{k}-M_T}
S_{\sigma\rho}(p^\prime-k)\,
\Theta^{\rho\lambda} (k-\Delta)_\lambda   
\frac{2i k^+}{-2k \cdot \Delta + \Delta^2}
\big( \widetilde{F}(k-\Delta) - \widetilde{F}(k) \big)
\nonumber\\
&& +\, (k+\Delta)_\nu \Theta^{\nu\sigma}
\frac{-2ik^+}{2k\cdot\Delta + \Delta^2}
\big( \widetilde{F}(k) - \widetilde{F}(k+\Delta) \big)
\frac{i}{\slashed{p}-\slashed{k}-M_T}
S_{\sigma\rho}(p-k)\,
\Theta^{\rho\lambda} k_\lambda
\bigg]
\nonumber\\
&\times&
\frac{i}{D_\phi(k)}\,
\widetilde{F}(k)\,
\delta\Big(y-\frac{k^+}{p^+}\Big)\,
u(p).   
\end{eqnarray}
where the gauge link dependent KR splitting functions $\delta f^{\rm (KR)}_{T\phi}(y,t)$ and $\delta g_{T\phi}^{\rm (KR)}(y,t)$ are given by 
\begin{subequations}
\label{eq.fg_delKRTphi}
\begin{eqnarray}
\delta f^{({\rm KR})}_{T\phi}(y,t)
&=& -\frac{C^2_{T\phi}}{f^2}
\int\!\frac{\dd[4]{k}}{(2\pi)^4}\,
\widetilde{F}(k)\,
\frac{1}{D_\phi(k)}\,
\delta\Big(y-\frac{k^+}{p^+}\Big)
\nonumber \\
&& \times
\bigg[
  \frac{i \delta F^{\rm (KR\, 1)}_{T\phi}}{D_T(p'-k)}
  \frac{\widetilde{F}(k)-\widetilde{F}(k-\Delta)}{2k\cdot\Delta-\Delta^2} 
+ \frac{i \delta F^{\rm (KR\, 2)}_{T\phi}}{D_T(p-k)}
  \frac{\widetilde{F}(k)-\widetilde{F}(k+\Delta)}{2k\cdot\Delta+\Delta^2}
\bigg],
\\
\delta g^{({\rm KR})}_{T\phi}(y,t)
&=& -\frac{C^2_{T\phi}}{f^2}
\int\!\frac{\dd[4]{k}}{(2\pi)^4}\,
\widetilde{F}(k)\,
\frac{1}{D_\phi(k)}\,
\delta\Big(y-\frac{k^+}{p^+}\Big)
\nonumber \\
&& \times
\bigg[
  \frac{i \delta G^{\rm (KR\, 1)}_{T\phi}}{D_T(p'-k)}
  \frac{\widetilde{F}(k)-\widetilde{F}(k-\Delta)}{2k\cdot\Delta-\Delta^2} 
+ \frac{i \delta G^{\rm (KR\, 2)}_{T\phi}}{D_T(p-k)}
  \frac{\widetilde{F}(k)-\widetilde{F}(k+\Delta)}{2k\cdot\Delta+\Delta^2}
\Bigg],~~~~
\end{eqnarray}
\end{subequations}
The complete expressions for the numerator functions $\delta F^{\rm (KR\, 1,2)}_{T\phi}$ and $\delta G^{\rm (KR\, 1,2)}_{T\phi}$ are presented in Eq.~(\ref{eq.FG_delKRTphi}) of Appendix~\ref{sec.appendix}.

% The expressions for $F_{\phi T}^{({\rm rbw})}$,~$G_{\phi T}^{({\rm rbw})}$,~  $F_{T\phi }^{({\rm rbw})}$,~$G_{T\phi }^{({\rm rbw})}$,~$F_{\phi T}^{({\rm {rbw,mag}})}$,~$G_{\phi T}^{({\rm {rbw,mag}})}$,~$F_{BT}^{({\rm {rbw,mag1}})}$,~$G_{BT}^{({\rm {rbw,mag1}})}$,
% ~$F_{BT}^{({\rm {rbw,mag2}})}$,~$G_{BT}^{({\rm {rbw,mag2}})}$ are placed in the Appendix~\ref{sec.appendix}.

\clearpage
%%%%%%%%%%%%%%%%%%%%%%%%%%%%%%%%%%%%%%%%%%%%%%%%%%%%%%%%%%%%%%%%%%%%%%%%%%%%
\section{Nonanalytic behavior of splitting functions}
\label{sec.lna}

While certain features of the hadronic splitting functions derived in the previous section depend on short-distance physics that in the current framework is controlled by the regulator function $\widetilde{F}$, some aspects of their calculation are in fact model independent.
In particular, the moments of the splitting functions, which can be expanded as a series in the pseudoscalar meson mass, $m_\phi$, contain terms that do not depend on the regularization method~\cite{Thomas:2000ny, Arndt:2001ye, Chen:2001eg, Salamu:2018cny}.
These are the coefficients of the leading nonanalytic (LNA) terms, that are determined by the low-energy properties of the nucleon, such as the hadronic couplings and masses.
Since the LNA behavior is derived solely from the long-distance characteristics of the chiral effective theory, understanding these can place constraints on models of the splitting functions consistent with the symmetries of QCD.

To explore the LNA terms further, we define the lowest moments of the splitting functions $f^{\rm (x)}(y,t)$ and $g^{\rm (x)}(y,t)$ for the diagram in Fig.~\ref{diagrams}(``x''), where ``x'' = ``a'', $\ldots$ ``u'', as
\begin{subequations}
\begin{eqnarray}
F^{\rm (x)}_1(t) &=& \int_0^1 \dd{y} f^{\rm (x)}(y,t),    \\
F^{\rm (x)}_2(t) &=& \int_0^1 \dd{y} g^{\rm (x)}(y,t),    \\
\end{eqnarray}
\end{subequations}
which correspond to the Dirac and Pauli electromagnetic form factors of the nucleon, respectively.
Taking the values at zero four-momentum transfer squared, $t=0$, we expand the form factors $F^{\rm (x)}_{1,2}(t\!=\!0)$ in $m_\phi$, keeping only the nonanalytic (NA) terms in $m_\phi^2$, from which the LNA behavior can be extracted.

For the rainbow diagram in Fig.~\ref{diagrams}(a), the NA contributions can be written as 
\begin{subequations}
\begin{eqnarray}
F_1^{\rm (a)}(0)\Big|_{\rm NA}
&=&\frac{3 C^2_{B\phi}}{(4\pi f)^2}
\bigg[ \big( m_\phi^2 - 2\Delta_B^2\big) \log m_\phi^2
    + 2\Delta_B R_B \log\frac{\Delta_B-R_B}{\Delta_B+R_B}
\bigg],
\\
F_2^{\rm (a)}(0)\Big|_{\rm NA}
&=& 
\frac{4 C^2_{B\phi}}{(4\pi f)^2}
\bigg[ \big( M\Delta_B + 2\Delta^2_B + 2 R_B^2 \big) \log m_\phi^2 
\notag\\
&& \hspace*{3.7cm}
    -\, \big(M+4\Delta_B\big) R_B\, \log\frac{\Delta_B-R_B}{\Delta_B+R_B}
\bigg],
\end{eqnarray}
\end{subequations}
where $R_B = \sqrt{\Delta_B^2 - m_\phi^2}$ .
If $\Delta_B < m_\phi$, $R_B$ will be an imaginary number and the log term will become an arctangent, according to its definition~\cite{Salamu:2018cny}. 
The NA terms of $F_1^{\rm (a)}$ are the same as those reported in Eq.~(62) of Ref.~\cite{Salamu:2018cny} if one sums the NA contributions from the on-shell and $\delta$ terms of Eqs.~(114) and (117) of \cite{Salamu:2018cny}.

For the baryon rainbow diagram in Fig.~\ref{diagrams}(b), the NA behavior is given by
\begin{subequations}
\begin{eqnarray}
F_1^{\rm (b)}(0)\Big|_{\rm NA}
&=&~~\frac{3 C^2_{B\phi}}{(4\pi f)^2}
\bigg[ \big( m_\phi^2 - 2\Delta_B^2 \big) \log m_\phi^2 
    + 2\Delta_B R_B \log\frac{\Delta_B-R_B}{\Delta_B+R_B}
\bigg],
\\
F_2^{\rm (b)}(0)\Big|_{\rm NA}
&=& -\frac{4 C^2_{B\phi}}{(4\pi f)^2}
\bigg[ \big( m_\phi^2 - 2\Delta_B^2\big)\log m_\phi^2 
    + 2\Delta_B R_B \log\frac{\Delta_B-R_B}{\Delta_B+R_B}
\bigg].
\end{eqnarray}
\end{subequations}
Note that the NA terms of $F_1^{\rm (b)}(0)$ are identical to those of $F_1^{\rm (a)}(0)$.
For the diagram in Fig.~\ref{diagrams}(c), there is no NA term for $F_1^{\rm (c)}(0)$, while the NA part of $F_2^{\rm (c)}(0)$ can be written as 
\begin{eqnarray}
F_2^{\rm (c)}(0)\Big|_{\rm NA}
&=&-\frac{C^2_{B\phi}}{(4\pi f)^2}
\bigg[\big(m_\phi^2-2\Delta_B^2\big)\log m_\phi^2
    + 2\Delta_B R_B \log\frac{\Delta_B-R_B}{\Delta_B+R_B}
\bigg].
\end{eqnarray}

For the KR diagrams in Figs.~\ref{diagrams}(d) and \ref{diagrams}(e), the NA contributions to the Dirac form factor is given by
\begin{eqnarray}
F_1^{\rm (d)+(e)}(0)\Big|_{\rm NA}
&=& \frac{2 C^2_{B\phi}}{(4\pi f)^2} 
\frac{R_B^3}{M} \log\frac{\Delta_B-R_B}{\Delta_B+R_B}.
\end{eqnarray}
There is no NA contribution from these two diagrams to the Pauli form factor.
For the tadpole diagrams in Fig.~\ref{diagrams}(h) and \ref{diagrams}(j), the NA terms contain only the LNA contributions,
\begin{eqnarray}
\label{eq.LNA-F1h}
F_1^{\rm (h)}(0)\Big|_{\rm LNA}
&=& \frac{C_{\phi\phi^\dagger}}{(4\pi f)^2}\, m_\phi^2 \log m_\phi^2,
\\
F_2^{\rm (j)}(0)\Big|_{\rm LNA}
&=&\frac{C^{\rm mag}_{\phi\phi^\dagger}}{(4\pi f)^2}\, m_\phi^2\log m_\phi^2,
\end{eqnarray}
for the Dirac an Pauli form factors, respectively.
For the bubble diagram in Fig.~\ref{diagrams}(k), the NA behavior also reflects the simple LNA form for the Dirac form factor,
\begin{eqnarray}
\label{eq.LNA-F1k}
F_1^{\rm (k)}(0)\Big|_{\rm LNA}
&=& \frac{C_{\phi\phi^\dagger}}{(4\pi f)^2}\, m_\phi^2 \log m_\phi^2,
\end{eqnarray}
and is in fact identical to the LNA contributions from the tadpole diagram, Eq.~(\ref{eq.LNA-F1h}), as required by gauge invariance.
The bubble diagram with the magnetic interaction in Fig.~\ref{diagrams}(l) gives rise to a contribution to the Pauli form factor given by
\begin{eqnarray}
\label{eq.LNA-F2h}
F_2^{\rm (l)}(0)\Big|_{\rm LNA}
&=& -\frac{C'_{\phi\phi^\dagger}}{(4\pi f)^2}\, M m_\phi^2\log m_\phi^2.
\end{eqnarray}
Note that the mass dimensions of the couplings $C'_{\phi\phi^\dagger}$ here are inverse mass, in contrast to all the other couplings which are dimensionless (see Table~\ref{tab:C} in Sec.~\ref{sec.framework} above).
As discussed in Ref.~\cite{Salamu:2018cny}, the additional diagrams in Fig.~\ref{diagrams}(f), \ref{diagrams}(g) and \ref{diagrams}(i) generated from the gauge link yield no NA terms for either the Dirac or Pauli form factors.

The NA terms for the splitting functions involving intermediate decuplet baryon states can be obtained in a similar manner.
For the rainbow diagram Fig.~\ref{diagrams}(m), the NA behavior is given by
\begin{subequations}
\begin{eqnarray}
F_1^{\rm (m)}(0)\Big|_{\rm NA}
&=&\ \ \ \frac{2 C^2_{T\phi}}{(4\pi f)^2}
\bigg[ \big( m_\phi^2 - 2 \Delta_T^2 \big) \log m_\phi^2 
    + 2 \Delta_T R_T \log\frac{\Delta_T-R_T}{\Delta_T+R_T}
\bigg],
\\
F_2^{\rm (m)}(0)\Big|_{\rm NA}
&=&\!-\frac{C^2_{T\phi}}{3(4\pi f)^2} 
\bigg[ \big( 11 m_\phi^2 + 4 M \Delta_T - 6 \Delta_T^2 \big) \log m_\phi^2 
    - 4 M R_T \log\frac{\Delta_T-R_T}{\Delta_T+R_T}
\bigg].~~~~
\end{eqnarray}
\end{subequations}
The NA behavior for the decuplet baryon coupling rainbow diagram in Fig.~\ref{diagrams}(n) can be analogously written as
\begin{subequations}
\begin{eqnarray}
F_1^{\rm (n)}(0)\Big|_{\rm NA}
&=&\ \ \ \frac{2 C^2_{T\phi}}{(4\pi f)^2}
\bigg[ \big( m_\phi^2 - 2 \Delta_T^2 \big) \log m_\phi^2 
    + 2 \Delta_T R_T \log\frac{\Delta_T-R_T}{\Delta_T+R_T}
\bigg], 
\\
F_2^{\rm (n)}(0)\Big|_{\rm NA}
&=&\!-\frac{8 C^2_{T\phi }}{9(4\pi f)^2}
\bigg[ \big( m_\phi^2 - 2 \Delta_T^2 \big) \log m_\phi^2
    + 2 \Delta_T R_T \log\frac{\Delta_T-R_T}{\Delta_T+R_T}
\bigg].~~~
\end{eqnarray}
\end{subequations}
As in the octet case, the NA term for the Dirac form factor contribution from Fig.~\ref{diagrams}(n) is identical to that from Fig.~\ref{diagrams}(m).
For the additional magnetic interaction decuplet baryon rainbow diagram in Fig.~\ref{diagrams}(o), the contribution to the $F_2$ form factor is given by
\begin{eqnarray}
F_2^{\rm (o)}(0)\Big|_{\rm NA}
&=& \frac{10\, C^2_{T\phi}}{9(4\pi f)^2}
\bigg[ \big( m_\phi^2 - 2 \Delta_T^2 \big) \log m_\phi^2 
    + 2 \Delta_T R_T \log\frac{\Delta_T-R_T}{\Delta_T+R_T}
\bigg],
\end{eqnarray}
while there is no NA term for the Dirac $F_1$ form factor.

For the magnetic octet-decuplet baryon transition diagrams in Fig.~\ref{diagrams}(p) and \ref{diagrams}(q), the combined NA contribution to the Pauli form factor can be written as
\begin{eqnarray}
F_2^{\rm (p)+(q)}(0)\Big|_{\rm NA}
&=& \frac{C_{B\phi} C_{T\phi}}{6(4\pi f)^2}
\bigg[ \big( 8m_\phi^2 - 5\Delta_B^2 - 6\Delta_B\Delta_T - 5\Delta_T^2 \big)\,
\log m_\phi^2
\nonumber\\
&& \hspace*{1.5cm}
+\, R_T \frac{ 6m_\phi^2-\Delta_T(\Delta_B+5\Delta_T)}{\Delta_B-\Delta_T}
\log\frac{\Delta_T-R_T}{\Delta_T+R_T}
\nonumber\\
&& \hspace*{1.5cm}
-\, R_B \frac{ 6m_\phi^2-\Delta_B(\Delta_T+5\Delta_B)}{\Delta_B-\Delta_T}
\log\frac{\Delta_B-R_B}{\Delta_B+R_B}
\bigg],
\end{eqnarray}
with again no NA contributions to $F_1(0)$.
Finally, for the KR diagrams with decuplet intermediate states in Fig.~\ref{diagrams}(r) and \ref{diagrams}(s) the NA behavior of the Dirac and Pauli form factors is given by
\begin{subequations}
\begin{eqnarray}
F_1^{\rm (r)+(s)}(0)\Big|_{\rm NA}
&=& \ \ \frac{4 C^2_{T\phi}}{3M (4\pi f)^2}
\bigg[ 
    \Delta_T\, \big( \, m_\phi^2 \,-\,  2\Delta_T^2 \big) \log m_\phi^2 
    + 2 R_T^3 \log\frac{\Delta_T-R_T}{\Delta_T+R_T}
\bigg],
\\
F_2^{\rm (r)+(s)}(0)\Big|_{\rm NA}
&=&\!-\frac{16 C^2_{T\phi}}{9M (4\pi f)^2}
\bigg[
    \Delta_T\, \big( 3 m_\phi^2 - 2 \Delta_T^2 \big) \log m_\phi^2 
    + 2 R_T^3\log\frac{\Delta_T-R_T}{\Delta_T+R_T}
\bigg].~~~~~
\end{eqnarray}
\end{subequations}
As for the octet baryon case, the additional diagrams derived from the gauge link for the decuplet baryons do not give rise to any NA terms.
Our results for the LNA terms of the various loop diagrams are consistent with the results for the Dirac form factor discussed in Ref.~\cite{Salamu:2018cny}. There have also been some attempts made to resum the leading chiral logarithms to all orders~\cite{Kivel:2008ry, Kivel:2008mf}.

%%%%%%%%%%%%%%%%%%%%%%%%%%%%%%%%%%%%%%%%%%%%%%%%%%%%%%%%%%%%%%%%%%%%%%%%%%
\section{Convolution formalism}
\label{sec.convolution}

Having derived the full set of splitting functions for the diagrams in Fig.~\ref{diagrams} involving the SU(3) octet and decuplet baryon intermediate states, in this section we discuss the calculation of the GPDs in the proton arising from these contributions.
We derive expressions for the GPDs in terms of convolutions of the splitting functions and GPDs of quarks in the various hadronic configurations.
Using flavor symmetry constraints, we discuss relations for the GPDs in the hadronic configurations among the various SU(3) baryons.

% ........................................................................
\subsection{GPDs as convolutions}
\label{ssec.gpdconv}

The $n$-th Mellin moments of the generalized quark distributions $H^q(x,\xi,t)$ and $E^q(x,\xi,t)$ are given by~\cite{Ji:1998pc}
\begin{subequations}
\begin{eqnarray}
H_q^{(n)}(\xi,t) 
\equiv \int_{-1}^1 \dd{x} x^{n-1} H^q(x,\xi,t) 
&=& \sum_{i=0,\,{\rm even}}^{n-1} \big(\!-2\xi\big)^i A^{(n)i}_q(t) 
+ \big(\!-2\xi\big)^n C_q^{(n)}(t)\Big|_{n\,{\rm even}},~~~~~~~~
\\
E_q^{(n)}(\xi,t) 
\equiv \int_{-1}^1 \dd{x} x^{n-1} E^q(x,\xi,t) 
&=& \sum_{i=0,\,{\rm even}}^{n-1} \big(\!-2\xi\big)^i B^{(n)i}_q(t) 
- \big(\!-2\xi\big)^n C_q^{(n)}(t)\Big|_{n\,{\rm even}},~~~~~~~~
\end{eqnarray}
\label{eq:polynomial}
\end{subequations}
where $A_q^{(n)i}(t)$, $B_q^{(n)i}(t)$ and $C_q^{(n)}(t)$ are the generalized form factors of rank $n$. 
The generalized form factors can be related to the matrix elements of local twist-2 operators ${\cal O}_q^{\{\mu\mu_1 \cdots \mu_{n-1}\}}$ between nucleon states~\cite{Ji:1998pc},
\begin{eqnarray}
\langle N(p') | \mathcal{O}_{q}^{\{ \mu \mu_1 \cdots \mu_{n-1} \}} | N(p) \rangle
&=& \bar{u}(p')
\bigg[ \
\sum_{i=0,\,\rm even}^{n-1} A_q^{(n)i}(t)\,
\gamma^{ \{ \mu } \Delta^{\mu_1} \cdots \Delta^{\mu_i} 
P^{\mu_i+1}\cdots P^{\mu_{n-1} \} }
\notag\\
& & \hspace*{0.2cm}
- \frac{i}{2M}
\sum_{i=0,\,\rm even}^{n-1} B_q^{(n)i}(t)\,
\Delta_\nu \sigma^{ \nu \{ \mu} \Delta^{\mu_1} \cdots \Delta^{\mu_i} P^{\mu_i+1}\cdots P^{\mu_{n-1} \} }~~~~~
\notag\\
& & \hspace*{0.2cm}
+ \frac{1}{M}\ C_q^{(n)} (t)\Big|_{n\,{\rm even}}\
\Delta^{\{ \mu} \cdots \Delta^{\mu_{n-1} \}}
\bigg]
u(p),
\end{eqnarray}
where the symmetric and traceless operators are defined as
\begin{equation}\label{eq:O-q}
\mathcal{O}^{\{ \mu\mu_1 \cdots \mu_{n-1} \}}_q
= i^{n-1}\, \bar{\psi}_q\, \gamma^{ \{\mu }
  \overleftrightarrow{D}^{\mu_1} \cdots
  \overleftrightarrow{D}^{\mu_{n-1}\} }\, \psi_q,
\end{equation}
with $\overleftrightarrow{D} = \frac{1}{2} \big( \overrightarrow{D} - \overleftarrow{D} \big)$.
The braces $\{ \cdots \}$ represent symmetrization over the indices $\mu\mu_1 \cdots \mu_n$ and subtraction of traces.

In an effective field theory, these quark operators are matched to hadronic operators with the corresponding set of quantum numbers~\cite{Chen:2001eg},
\begin{equation}\label{eq:operator-match}
\mathcal{O}^{\{\mu\mu_1\cdots \mu_{n-1} \}}_q
= \sum_j Q^{(n-1)}_j\,
        \mathcal{O}^{\{\mu\mu_1 \cdots \mu_{n-1}\}}_j,
\end{equation}
where the subscript $j$ labels different types of hadronic operators.
The coefficients $Q^{(n-1)}_j$ can be defined through the $n$-th moments of the generalized parton distributions in the hadronic configuration $j$ (see Sec.~\ref{ssec.gpdhad} below).
Matrix elements of the hadronic operators $\mathcal{O}^{\{\mu\mu_1 \cdots \mu_{n-1}\}}_j$ can be used to define the moments of the Dirac-like and Pauli-like hadronic splitting functions $f_j$ and $g_j$, respectively, introduced in Eq.~(\ref{eq.Gamma+def}),
\begin{subequations}
\begin{eqnarray}
f^{(n)}_j &=& \int^{1}_{-1} \dd{y} y^{(n-1)} f_j(y,t), \\
g^{(n)}_j &=& \int^{1}_{-1} \dd{y} y^{(n-1)} g_j(y,t),
\end{eqnarray}
\end{subequations}
where $y$ is the light-cone momentum fraction of the nucleon carried by the hadronic state $j$.
Taking the matrix elements of the matched operators in Eq.~(\ref{eq:operator-match}) between nucleon states with unequal initial and final momenta, and contracting both sides with light-cone vector $n_\mu n_{\mu_1}\cdots n_{\mu_{n-1}}$, we then arrive at a relation for the GPD moments in terms of the moments of the hadronic splitting functions,
\begin{eqnarray}
\label{eq:fj}
& & H_q^{(n)}\, 
  \bar{u}(p') \slashed {n} u(p) 
+ \frac{i}{2M} 
  E_q^{(n)}\, 
  \bar{u}(p') n_{\mu}\sigma^{\mu\nu}\Delta_\nu\, u(p) \notag\\
& & \hspace*{-2.35cm}
= \sum_j Q^{(n-1)}_j
\bigg[ 
   f^{(n)}_j\, 
   \bar{u}(p') \slashed {n} u(p) 
 + \frac{i}{2M}\, 
   g^{(n)}_j\ 
   \bar{u}(p') n_{\mu}\sigma^{\mu\nu}\Delta_\nu\, u(p) 
\bigg].
\end{eqnarray}
Since Eq.~(\ref{eq:fj}) is valid for all moments $n$, we deduce that the corresponding convolution relation exists also for the GPDs as a function of $x$.
Specializing to the case of zero skewness, $\xi=0$, we can write the contributions to the $H^q$ and $E^q$ GPDs from the hadronic configurations in the form
\begin{subequations}
\label{eq:convolution}
\begin{eqnarray} 
H^q(x,t)\, \equiv\, H^q(x,\xi\!=\!0,t)
&=& \sum_j \big[f_j \otimes q_j^v\big](x,t)
\notag\\
&\equiv&
    \sum_j \int_0^1 \dd{y} \int_0^1 \dd{z} 
    \delta(x-yz)\, f_j(y,t)\, q_j^v(z,t),
\\
E^q(x,t)\, \equiv\, E^q(x,\xi\!=\!0,t)
&=& \sum_j \big[g_j \otimes q_j^v\big](x,t)
\notag\\
&\equiv&
    \sum_j \int_0^1 \dd{y} \int_0^1 \dd{z} 
    \delta(x-yz)\, g_j(y,t)\, q_j^v(z,t),
\end{eqnarray} 
\end{subequations}
where we define 
    $q^v_j(x,t) \equiv q_j(x,\xi\!=\!0,t) - \bar{q}_j(x,\xi\!=\!0,t)$
as the GPD of the valence quark $q$ in the hadronic configuration $j$ evaluated at zero skewness. We should mention that the above convolutions are only valid when $\xi=0$. The nonzero skewness GPDs have additional properties, such as $H^q(x,\xi,t)$ and $E^q(x,\xi,t)$ being even under the transformation $\xi \to -\xi$.  The proof of the polynomiality properties in Eq.~(\ref{eq:polynomial}) will also be an important check for the convolution equations in the nonzero skewness case.  Such a proof has been demonstrated in some nonperturbative approaches~\cite{Schweitzer:2002nm}.

Since the total $H^q$ and $E^q$ GPDs can each receive contributions from both Dirac-like and Pauli-like GPDs of the hadronic configurations, the sum over $j$ in Eqs.~(\ref{eq:convolution}) includes both electric and magnetic couplings, $q_j \to H^q_j$ or $E^q_j$.
Note also that crossing symmetry,
    $q_j(-x,\xi\!=\!0,t) = -\bar{q}_j(x,\xi\!=\!0,t)$,
in direct analogy to that for forward ($t=0$) parton distributions, has been used to write the integrals in Eqs.~(\ref{eq:convolution}) over the interval 0 to 1.

To illustrate the application of Eqs.~(\ref{eq:convolution}) to chiral loop contributions to sea quark GPDs in the proton, in this paper we focus on the asymmetries between the GPDs for $\bar{d}$ and $\bar{u}$ quarks, and between the $s$ and $\bar{s}$ quark flavors.
Assuming that the intermediate state octet and decuplet baryons are flavor symmetric, with mesons $\phi$ the only source of antiquarks, the convolution form for the antiquark electric and magnetic GPDs in the proton involves contributions only from the diagrams in Fig.~\ref{diagrams}(a), (k), (l) and~(m).
Specifically, for the $H^{\bar{q}}$ and $E^{\bar{q}}$ GPDs at zero skewness, we have
\begin{subequations}
\label{eq:HEqbar}
\begin{eqnarray}
H^{\bar{q}}(x,t)
&=& \sum_{\phi B T} 
    \Big[ \big( f_{\phi B}^{(\rm rbw)} 
              + f_{\phi T}^{(\rm rbw)} 
              + f^{(\rm bub)}_\phi 
          \big) \otimes H^{\bar{q}}_\phi
    \Big](x,t),
\\
E^{\bar{q}}(x,t)
&=& \sum_{\phi B T} 
    \Big[ \big( g_{\phi B}^{(\rm rbw)}
              + g_{\phi T}^{(\rm rbw)}
              + g'^{(\rm bub)}_\phi 
           \big) \otimes H^{\bar{q}}_\phi
    \Big](x,t),
\end{eqnarray}
\end{subequations}
where $H^{\bar{q}}_\phi$ is the electric GPD for quark flavor $\bar{q}$ in the meson $\phi$.
The splitting functions 
    $f_{\phi B}^{(\rm rbw)}$ and $g_{\phi B}^{(\rm rbw)}$
for Fig.~\ref{diagrams}(a) are given in Eqs.~(\ref{eq:split_rbw_fphiB}) and (\ref{eq:split_rbw_gphiB}), respectively, the decuplet recoil splitting functions
        $f_{\phi T}^{(\rm rbw)}$ and $g_{\phi T}^{(\rm rbw)}$
for Fig.~\ref{diagrams}(m) are given in Eqs.~(\ref{eq.f_phiTrbw}) and (\ref{eq.g_phiTrbw}), respectively, and the functions $f^{(\rm bub)}_\phi$ and $g'^{(\rm bub)}_\phi$ for Figs.~\ref{diagrams}(k) and (l) are given in Eqs.~(\ref{eq:f_phi_bub}) and (\ref{eq:gpr_phi_bub}), respectively.

The convolution form for the quark GPDs received contributions from all other diagrams in Fig.~\ref{diagrams}, and so has a more complicated structure,
\begin{subequations}
\label{eq:HEq}
\begin{eqnarray}
\label{eq:Hq}
H^q(x,t)
&=& Z_2\, H^q_0(x,t)\
 +\ \sum_{\phi B T}
\Big[ 
   \big( f_{\phi B}^{(\rm rbw)} + f_{\phi T}^{(\rm rbw)} + f^{(\rm bub)}_\phi \big) \otimes H^q_\phi 
\nonumber\\
& & \hspace*{3.3cm}
+\ \bar{f}_{B\phi}^{(\rm rbw)} \otimes H^q_B\
+\ \bar{f}^{(\rm KR)}_{B\phi} \otimes H^{q({\rm KR})}_B\
+\ \delta \bar{f}^{(\rm KR)}_{B} \otimes H^{q({\rm KR})}_B\
\nonumber\\
& & \hspace*{3.3cm}
+\ \bar{f}_{T\phi}^{(\rm rbw)} \otimes H^q_T\
+\ \bar{f}^{(\rm KR)}_{T\phi} \otimes H^{q({\rm KR})}_T\
+\ \delta\bar{f}^{(\rm KR)}_{T\phi} \otimes H^{q({\rm KR})}_T\
\nonumber\\
& & \hspace*{3.3cm}
+\ \bar{f}_{B\phi}^{(\rm rbw\,mag)} \otimes E^q_B\
+\ \bar{f}_{T\phi}^{(\rm rbw\,mag)} \otimes E^q_T\
+\ \bar{f}_{BT}^{(\rm rbw\,mag)} \otimes E^q_{BT}~~~~~~
\nonumber\\
& & \hspace*{3.3cm}
+\ \bar{f}^{(\rm tad)}_\phi \otimes H^{q({\rm tad})}_{\phi\phi^\dag}\
+\ \delta\bar{f}^{(\rm tad)}_\phi \otimes H^{q({\rm tad})}_{\phi\phi^\dag}
\Big](x,t),
\nonumber\\
& &
\\
\label{eq:Eq}
E^q(x,t)
&=& Z_2\, E^q_0(x,t)\
 +\ \sum_{\phi B T}
\Big[ 
   \big( g_{\phi B}^{(\rm rbw)} + g_{\phi T}^{(\rm rbw)} + g'^{(\rm bub)}_\phi \big) \otimes H^q_\phi
\nonumber\\
& & \hspace*{3.3cm}
+\ \bar{g}_{B\phi}^{(\rm rbw)} \otimes H^q_B\
+\ \bar{g}^{(\rm KR)}_{B\phi} \otimes H^{q({\rm KR})}_B\
+\ \delta \bar{g}^{(\rm KR)}_{B} \otimes H^{q({\rm KR})}_B\
\nonumber\\
& & \hspace*{3.3cm}
+\ \bar{g}_{T\phi}^{(\rm rbw)} \otimes H^q_T\
+\ \bar{g}^{(\rm KR)}_{T\phi} \otimes H^{q({\rm KR})}_T\
+\ \delta\bar{g}^{(\rm KR)}_{T\phi} \otimes H^{q({\rm KR})}_T\
\nonumber\\
& & \hspace*{3.3cm}
+\ \bar{g}_{B\phi}^{(\rm rbw\,mag)} \otimes E^q_B\
+\ \bar{g}_{T\phi}^{(\rm rbw\,mag)} \otimes E^q_T\
+\ \bar{g}_{BT}^{(\rm rbw\,mag)} \otimes E^q_{BT}~~~~~~
\nonumber\\
& & \hspace*{3.3cm}
+\ \bar{g}^{(\rm tad\,mag)}_\phi \otimes E^{q({\rm tad})}_{\phi\phi^\dag}
\Big](x,t),
\end{eqnarray}
\end{subequations}
where $H^q_0$ and $E^q_0$ are the quark GPDs of the bare proton, and $Z_2$ is the wave function renormalization constant associated with the dressing of the bare proton by the meson loops in Fig.~\ref{diagrams}.
As shorthand, in Eqs.~(\ref{eq:HEq}) we use the notation $\bar{f}_j(y) \equiv f_j(1-y)$ and $\bar{g}_j(y) \equiv g_j(1-y)$ for the electric and magnetic splitting functions involving couplings to baryons.
Note that both the electric and magnetic operators contribute to $H^q(x,t)$ and $E^q(x,t)$ at zero and finite momentum transfer.
At zero momentum transfer, however, there is no contribution from the magnetic term to the matrix element, even though the GPD $E^q(x,0)$ itself is nonzero.

The expressions for the quark and antiquark GPDs in Eqs.~(\ref{eq:HEqbar}) and (\ref{eq:HEq}) form the basis for our calculations of the meson loop contributions to the GPD flavor asymmetries.
For the case of $u$ and $d$ quarks, the intermediate states include the nucleon and $\Delta$ baryons and $\pi$ mesons.
For the strange quark, on the other hand, the intermediate states that contribute are the hyperons $\Lambda$, $\Sigma$ and $\Sigma^*$ and $K$ mesons.
To compute the quark and antiquark GPDs numerically, we require information on the GPDs for the various hadronic configurations that contribute in Eqs.~(\ref{eq:HEqbar})--(\ref{eq:HEq}), which we turn to next.

% ........................................................................
\subsection{GPDs of hadronic configurations}
\label{ssec.gpdhad}

% The splitting functions were calculated in section V. 
% Now we derive the relationships for the input GPDs of different quark flavours. %
The twist-two operators associated with the PDFs of the hadronic configurations have been discussed in detail in Refs.~\cite{Wang:2016ndh, Salamu:2019dok}.
For GPDs at finite momentum transfer $t$, the operators are somewhat more complicated, and we consider first the intermediate octet baryons as the hadronic configuration as an example. 
The relevant operator here can be written as 
\begin{eqnarray}
\label{opB}
{\cal O}_{q\,B}^{\mu\mu_1\ldots\mu_{n-1}}
&=& \sum_{i=0,\rm even}^{n-1}
\frac12
\Big(
    \alpha_A^{(n)i}\,
{\rm Tr}
\left[ \bar{B} \gamma^\mu \left\{(u^\dag \lambda_q u + u \lambda_q u^\dag),B\right\}
\right] 
\nonumber\\
& & \hspace{0.73cm}
 +\ \beta_A^{(n)i}\,
{\rm Tr}
\left[ \bar{B} \gamma^\mu \left[(u^\dag \lambda_q u + u
\lambda_q u^\dag),B \right] \right]
\nonumber\\
& & \hspace{0.73cm}
 +\ \sigma_A^{(n)i}\,
{\rm Tr}
\left[\bar{B} \gamma^\mu B \right]
{\rm Tr}
\left[(u^\dag \lambda_q u + u \lambda_q u^\dag) \right]
\Big)\,
\Delta^{\mu_1} \cdots \Delta^{\mu_i} P^{\mu_{i+1}}\cdots P^{\mu_{n-1}}
\nonumber\\
&+& \sum_{i=0,\rm even}^{n-1}
\frac{i}{4M_B}
\Big(
\alpha_B^{(n)i}\,
{\rm Tr}
\left[ \bar{B} \sigma^{\mu\nu} \left\{(u^\dag \lambda_q u + u \lambda_q u^\dag),B\right\}
\right]
\nonumber\\
& & \hspace{0.73cm}
 +\ \beta_B^{(n)i}\,
{\rm Tr}
\left[ \bar{B} \sigma^{\mu\nu} \left[(u^\dag \lambda_q u + u \lambda_q u^\dag),B \right]\right]
\nonumber\\
& & \hspace{0.73cm}
 +\ \sigma_B^{(n)i}\,
{\rm Tr}
\left[ \bar{B} \sigma^{\mu\nu}B \right]
{\rm Tr}
\left[ (u^\dag \lambda_q u + u \lambda_q u^\dag) \right]
\Big)\,
\Delta^\nu \Delta^{\mu_1} \cdots \Delta^{\mu_i} P^{\mu_{i+1}}\cdots P^{\mu_{n-1}}     \nonumber\\
&+& \frac{1}{2M_B} 
\Big(
\alpha_C^{(n)}\big|_{n=\rm even}\,
{\rm Tr}
\left[ \bar{B} \left\{(u^\dag \lambda_q u + u \lambda_q u^\dag),B\right\} \right] \nonumber\\
& & \hspace{0.73cm}
 +\ \beta_C^{(n)}\big|_{n=\rm even}\,
{\rm Tr}
\left[ \bar{B} \left[(u^\dag \lambda_q u + u \lambda_q u^\dag),B \right] \right]   \nonumber\\
& & \hspace{0.73cm}
 +\ \sigma_C^{(n)}\big|_{n=\rm even}\,
{\rm Tr}
\left[ \bar{B} B \right]
{\rm Tr}
\left[ (u^\dag \lambda_q u + u \lambda_q u^\dag) \right]
\Big)\,
\Delta^\mu\Delta^{\mu_1} \cdots \Delta^{\mu_{n-1}}.
\end{eqnarray}
Contracting both sides of Eq.~(\ref{opB}) with the light-front unit vectors $n_\mu n_{\mu_1} \cdots n_{\mu_{n-1}}$ then gives
\begin{eqnarray}
\label{eq:octgpds}
n_\mu n_{\mu_1}\cdots n_{\mu_{n-1}}{\cal O}_{ q\,B}^{\mu\mu_1\ldots\mu_{n-1}}
&=& \frac12 \alpha^{(n)}\,
{\rm Tr}
\left[ \bar{B} \slashed{n} \left\{(u^\dag \lambda_q u + u \lambda_q u^\dag),B\right\} 
\right] 
\nonumber\\
&+& \frac12 \beta^{(n)}\,
{\rm Tr}
\left[ \bar{B} \slashed{n} \left[(u^\dag \lambda_q u + u \lambda_q u^\dag),B \right]
\right] 
\nonumber\\
&+& \frac12 \sigma^{(n)}\,
{\rm Tr}
\left[ \bar{B} \slashed{n} B \right]
{\rm Tr}
\left[ (u^\dag \lambda_q u + u \lambda_q u^\dag) \right]
\nonumber\\
&+& \frac{i\alpha_{\rm mag}^{(n)}}{8M_B}\,
{\rm Tr}\left[\bar{B} n_\mu\sigma^{\mu\nu}\Delta_\nu
\left\{(u^\dag \lambda_q u + u
\lambda_q u^\dag),B\right\}\right] 
\nonumber\\
&+& \frac{i\beta_{\rm mag}^{(n)}}{8M_B}\,
{\rm Tr}\left[\bar{B}
n_\mu\sigma^{\mu\nu}\Delta_\nu \left[(u^\dag \lambda_q u + u
\lambda_q u^\dag),B \right]\right]  
\nonumber\\
&+& \frac{i\sigma_{\rm mag}^{(n)}}{8M_B}\,
{\rm Tr}\left[\bar{B}
n_\mu\sigma^{\mu\nu}\Delta_\nu B \right]
{\rm Tr}
\left[ (u^\dag \lambda_q u + u \lambda_q u^\dag)
\right],
\end{eqnarray}
where for shorthand we define by $X^{(n)}$ and $X_{\rm mag}^{(n)}$, where $X=\alpha$, $\beta$ or $\sigma$, the following combinations,
\begin{subequations}
\begin{eqnarray}
X^{(n)}
&=& \sum_{i=0,{\rm even}}^{n-1} 
  (-2\xi)^i X_A^{(n)i} 
+ (-2\xi)^n X_C^{(n)}\Big|_{n={\rm even}},
\\
X_{\rm mag}^{(n)}
&=& \sum_{i=0,{\rm even}}^{n-1} 
  (-2\xi)^i X_B^{(n)i} 
- (-2\xi)^n X_C^{(n)}\Big|_{n={\rm even}}.
\end{eqnarray}
\end{subequations}
The $n$-th moments of the GPDs in octet baryons can then be related to the coefficients $X^{(n)}$ and $X^{(n)}_{\rm mag}$.
By expanding the matrices in Eq.~(\ref{eq:octgpds}), one can derive relations between the quark GPDs of different flavors in the octet baryons.
Since such relations are independent of the momentum transfer and structure of the $\gamma$-matrices of the operator, the relations between the Dirac-like GPDs $H^q(x,t)$ will be the same as those for PDFs obtained in Refs.~\cite{Wang:2016ndh, Salamu:2019dok}. 
In the following, we therefore focus on the derivation of the relations for the spin-flip GPDs, $E^q(x,t)$, and examine whether the relations between the different flavors are the same as those for the PDFs.

\clearpage
For the magnetic operator, the contraction with the light-front unit vectors gives
\begin{eqnarray}
\label{eq.opmag}
n_\mu n_{\mu_1}\cdots n_{\mu_{n-1}} {\cal O}_{q\,\rm mag}^{\mu\mu_1\ldots\mu_{n-1}}
&=& \frac{i\alpha_{\rm mag}^{(n)}}{4M_B}\,
{\rm Tr}
\left[ \bar{B} n_\mu \sigma^{\mu\nu}\Delta_\nu \left\{(u^\dag \lambda_q u + u
\lambda_q u^\dag),B\right\}
\right] 
\nonumber\\
&+& \frac{i\beta_{\rm mag}^{(n)}}{4M_B}\,
{\rm Tr}
\left[ \bar{B} n_\mu \sigma^{\mu\nu}\Delta_\nu \left[(u^\dag \lambda_q u + u
\lambda_q u^\dag),B \right]
\right]  
\nonumber\\
&+& \frac{i\sigma_{\rm mag}^{(n)}}{4M_B}\,
{\rm Tr}
\left[ \bar{B} n_\mu \sigma^{\mu\nu}\Delta_\nu B 
\right]
{\rm Tr}
\left[ (u^\dag \lambda_q u + u \lambda_q u^\dag)
\right]
\nonumber\\
&-& \frac{i\theta_{\rm mag}^{(n)}}{2M_T}\,
\overline{T}_\alpha^{abc} n_\mu \sigma^{\mu\nu} \Delta_\nu (\lambda_q^+)_{ae} T_\alpha^{ebc}
\nonumber\\
&-& \frac{i\rho_{\rm mag}^{(n)}}{2M_T}\,
\overline{T}_\alpha^{abc} n_\mu \sigma^{\mu\nu} \Delta_\nu T_\alpha^{abc}\,
{\rm Tr}[\lambda_q^+]
\notag\\
&+& \frac{\omega_{\rm mag}^{(n)}}{4M_B}\,
\Big[ 
  \epsilon_{ijk}(\lambda_q^+)_{il} \bar{B}^j_m \slashed{\Delta} \gamma_5 n_\mu T^{\mu,klm} 
\notag\\
& & \hspace*{0.9cm}
- \epsilon_{ijk}(\lambda_q^+)_{il} \bar{B}^j_m \slashed{n} \gamma_5 \Delta_\nu T^{\nu,klm}
\Big] 
+ {\rm H.c.}
\end{eqnarray}
The coefficients $\big\{ \alpha_{\rm mag}^{(n)}$, $\beta_{\rm mag}^{(n)}$, $\sigma_{\rm mag}^{(n)}$, $\theta_{\rm mag}^{(n)}$, $\rho_{\rm mag}^{(n)}$, $\omega_{\rm mag}^{(n)} \big\}$ are related to the $n$-th magnetic moments of the quark distributions in the corresponding hadronic configurations.
With the simplification of the flavor matrices, the contracted magnetic operator can be rewritten in the form
\begin{eqnarray}
&& \hspace*{-0.5cm}
n_\mu n_{\mu_1}\cdots n_{\mu_{n-1}}{\cal O}_{q\,\rm mag}^{\mu\mu_1\cdots\mu_{n-1}}
\notag\\
&& \hspace*{0.7cm}
=\, Q_{B\, \rm mag}^{(n-1)}\, {\cal O}_{B\, \rm mag}\,
+\, Q_{T\, \rm mag}^{(n-1)}\, {\cal O}_{T\, \rm mag}\,
+\, Q_{BT\, \rm mag}^{(n-1)}\, {\cal O}_{BT\, \rm mag}\,
+\, Q_{\phi\phi^\dag\, \rm mag}^{(n-1)}\, {\cal O}_{\phi\phi^\dag\, \rm mag},
\end{eqnarray}
where the hadronic operators are given by
\begin{subequations}
\begin{eqnarray}
{\cal O}_{B\, \rm mag}
&=& \frac{i}{2M_B}\,
\bar{B} n_\mu\sigma^{\mu\nu} \Delta_\nu B,
\\
{\cal O}_{T\, \rm mag}
&=& -\frac{i}{2M_T}\,
\overline{T}_\alpha n_\mu \sigma^{\mu\nu} \Delta_\nu T_\alpha,
\\
{\cal O}_{BT\, \rm mag}
&=& -\frac{\sqrt{3}}{2\sqrt{2}M_B}\,
\big( \bar{B} \slashed{\Delta}\gamma_5 n_\mu T^\mu-\bar{B} \slashed{n}\gamma_5 \Delta_\nu T^\nu
\big),
\\
{\cal O}_{\phi\phi^\dag\, \rm mag}
&=&-\frac{i}{2M_Bf^2}\,
\bar{B} n_\mu\sigma^{\mu\nu} \Delta_\nu B\, \phi\phi^\dag.
\end{eqnarray}
\end{subequations}
The coefficients $Q^{(n-1)}_{j\,\rm mag}$ of each of the operators are defined in terms of the Mellin moments of the corresponding distribution functions in the intermediate hadron states,
\begin{subequations}
\begin{eqnarray}
Q^{(n-1)}_{B\,\rm mag}
&=& \int_{-1}^1 \dd{x} x^{n-1}\, E^q_B(x,t), 
\\
Q^{(n-1)}_{T\,\rm mag}
&=& \int_{-1}^1 \dd{x} x^{n-1}\, E^q_T(x,t), 
\\
Q^{(n-1)}_{BT\,\rm mag}
&=& \int_{-1}^1 \dd{x} x^{n-1}\, E^q_{BT}(x,t),
\\
Q^{(n-1)}_{\phi\phi^\dag\,\rm mag}
&=& \int_{-1}^1 \dd{x} x^{n-1}\, E^q_{\phi\phi^\dag}(x,t),
\end{eqnarray}
\end{subequations}
where the GPDs correspond to those appearing in the convolution expressions on the right-hand-sides of Eqs.~(\ref{eq:HEq}).

\begin{table}[b]
\begin{center}
\caption{Moments $Q^{(n-1)}_{B\,\rm mag}$ of the $u$, $d$ and $s$ quark GPDs in octet baryons arising from the magnetic interaction.\\}
\begin{tabular}{l|c|c|c} \hline
& & &
\\
~$B$ &
$U^{(n-1)}_{B\,\rm mag}$ &
$D^{(n-1)}_{B\,\rm mag}$ & 
$S^{(n-1)}_{B\,\rm mag}$ 
\\ 
& & &
\\ \hline
& & &
\\
~\,$p$ &
~$\alpha_{\rm mag}^{(n)}+\beta_{\rm mag}^{(n)}+\sigma_{\rm mag}^{(n)}$~ &
$\sigma_{\rm mag}^{(n)}$ &
~$\alpha_{\rm mag}^{(n)}-\beta_{\rm mag}^{(n)}+\sigma_{\rm mag}^{(n)}$~
\\
& & &
\\
~\,$n$ &
$\sigma_{\rm mag}^{(n)}$ &
~$\alpha_{\rm mag}^{(n)}+\beta_{\rm mag}^{(n)}+\sigma_{\rm mag}^{(n)}$~ &
$\alpha_{\rm mag}^{(n)}-\beta_{\rm mag}^{(n)}+\sigma_{\rm mag}^{(n)}$
\\
& & &
\\
~$\Sigma^+$ &
$\alpha_{\rm mag}^{(n)}+\beta_{\rm mag}^{(n)}+\sigma_{\rm mag}^{(n)}$ &
$\alpha_{\rm mag}^{(n)}-\beta_{\rm mag}^{(n)}+\sigma_{\rm mag}^{(n)}$ &
$\sigma_{\rm mag}^{(n)}$
\\
& & &
\\
~$\Sigma^0$ &
$\alpha_{\rm mag}^{(n)}+\sigma_{\rm mag}^{(n)}$ &
$\alpha_{\rm mag}^{(n)}+\sigma_{\rm mag}^{(n)}$ &
$\sigma_{\rm mag}^{(n)}$ 
\\
& & &
\\
~$\Sigma^-$ &
$\alpha_{\rm mag}^{(n)}-\beta_{\rm mag}^{(n)}+\sigma_{\rm mag}^{(n)}$ &
$\alpha_{\rm mag}^{(n)}+\beta_{\rm mag}^{(n)}+\sigma_{\rm mag}^{(n)}$ &
$\sigma_{\rm mag}^{(n)}$ 
\\
& & &
\\
~$\Lambda$ &
$\frac13\alpha_{\rm mag}^{(n)}+\sigma_{\rm mag}^{(n)}$ &
$\frac13\alpha_{\rm mag}^{(n)}+\sigma_{\rm mag}^{(n)}$ &
$\frac43\alpha_{\rm mag}^{(n)}+\sigma_{\rm mag}^{(n)}$ 
\\
& & &
\\
$\Lambda\Sigma^0$~ &
$\frac{1}{\sqrt3} \alpha_{\rm mag}^{(n)}$ &
$-\frac{1}{\sqrt3}\alpha_{\rm mag}^{(n)}$ &
$0$ 
\\ 
& & &
\\
\hline
\end{tabular}
\label{tab:BB}
\end{center}
\end{table}

The moments $U_{B\,\rm mag}^{(n-1)}$, $D_{B\,\rm mag}^{(n-1)}$ and $S_{B\,\rm mag}^{(n-1)}$ of the $u$, $d$ and $s$ quark GPDs in the octet baryons can be expressed in terms of the coefficients $\alpha_{\rm mag}^{(n)}$, $\beta_{\rm mag}^{(n)}$ and $\sigma_{\rm mag}^{(n)}$, as listed in Table~\ref{tab:BB}.
Solving for the coefficients, one can write these as linear combinations of the quark GPD moments in the proton,
\begin{subequations}
\label{eq.alpha-beta-sigma-mag}
\begin{eqnarray}
\alpha_{\rm mag}^{(n)}
&=& \frac12 
\Big( U_{p\,\rm mag}^{(n-1)} + S_{p\,\rm mag}^{(n-1)} \Big)
- D_{p\,\rm mag}^{(n-1)},
\\
\beta_{\rm mag}^{(n)}
&=& \frac12
\Big( U_{p\,\rm mag}^{(n-1)} - S_{p\,\rm mag}^{(n-1)} \Big),
\\
\sigma_{\rm mag}^{(n)}
&=& D_{p\,\rm mag}^{(n-1)}.
\end{eqnarray}
\end{subequations}
Furthermore, assuming the strangeness in the bare nucleon state to be zero, we have
\begin{equation}
\label{eq.sig-beta-alpha-mag}
\sigma_{\rm mag}^{(n)} = \beta_{\rm mag}^{(n)} - \alpha_{\rm mag}^{(n)}.
\end{equation}
In particular, for $n=1$ the relation Eq.~(\ref{eq.sig-beta-alpha-mag}) is consistent with the Lagrangian for the magnetic interaction, Eq.~(\ref{lomag}), where $c_3=c_2-c_1$.

\begin{table}[b]
\begin{center}
\caption{Moments $Q^{(n-1)}_{T\,\rm mag}$ of the $u$, $d$ and $s$ quark GPDs in decuplet baryons arising from the magnetic interaction.\\}
\begin{tabular}{l|c|c|c} \hline
& & &
\\
~~$T$ & 
$U^{(n-1)}_{T\,\rm mag}$ &
$D^{(n-1)}_{T\,\rm mag}$ &
$S^{(n-1)}_{T\,\rm mag}$
\\ 
& & &
\\ \hline
& & &
\\
~$\Delta^{++}$~ & 
$\theta_{\rm mag}^{(n)}+\rho_{\rm mag}^{(n)}$ &
$\rho_{\rm mag}^{(n)}$ & $\rho_{\rm mag}^{(n)}$ 
\\
& & &
\\
~$\Delta^+$ &
$\frac23\theta_{\rm mag}^{(n)}+\rho_{\rm mag}^{(n)}$ & 
$\frac13\theta_{\rm mag}^{(n)}+\rho_{\rm mag}^{(n)}$ &
$\rho_{\rm mag}^{(n)}$ 
\\
& & &
\\
~$\Delta^0$ &
$\frac13\theta_{\rm mag}^{(n)}+\rho_{\rm mag}^{(n)}$ & 
$\frac23\theta_{\rm mag}^{(n)}+\rho_{\rm mag}^{(n)}$ & 
$\rho_{\rm mag}^{(n)}$ 
\\
& & &
\\
~$\Delta^-$ & 
$\rho_{\rm mag}^{(n)}$ & 
$\theta_{\rm mag}^{(n)}+\rho_{\rm mag}^{(n)}$ & 
$\rho_{\rm mag}^{(n)}$ 
\\
& & &
\\
~$\Sigma^{*+}$ &
$\frac23\theta_{\rm mag}^{(n)}+\rho_{\rm mag}^{(n)}$ & 
$\rho_{\rm mag}^{(n)}$ &
$\frac13\theta_{\rm mag}^{(n)}+\rho_{\rm mag}^{(n)}$
\\
& & &
\\
~$\Sigma^{*0}$ & 
~$\frac13\theta_{\rm mag}^{(n)}+\rho_{\rm mag}^{(n)}$~ &
~$\frac13\theta_{\rm mag}^{(n)}+\rho_{\rm mag}^{(n)}$~ & 
~$\frac13\theta_{\rm mag}^{(n)}+\rho_{\rm mag}^{(n)}$~ 
\\
& & &
\\
~$\Sigma^{*-}$ & 
$\rho_{\rm mag}^{(n)}$ &
$\frac23\theta_{\rm mag}^{(n)}+\rho_{\rm mag}^{(n)}$ &
$\frac13\theta_{\rm mag}^{(n)}+\rho_{\rm mag}^{(n)}$ 
\\
& & &
\\ \hline
\end{tabular}
\label{tab:TT}
\end{center}
\end{table}

The moments of the GPDs in the decuplet baryons, $Q_{T\,\rm mag}^{(n-1)}$, for $u$, $d$ and $s$ quarks can be expressed in terms of the coefficients $\theta_{\rm mag}^{(n)}$ and $\rho_{\rm mag}^{(n)}$, which are listed in Table~\ref{tab:TT}.
Solving for the coefficients $\theta_{\rm mag}^{(n)}$ and $\rho_{\rm mag}^{(n)}$ in terms of the GPD moments in the $\Delta^+$ baryon, one has
\begin{subequations}
\begin{eqnarray}
\theta_{\rm mag}^{(n)}
&=& 3 \Big( D_{\Delta^+}^{(n-1)} - S_{\Delta^+}^{(n-1)} \Big)
= \frac32 \Big( U_{\Delta^+}^{(n-1)} - S_{\Delta^+}^{(n-1)} \Big),
\\
\rho_{\rm mag}^{(n)}
&=& S_{\Delta^+}^{(n-1)}.
\end{eqnarray}
\end{subequations}
Proceeding in analogy with the nucleon case, we assume no strange quark contribution in the undressed $\Delta$ baryons, which enables the decuplet coefficients to be written in terms of the $u$ and $d$ quark GPD moments in the $\Delta^+$,
\begin{eqnarray}
\theta_{\rm mag}^{(n)}
&=& 3D_{\Delta^+}^{(n-1)}\,
 =\, \frac32 U_{\Delta^+}^{(n-1)},
\qquad
\rho_{\rm mag}^{(n)}\,
 =\, 0.
\end{eqnarray}
This is also consistent with the effective Lagrangian of Eq.~(\ref{lomag}), where only one term for the decuplet magnetic interaction is included.

\begin{table}[b]
\begin{center}
\caption{Moments $Q^{(n-1)}_{BT\,\rm mag}$ of the $u$, $d$ and $s$ quark distributions arising from the magnetic interactions for the octet-decuplet transition.\\}
\begin{tabular}{l|c|c|c} \hline
& & &
\\
~~$BT$ &
$U^{(n-1)}_{BT\,\rm mag}$ & 
$D^{(n-1)}_{BT\,\rm mag}$ & 
$S^{(n-1)}_{BT\,\rm mag}$ 
\\ 
& & &
\\ \hline
& & &
\\
$~~p\Delta^{+}$ & 
$\frac{1}{3\sqrt2}\, \omega_{\rm mag}^{(n)}$ &
~$-\frac{1}{3\sqrt2}\, \omega_{\rm mag}^{(n)}$~ &
$0$ 
\\
& & &
\\
$~~n\Delta^0$ &
$\frac{1}{3\sqrt2}\, \omega_{\rm mag}^{(n)}$ &
$-\frac{1}{3\sqrt2}\, \omega_{\rm mag}^{(n)}$ &
$0$
\\
& & &
\\
~$\Sigma^+\Sigma^{*+}$ &
~$-\frac{1}{3\sqrt2}\, \omega_{\rm mag}^{(n)}$~ &
$0$ &
$\frac{1}{3\sqrt2}\, \omega_{\rm mag}^{(n)}$
\\
& & &
\\
~$\Sigma^0\Sigma^{*0}$ & 
$\frac{1}{6\sqrt2}\, \omega_{\rm mag}^{(n)}$ & 
$\frac{1}{6\sqrt2}\, \omega_{\rm mag}^{(n)}$ &
$-\frac{1}{3\sqrt2}\, \omega_{\rm mag}^{(n)}$
\\
& & &
\\
~$\Sigma^-\Sigma^{*-}$ &
$0$ &
$\frac{1}{3\sqrt2}\, \omega_{\rm mag}^{(n)}$ &
$-\frac{1}{3\sqrt2}\, \omega_{\rm mag}^{(n)}$
\\
& & &
\\
~~$\Lambda\Sigma^{*0}$ & 
~$-\frac{1}{\sqrt{24}}\, \omega_{\rm mag}^{(n)}$~ &
$\frac{1}{\sqrt{24}}\, \omega_{\rm mag}^{(n)}$ & 
$0$
\\
& & &
\\ \hline
\end{tabular}
\label{tab:BT}
\end{center}
\end{table}

The moments of quark GPDs for the octet-decuplet transition, $Q_{BT\,\rm mag}^{(n-1)}$, can be expressed in terms of the coefficient $\omega_{\rm mag}^{(n)}$ defined in Eq.~(\ref{eq.opmag}), and are given in Table~\ref{tab:BT} for the allowed configurations.
One can write $\omega_{\rm mag}^{(n)}$ in terms of the proton--$\Delta^+$ transition GPD moments as
\begin{equation}
\omega_{\rm mag}^{(n)}
= 3\sqrt2\, U_{p\Delta^+}^{(n-1)} = -3\sqrt2\, D_{p\Delta^+}^{(n-1)},
\end{equation}
with relations for the other octet-decuplet transitions in Table~\ref{tab:BT}.

The moments of the distributions generated by the tadpole vertex are listed in Table~\ref{tab:tad}, where the corresponding moments $Q^{(n-1)}_{ \phi\phi^\dag\,\rm mag}$ of the $u$, $d$ and $s$ quark GPDs are expressed in terms of the $\alpha_{\rm mag}^{(n)}$ and $\beta_{\rm mag}^{(n)}$ coefficients in Eq.~(\ref{eq.alpha-beta-sigma-mag}).
Note that combinations involving $K^0 \overline {K}^0$ mesons do not contribute to the $u$-quark moments, while those involving $K^+ K^-$ do not contribute to the $d$-quark moments, and contributions from $\pi^+ \pi^-$ configurations to the $s$-quark moments also vanish.

\begin{table}[t]
\begin{center}
\caption{Moments $Q^{(n-1)}_{B\phi\phi\,\rm mag}$ of the $u$, $d$ and $s$ quark GPDs arising from the magnetic interactions involving the $BB\phi\phi$ tadpole vertex. For short-hand we define 
$\Gamma_\pm^{(n)} \equiv \alpha_{\rm mag}^{(n)} \pm \beta_{\rm mag}^{(n)}$.\\}
\begin{tabular}{l|cc|cc|cc} \hline
& \multicolumn{2}{c|}{} & \multicolumn{2}{c|}{} &
\\
& \multicolumn{2}{c|}{$U^{(n-1)}_{\phi\phi\,\rm mag}$}
& \multicolumn{2}{c|}{$D^{(n-1)}_{\phi\phi\,\rm mag}$}
& \multicolumn{2}{c}{$S^{(n-1)}_{\phi\phi\,\rm mag}$}
\\
~~$B$ & & & & &
\\
& $\pi^+ \pi^-$ & $K^+ K^-$ 
& $\pi^+ \pi^-$ & $K^0 \overline{K}^0$
& $K^0 \overline{K}^0$
& $K^+ K^-$ 
\\ 
& & & & &
\\ \hline
& & & & &
\\
~~\,$p$ &
~$\frac12 \Gamma_+^{(n)}$ & 
$\beta_{\rm mag}^{(n)}$ & 
$-\frac12 \Gamma_+^{(n)}$ &
$-\frac12 \Gamma_-^{(n)}$ & 
$\frac12 \Gamma_-^{(n)}$ &
$-\beta_{\rm mag}^{(n)}$ 
\\
& & & & &
\\
~~\,$n$ &
$-\frac12 \Gamma_+^{(n)}$ &
$-\frac12 \Gamma_-^{(n)}$ &
$\frac12 \Gamma_+^{(n)}$ &
$\beta_{\rm mag}^{(n)}$ &
$-\beta_{\rm mag}^{(n)}$ &
$\frac12 \Gamma_-^{(n)}$
\\
& & & & &
\\
~~$\Sigma^+$ &
$\beta_{\rm mag}^{(n)}$ &
$\frac12 \Gamma_+^{(n)}$ &
$-\beta_{\rm mag}^{(n)}$ &
$\frac12 \Gamma_-^{(n)}$ &
$-\frac12 \Gamma_-^{(n)}$ &
$-\frac12 \Gamma_+^{(n)}$
\\
& & & & &
\\
~~$\Sigma^0$ & 
$0$ &
$\frac12 \alpha_{\rm mag}^{(n)}$ &
$0$ &
$\frac12 \alpha_{\rm mag}^{(n)}$ &
$-\frac12 \alpha_{\rm mag}^{(n)}$ &
$-\frac12 \alpha_{\rm mag}^{(n)}$ 
\\
& & & & &
\\
~~$\Sigma^-$ &
$-\beta_{\rm mag}^{(n)}$ &
$\frac12 \Gamma_-^{(n)}$ &
$\beta_{\rm mag}^{(n)}$ &
$\frac12 \Gamma_+^{(n)}$ & 
$-\frac12 \Gamma_+^{(n)}$ & 
$-\frac12 \Gamma_-^{(n)}$  
\\
& & & & &
\\
~~$\Lambda$ & 
$0$ &
$-\frac12 \alpha_{\rm mag}^{(n)}$ &
$0$ &
$-\frac12 \alpha_{\rm mag}^{(n)}$ &
$\frac12 \alpha_{\rm mag}^{(n)}$ &
$\frac12 \alpha_{\rm mag}^{(n)}$ 
\\
& & & & &
\\
~$\Lambda\Sigma^0$~ &
~$\frac{1}{\sqrt3}\, \alpha_{\rm mag}^{(n)}$~ &
~$\frac{1}{2\sqrt3}\, \alpha_{\rm mag}^{(n)}$~ &
~$-\frac{1}{\sqrt3}\, \alpha_{\rm mag}^{(n)}$~ &
~$-\frac{1}{2\sqrt3}\, \alpha_{\rm mag}^{(n)}$~ &
~$\frac{1}{2\sqrt3}\, \alpha_{\rm mag}^{(n)}$~ &
~$-\frac{1}{2\sqrt3}\, \alpha_{\rm mag}^{(n)}$~
\\ 
& & & & &
\\ \hline
\end{tabular}
\label{tab:tad}
\end{center}
\end{table}

% We have discussed the $n$-th magnetic moments in different hadron configurations which do not appear for the PDFs. 
% The other moments have been discussed in detail in Refs.~\cite{Wang:2016ndh, Salamu:2019dok}. 
Since the above relations for the GPD moments are valid for all values of $n$, one can derive from the moments explicit relations between the input valence GPDs for different quark flavors in various hadrons.
Focusing still on the magnetic GPD $E^q$, we relate the valence GPDs in the octet baryons to the GPDs in the proton, denoted by
        $E^q \equiv E^q_p(x,t) \equiv E^q_p(x,\xi=0,t)$,
using the results from Table~\ref{tab:BB}, 
%
%
%\begin{eqnarray}
%E^u_n (x,t) &=& E^d(x,t),~~~~ E^d_n(x,t)= E^u(x,t),~~~~E^s_n(x,t) = E^s(x,t),
%\nonumber \\
%E^u_{\Sigma^+} (x,t) &=& E^u(x,t),~~~~ E^d_{\Sigma^+}(x,t)= %E^s(x,t),~~~~E^s_{\Sigma^+}(x,t) = E^d(x,t), \nonumber \\
%E^u_{\Sigma^0} (x,t) &=& \frac12[E^u(x,t)+E^s(x,t)],~~~~ E^d_{\Sigma^0}(x,t)= %E^u_{\Sigma^0}(x,t), ~~~~ E^s_{\Sigma^0}(x,t) = E^d(x,t), \nonumber \\
%E^u_{\Sigma^-} (x,t) &=& E^s(x,t),~~~~ E^d_{\Sigma^-}(x,t)= %E^u(x,t),~~~~E^s_{\Sigma^-}(x,t) = E^d(x,t), \nonumber \\
%E^u_{\Lambda} (x,t) &=& \frac16[E^u(x,t)+4E^d(x,t)+E^s(x,t)],~~~~ %E^d_{\Lambda}(x,t)= E^u_\Lambda(x,t), \nonumber \\
%E^s_{\Lambda}(x,t) &=& \frac13[2E^u(x,t)-E^d(x,t)+2E^s(x,t)],\nonumber \\
%E^u_{\Lambda\Sigma^0}(x,t) &=& %\frac{\sqrt{3}}{6}[E^u(x,t)-2E^d(x,t)+E^s(x,t)],~~~~ %E^d_{\Lambda\Sigma^0}(x,t)= -E^u_{\Lambda\Sigma^0}(x,t), \nonumber \\
%E^s_{\Lambda\Sigma^0}(x,t) &=& 0.
%\end{eqnarray}
%
\begin{subequations}
\begin{alignat}{3}
 & E^u_n\ = E^d,
&& E^d_n\ = E^u,
&& E^s_n\ = E^s,
\\
 & E^u_{\Sigma^+}\!= E^u,
&& E^d_{\Sigma^+}\!= E^s,
&& E^s_{\Sigma^+}\!= E^d,
\\
 & E^u_{\Sigma^0} = \frac12 \big( E^u + E^s \big),
&& E^d_{\Sigma^0} = E^u_{\Sigma^0},
&& E^s_{\Sigma^0} = E^d,
\\
 & E^u_{\Sigma^-}\!= E^s,
&& E^d_{\Sigma^-}\!= E^u,
&& E^s_{\Sigma^-}\!= E^d,
\\
 & E^u_{\Lambda}\ = \frac16 \big( E^u+4E^d+E^s \big),
&& E^d_{\Lambda}\ = E^u_\Lambda,
&& E^s_{\Lambda}\ = \frac13 \big( 2E^u-E^d+2E^s \big),
\\
 & E^u_{\Lambda\Sigma^0}\!= \frac{1}{2\sqrt3} \big( E^u-2E^d+E^s \big),~~~~~
&& E^d_{\Lambda\Sigma^0}\!= -E^u_{\Lambda\Sigma^0},~~~~~ 
&& E^s_{\Lambda\Sigma^0}\!= 0.
\end{alignat}
\end{subequations}
For the GPDs in the SU(3) decuplet baryons, using the relations in Table~\ref{tab:TT} these can be written in terms of the GPDs in the $\Delta^+$ baryon, which we denote by $E^q_\Delta \equiv E^q_{\Delta^+}(x,t) \equiv E^q_{\Delta^+}(x,\xi=0,t)$,
\begin{subequations}
\begin{alignat}{3}
 & E^u_{\Delta^{++}}\!= E^u_{\Delta} + E^d_{\Delta}- E^s_{\Delta},~~~~~~~~~~
&& E^d_{\Delta^{++}}\!= E^s_{\Delta},~~~~~~~~~~~
&& E^s_{\Delta^{++}}\!= E^s_{\Delta},~~~~~~~~~~~
\\
 & E^u_{\Delta^{0}}\,\: = E^d_{\Delta},
&& E^d_{\Delta^{0}}\,\: = E^u_{\Delta},
&& E^s_{\Delta^{0}}\,\: = E^s_{\Delta},
\\
 & E^u_{\Sigma^{*+}} = E^u_{\Delta},
&& E^d_{\Sigma^{*+}} = E^s_{\Delta},
&& E^s_{\Sigma^{*+}} = E^d_{\Delta},
\\
 & E^u_{\Sigma^{*0}}\, = E^d_{\Delta},
&& E^d_{\Sigma^{*0}}\, = E^d_{\Delta},
&& E^s_{\Sigma^{*0}}\, = E^d_{\Delta},
\\
 & E^u_{\Sigma^{*-}} = E^s_{\Delta},
&& E^d_{\Sigma^{*-}} = E^u_{\Delta},
&& E^s_{\Sigma^{*-}} = E^d_{\Delta}.
\end{alignat}
\end{subequations}
Interestingly, the relations between the $E^q$ GPDs between different flavors in the SU(3) octet and decuplet baryon configurations are identical to those for the PDFs derived in Refs.~\cite{Wang:2016ndh, Salamu:2019dok}.
Similar relations will therefore follow also for the electric $H^q$ GPDs.

\clearpage
For the octet-decuplet transition GPDs, from Table~\ref{tab:BT} one can write each of the transition GPDs in terms of the $u$-quark GPD for the $p$--$\Delta^+$ transition,
\begin{subequations}
\begin{alignat}{3}
 & E^u_{n\Delta^0}\;\ = E^u_{p\Delta^+},
&& E^d_{n\Delta^0}\;\ = -E^u_{p\Delta^+},
&& E^s_{n\Delta^0}\;\ = 0,
\\
 & E^u_{\Sigma^+\Sigma^{*+}}\!\!= - E^u_{p\Delta^+},
&& E^d_{\Sigma^+\Sigma^{*+}}\!\!= 0,
&& E^s_{\Sigma^+\Sigma^{*+}}\!\!= E^u_{p\Delta^{+}} ,
\\
 & E^u_{\Sigma^0\Sigma^{*0}} = \frac12 E^u_{p\Delta^+},
&& E^d_{\Sigma^0\Sigma^{*0}} = \frac12 E^u_{p\Delta^+},
&& E^s_{\Sigma^0\Sigma^{*0}} = -E^u_{p\Delta^{+}} ,
\\
 & E^u_{\Sigma^-\Sigma^{*-}}\!\!= 0,
&& E^d_{\Sigma^-\Sigma^{*-}}\!\!= E^u_{p\Delta^+},
&& E^s_{\Sigma^-\Sigma^{*-}}\!\!= -E^u_{p\Delta^{+}} ,
\\
 & E^u_{\Lambda\Sigma^{*0}}\,\:= -\frac{\sqrt{3}}{2} E^u_{p\Delta^+},~~~~~~~~~
&& E^d_{\Lambda\Sigma^{*0}}\,\:= \frac{\sqrt{3}}{2} E^u_{p\Delta^+},~~~~~~~~~
&& E^s_{\Lambda\Sigma^{*0}}\,\:= 0.
\end{alignat}
\end{subequations}
Finally, for the GPDs associated with the tadpoles, the distributions can be expressed in terms of GPDs in the proton.
For the case of the nucleon, from Table~\ref{tab:tad} we have
\begin{subequations}
\begin{alignat}{3}
 & E^{u(\rm tad)}_{\pi^+\pi^-}\, = \frac12 \big( E^u-E^d \big),~~~~~~~
&& E^{d(\rm tad)}_{\pi^+\pi^-}\, = \frac12 \big( E^d-E^u \big),~~~~~~
&& E^{s(\rm tad)}_{\pi^+\pi^-}\, = 0,~~
\\
 & E^{u(\rm tad)}_{K^+K^-}\! = \frac12 \big( E^u-E^s \big),
&& E^{d(\rm tad)}_{K^+K^-}\! = 0,
&& E^{s(\rm tad)}_{K^+K^-}\! = \frac12 \big( E^s-E^u \big),
\\
 & E^{u(\rm tad)}_{K^0\overline{K}^0}\, = 0,
&& E^{d(\rm tad)}_{K^0\overline{K}^0}\, = \frac12 \big( E^d-E^s \big),
&& E^{s(\rm tad)}_{K^0\overline{K}^0}\, = \frac12 \big( E^s-E^d \big),
\end{alignat}
\end{subequations}
he tadpole GPDs for the other baryons can also be % similarly 
derived from the relations in Table~\ref{tab:tad}.

Turning now to the Dirac-like $H^q_j(x,t)$ GPDs for the various hadronic configurations $j$, we observe that these have the same relationships as for PDFs \cite{Wang:2016ndh, Salamu:2019dok}. 
Taking the strange quark flavor as an example, for the strange GPDs in octet baryons we have
\begin{subequations}
\begin{eqnarray}
H^s_{\Sigma^+}\! &=& H^s_{\Sigma^0} = H^s_{\Sigma^-} = H^d,
\\
H^s_{\Lambda}\,  &=& \frac13 \big( 2H^u-H^d+2H^s \big),
\end{eqnarray}
\end{subequations}
while for the strange GPDs in decuplet baryons, 
\begin{eqnarray}
H^s_{\Sigma^{*+}}\!= H^s_{\Sigma^{*0}}\!= H^s_{\Sigma^{*-}}\!= H^d_{\Delta},
\end{eqnarray}
with the strange GPDs in all other baryons vanishing.

For the strange GPDs associated with the tadpole diagram, we find
%
%\begin{subequations}
\begin{eqnarray}
%H^{s(\rm tad)}_{K^+K^-}\!&=& \frac12 \big( H^u-H^s \big),
H^{s(\rm tad)}_{K^+K^-} = \frac12 \big( H^u-H^s \big), \qquad
%\\
%H^{s(\rm tad)}_{K^0\overline{K}^0}\,&=& H^d-H^s. ~~
H^{s(\rm tad)}_{K^0\overline{K}^0} = H^d-H^s.
\end{eqnarray}
%\end{subequations}
%
The strange electric GPDs for the KR diagrams are related to the magnetic GPDs, which for the octet baryons are given by
\begin{subequations}
\begin{eqnarray}
H^{s(\rm KR)}_{\Sigma^+}
&=& H^{s(\rm KR)}_{\Sigma^0}
 = \frac{1}{F-D}\, \big(\widetilde{H}^d - \widetilde{H}^s \big),
\\
H^{s(\rm KR)}_{\Lambda}
&=& \frac{1}{3F+D}\,
    \big( 2\widetilde{H}^u - \widetilde{H}^d - \widetilde{H}^s \big),
\end{eqnarray}
\end{subequations}
and for the decuplet baryons by
\begin{eqnarray}
H^{s(\rm KR)}_{\Sigma^{*+}} 
= H^{s(\rm KR)}_{\Sigma^{*0}}
= \frac{1}{2D}
  \big( \widetilde{H}^u - 2\widetilde{H}^d + \widetilde{H}^s \big),
\end{eqnarray}
where $\widetilde{H}^q \equiv \widetilde{H}^q(x,\xi=0,t)$ are the corresponding spin-dependent GPDs in the proton.
Finally, for the antiquark GPDs in pions and kaons that enter in the convolution formulas, we have the relations
\begin{eqnarray} 
H^{\bar s}_{K^+}
&=& H^{\bar s}_{K^0} 
 =  H^{\bar{d}}_{\pi^+}
 =  H^d_{\pi^-}
 =  H^{\bar{u}}_{\pi^-}
 = 2H^{\bar{u}}_{\pi^0}
 = 2H^{\bar{d}}_{\pi^0}.
\end{eqnarray}

In our numerical calculations, % that will be presented in the next section, 
we will assume valence quark dominance for the undressed states, so that for the proton $H^s = E^s = 0$. 
The decuplet and transition GPDs $H^q_\Delta$, $E^q_\Delta$ and $E^q_{p\Delta^+}$ can be related to the proton GPDs $H^q$ and $E^q$ using SU(3) flavor symmetry, which constrains the coefficients according to
$\theta_{\rm mag}^{(n)} = \alpha_{\rm mag}^{(n)} + 3\beta_{\rm mag}^{(n)}$.
Since $H^q$ and $E^q$ have the same flavor symmetry, the decuplet GPDs can then be written as
\begin{eqnarray}
H^u_{\Delta} &=& 2 H^d_{\Delta} = \frac43 H^u - \frac23 H^d,
\\
E^u_{\Delta} &=& 2 E^d_{\Delta} = \frac43 E^u - \frac23 E^d.
\end{eqnarray}
Similarly, the constraint $\omega_{\rm mag}^{(n)}= 4 \alpha_{\rm mag}^{(n)}$ leads to the relations for the transition GPDs,
\begin{equation}
E^u_{p\Delta^+}
= -E^d_{p\Delta^+}
= \frac{\sqrt2}{3} \big( E^u - 2 E^d \big).
\end{equation}
With these relations, all the GPDs used in the calculation in the next section can be expressed in terms of the GPDs in the proton.
Note that because the magnetic coefficients $C_j^{\rm mag}$ ($j=B,T,BT$) are not included in the splitting functions, the GPDs $E^q_j(x,t=0)$ need to be normalized to their magnetic moments with unit charge obtained from $c_1$ and $c_2$.

%%%%%%%%%%%%%%%%%%%%%%%%%%%%%%%%%%%%%%%%%%%%%%%%%%%%%%%%%%%%%%%%%%%%%%%%%%%%
\section{Numerical results for sea quark GPDs}
\label{sec.numerical}

In this section we present the numerical results of our calculation of the meson loop contributions to the sea quark GPDs of the proton.
We begin first by summarizing the inputs used in the calculation, followed by discussions of the results for the light antiquark contributions and the strange quark contributions to the GPDs.

% ..........................................................................
\subsection{Theoretical inputs}

For the meson--baryon couplings in our numerical calculations we use the values $D = 0.76$ and $F = 0.5$ for the octet baryon couplings (with $g_A = D+F = 1.26$), and ${\cal C} = -2D$ for the octet-decuplet transition coupling under the assumption of SU(6) symmetry.
The loop integrals are regularized using a covariant regulator of dipole form,
\begin{equation}
\widetilde{F}(k) = \left(\frac{\Lambda^2-m^2_\phi}{\Lambda^2-k^2}\right)^2,
\end{equation} 
with $\Lambda$ a mass parameter.
Such a regulator can suppress the short distance physics and improve the chiral convergence~\cite{Young:2002ib}. 
On the other hand, the Taylor expansion of $\widetilde{F}(k)$ is a series in $k^2$, which can be regarded as the resummation of contributions from higher order meson--baryon interactions.
From previous analyses of nucleon electromagnetic and strange from factors, we take $\Lambda=1.0(1)$~GeV~\cite{He:2017viu, He:2018eyz}.
The parameters $c_1$ and $c_2$ are determined by fitting to the nucleon anomalous magnetic moments, and we find $c_1=1.40$ and $c_2=0.54$ reproduce the empirical values $\mu_p=2.79\, \mu_N$ and $\mu_n=-1.91\, \mu_N$ in units of the nucleon magneton, $\mu_N=e\hbar/2M$.

% ... and the expressions for the GPDs of $H^q$, $E^q$, $\widetilde{H}^q$ can be found in \cite{Diehl:2004cx, Martin:1998sq, Leader:2006xc}.
% The input valence quark distributions are from Refs.~\cite{Diehl:2004cx, Martin:1998sq, Leader:2006xc} at the scale $\mu_0 = 1$ GeV.

For the valence quark GPDs in the proton, we follow Diehl {\it et al.}~\cite{Diehl:2004cx} and parametrize the GPDs as products of forward distributions and $t$-dependent exponential factors,
\begin{subequations}
\begin{eqnarray}
H^q(x,t) &=& q_v(x)\, \exp\big[t\,f_q(x)\big],
\\
E^q(x,t) &=& e_q(x)\, \exp\big[t\,f_q(x)\big],
\\
\widetilde{H}^q(x,t) &=& \Delta q_v(x)\, \exp\big[t\,\tilde{f}_q(x)\big].
\end{eqnarray}
\end{subequations}
Here the unpolarized $q_v(x)$, helicity-flip $e_q(x)$, and helicity-dependent $\Delta q_v(x)$ PDFs for the valence quarks are taken from the parametrizations in Refs.~\cite{Martin:1998sq, Diehl:2004cx,  Leader:2006xc}.
The profile functions $f_q(x)$ and $\tilde{f}_q(x)$ parametrize the $x$ dependence of the average impact parameter of the corresponding quark distribution, which can be seen after performing a Fourier transform to coordinate space~\cite{Diehl:2004cx}.

The valence quark GPD in the pion is parametrized as a simple factorized product of a pion valence PDF and a $t$-dependent elastic form factor,
\begin{eqnarray}
H^q_{\pi}(x,t) = q^\pi_v(x)\, F_\pi(t),
\end{eqnarray}
where $q^\pi_v(x)$ is the pion valence quark PDF.
For illustration purposes we use the parametrization of $q^\pi_v$ from Ref.~\cite{Aicher:2010cb}, while more recent analyses have studied the large-$x$ behavior in the presence of next-to-leading-log threshold resummation effects~\cite{Barry:2021osv}.
For the pion elastic electromagnetic from factor $F_\pi(t)$ we use a monopole form,
\begin{eqnarray}
F_\pi(t) = \frac{1}{1-t/\Lambda_\pi^2}.
\end{eqnarray}
The cutoff parameter $\Lambda_\pi$ is tuned to be 0.79~GeV, corresponding to the average of the charge radii for the pion and kaon~\cite{ParticleDataGroup:2018ovx} (since we use the same inputs for all the meson valence quark GPDs).
The valence quark GPDs in other hadronic configurations are obtained through the SU(3) symmetry relations in Sec.~\ref{ssec.gpdhad}.
With the calculated splitting functions and the valence quark distributions as input, we can proceed to evaluate the GPDs of the sea quarks from the convolution expressions (\ref{eq:HEqbar})--(\ref{eq:HEq}).

% \begin{figure}[]
% \begin{center}
% \subfigure[] 
% {
% 	\begin{minipage}{8cm}
% 	\centering  
% 	\includegraphics[scale=0.8]{figures/gFig2a.pdf}  
% 	\end{minipage}
% }
% \subfigure[]
% {
% 	\begin{minipage}{8cm}
% 	\centering      
% 	\includegraphics[scale=0.8]{figures/gFig2b.pdf}   
% 	\end{minipage}
% }
% \caption{The 3D generalized parton distribution $xH^{\bar{u}}$ and $xE^{\bar{u}}$ versus momentum fraction $x$ and momentum transfer $-t$ with $\Lambda=1$~GeV. The initial scale is $\mu_0^2=1$~GeV$^2$.}
% \label{3dubar}
% \end{center}
% \end{figure}

% .........................................................................
\subsection{Light antiquark GPDs}

\begin{figure}[] % Fig. 2
\begin{center}
    \begin{minipage}{0.45\linewidth}
        \centering
        \centerline{
        \includegraphics[width=1\textwidth]{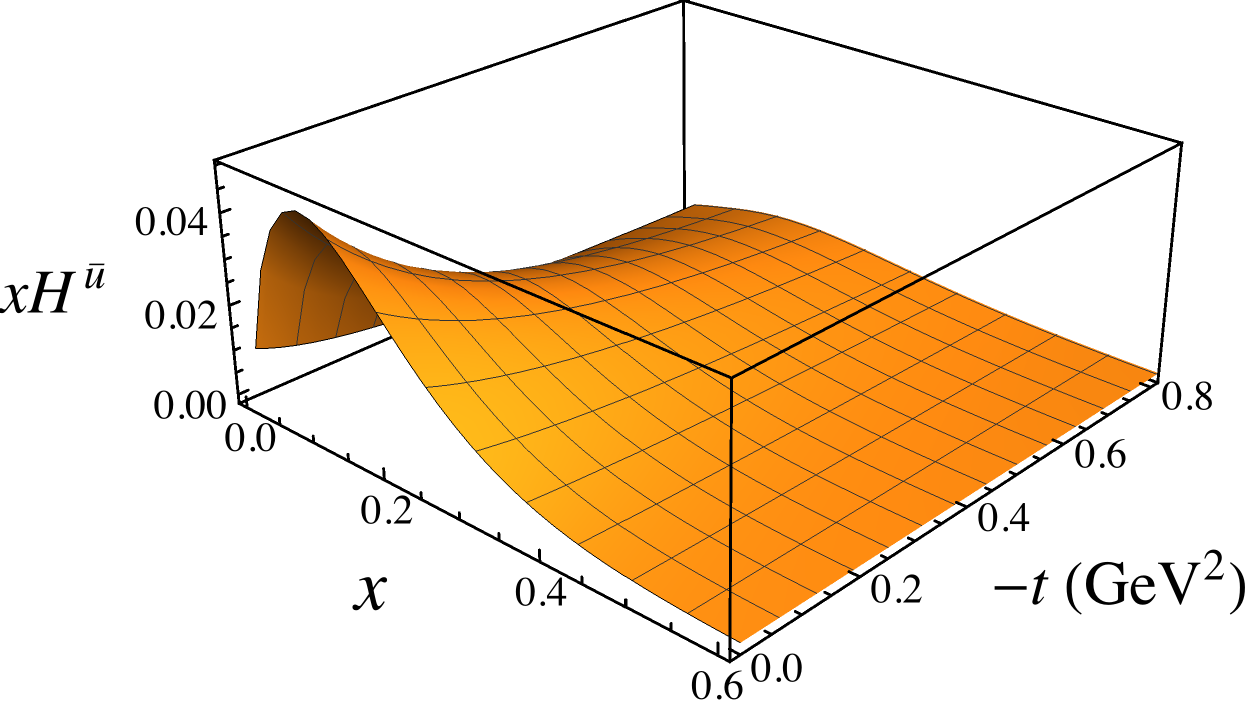}
        }
        \centerline{\small{\bf{(a)}}}
    \end{minipage}
    \begin{minipage}{0.45\linewidth}
        \centering
        \centerline{
        \includegraphics[width=1\textwidth]{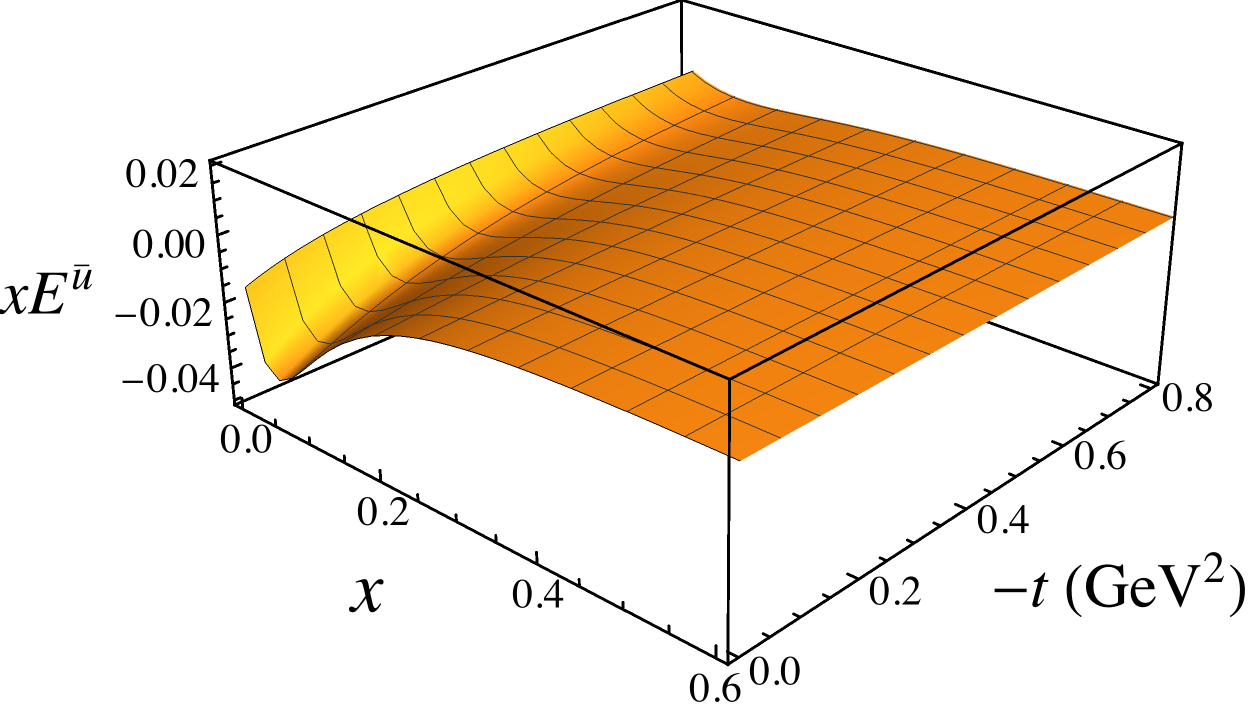}
        }
        \centerline{\small{\bf{(b)}}}
    \end{minipage}    
\\[0.4cm]
        \begin{minipage}{0.45\linewidth}
        \centering
        \centerline{
        \includegraphics[width=1\textwidth]{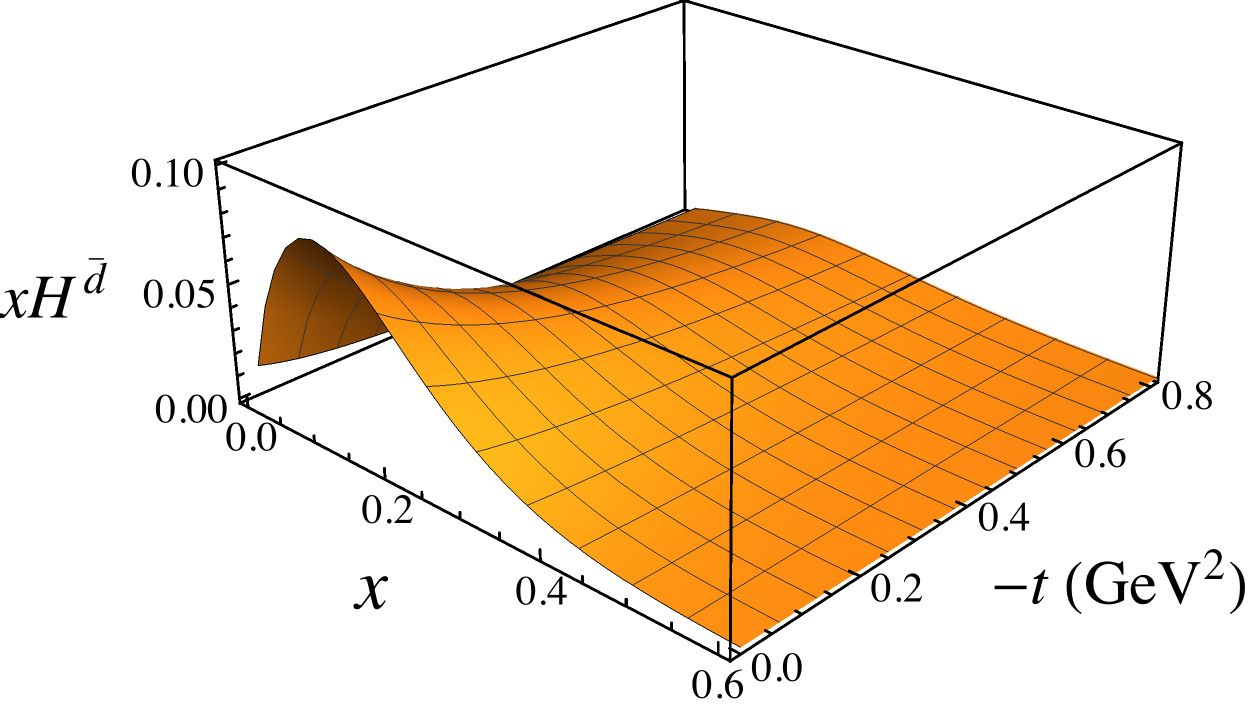}
        }
        \centerline{\small{\bf{(c)}}}
    \end{minipage}
        \begin{minipage}{0.45\linewidth}
        \centering
        \centerline{
        \includegraphics[width=1\textwidth]{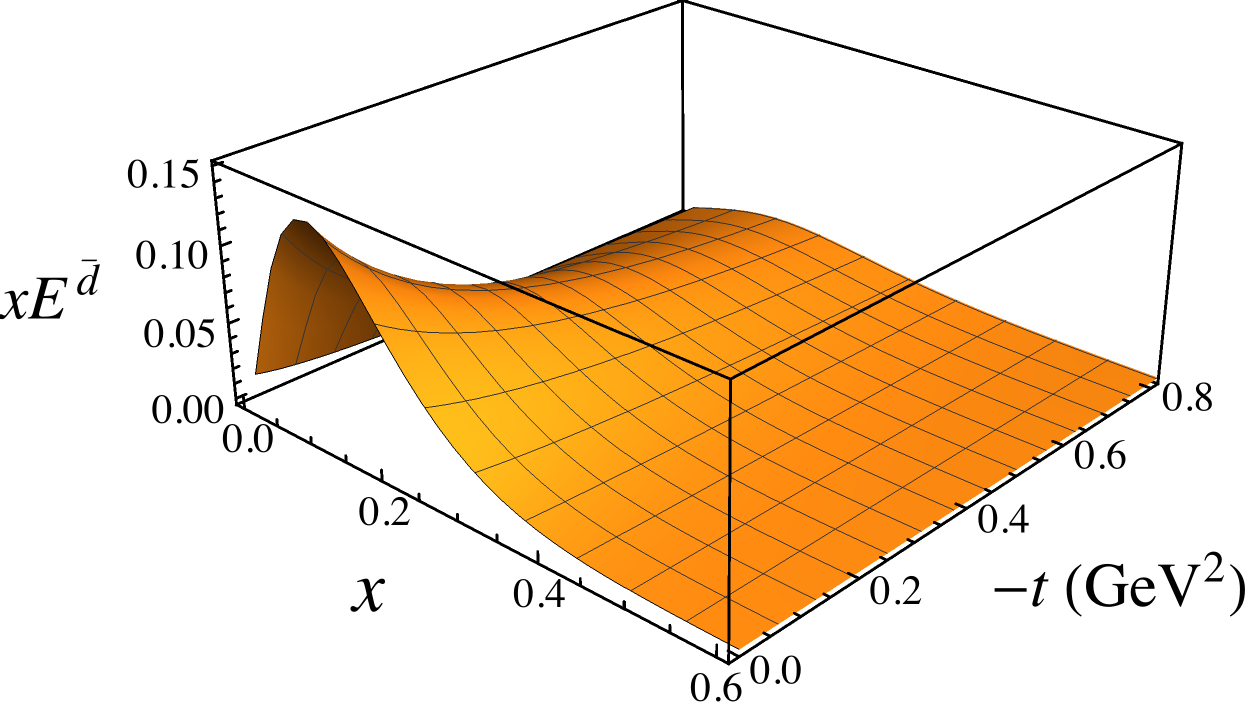}
        }
        \centerline{\small{\bf{(d)}}}
    \end{minipage}
\caption{Electric and magnetic GPDs for light antiquarks: 
{\bf (a)}~$xH^{\bar{u}}$, 
{\bf (b)}~$xE^{\bar{u}}$, 
{\bf (c)}~$xH^{\bar{d}}$, and 
{\bf (d)}~$xE^{\bar{d}}$, versus parton momentum fraction $x$ and four-momentum transfer squared $-t$, for cutoff mass $\Lambda=1$~GeV at a scale $Q_0=1$~GeV.}
\label{3dubar}
\end{center}
\end{figure}

The electric $H^{\bar q}$ and magnetic $E^{\bar q}$ GPDs for the light antiquarks in the proton arising from the meson loop diagrams in Fig.~\ref{diagrams} are presented in Fig.~\ref{3dubar} as a function of the parton momentum fraction $x$ and momentum transfer $-t$, for the $\bar q = \bar u$ and $\bar d$ flavors at the input scale $Q_0=1$~GeV.
%
% the 3-dimensional plots of the quark GPDs $xH^{\bar{u}}(x,t)$ and $xE^{\bar{u}}(x,t)$ are shown 
%
For $\bar u$ quarks, the function $xH^{\bar{u}}$ is positive and peaks at $x \approx 0.1$, roughly independent of the value of $t$.
For any fixed $x$ value, $xH^{\bar{u}}$ falls off monotonically with increasing values of $-t$.
In contrast, the magnetic $xE^{\bar{u}}$ distribution is negative, peaking in absolute value at slightly smaller $x$ compared with $xH^{\bar{u}}$, and again decreasing in magnitude with increasing $-t$ at fixed $x$.
For the $\bar d$ quarks, the shape of the $xH^{\bar{d}}$ GPD is similar to that of the $xH^{\bar{u}}$ distribution, although at any given  $x$ and $t$ the GPD for the $\bar d$ is larger.
This flavor asymmetry stems from the fact that the contribution to $H^{\bar{d}}$ comes from both the octet and decuplet intermediate states, while only the decuplet intermediate states contribute to the $H^{\bar{u}}$ GPD.

The shapes of the magnetic $E^{\bar{q}}$ GPDs reflect the important role played by the orbital angular momentum of the meson in the intermediate state.
For octet baryons, the meson orbital angular momentum tends to be $+1$, resulting in positive values of $E^{\bar{d}}$.
For $\bar u$ quarks, on the other hand, since the intermediate baryons can only be decuplets, the orbital angular momentum of the meson tends to be $-1$, resulting in negative values for $E^{\bar{u}}$.
The absolute value of $xE^{\bar{d}}$ is also much larger than $xE^{\bar{u}}$.
Note that the $\delta$-function term in the splitting functions does not contribute to the $H^{\bar{q}}$ and $E^{\bar{q}}$ GPDs, although it does contribute to the lowest moment of these functions.
% $x$ integrals of $H^{\bar{q}}$ and $E^{\bar{q}}$.

\begin{figure}[] % Fig. 3
\begin{center}
    \begin{minipage}{0.45\linewidth}
        \centering
        \centerline{
        \includegraphics[width=1\textwidth]{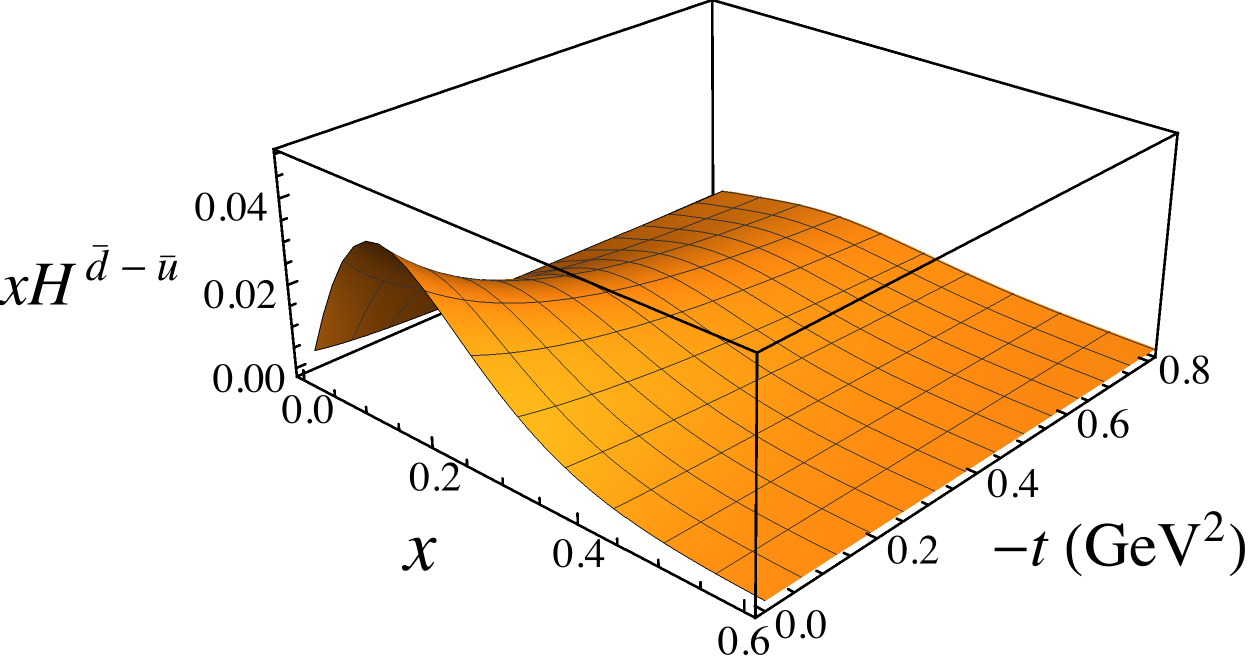}
        }
        \centerline{\small{\bf{(a)}}}
    \end{minipage}
    \begin{minipage}{0.45\linewidth}
        \centering
        \centerline{
        \includegraphics[width=1\textwidth]{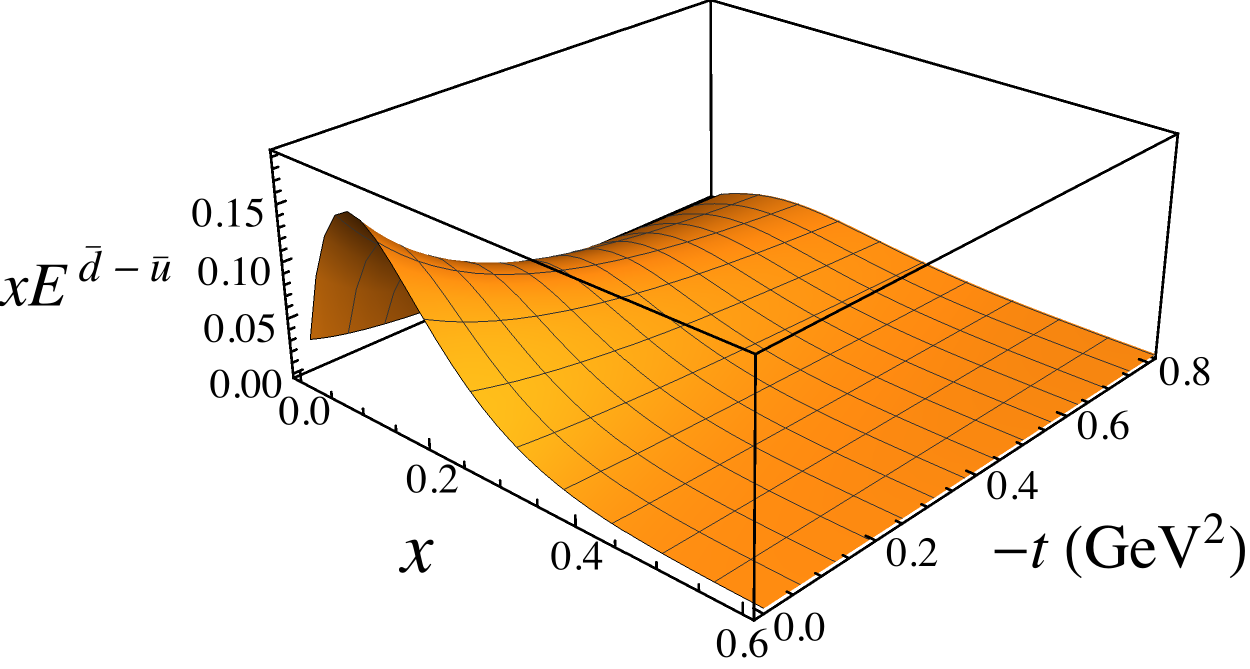}
        }
        \centerline{\small{\bf{(b)}}}
    \end{minipage}
\caption{Light antiquark flavor asymmetry for {\bf (a)}~the electric $xH^{\bar{d}-\bar{u}}$ and {\bf (b)}~magnetic $xE^{\bar{d}-\bar{u}}$ GPDs, versus parton momentum fraction $x$ and four-momentum transfer squared $-t$, for \mbox{$\Lambda = 1$~GeV.}} 
\label{3du-d} 
\end{center}
\end{figure}

Turning now to the light flavor asymmetry of the GPDs, in Fig.~\ref{3du-d} we show the distributions $xH^{\bar{d}-\bar{u}}$ and $xE^{\bar{d}-\bar{u}}$ versus $x$ and $-t$.
Both asymmetries are observed to be positive for all $x$ values, with a peak at $x \approx 0.1$ that decreases with increasing four-momentum transfer squared.
At the peak, the magnitude of the magnetic GPD asymmetry $xE^{\bar{d}-\bar{u}}$ is about 4 times larger than the electric asymmetry $xH^{\bar{d}-\bar{u}}$.

\begin{figure}[htbp] % Fig. 4
\begin{minipage}[b]{.45\linewidth}
\hspace*{-0.3cm}\includegraphics[width=1.1\textwidth, height=5.5cm]{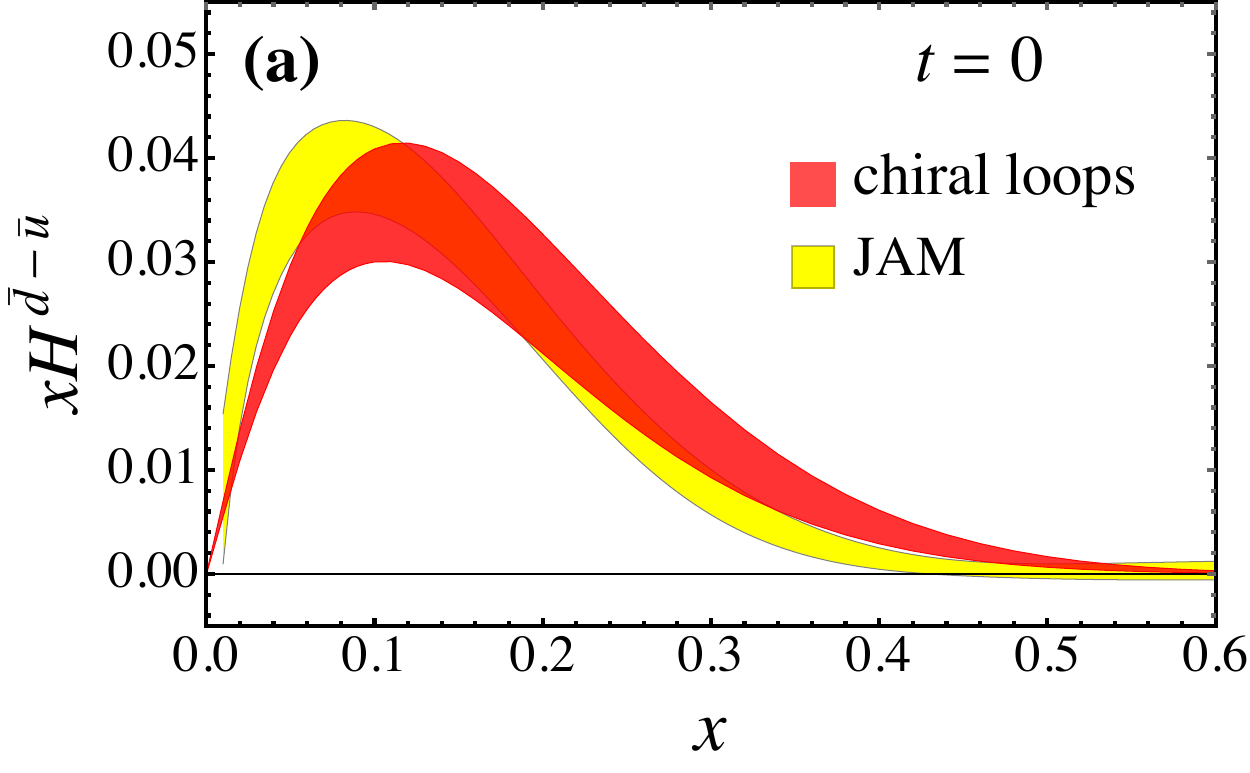}   
 \vspace{0pt}
\end{minipage}
\hfill
\begin{minipage}[b]{.45\linewidth}   
\hspace*{-0.95cm} \includegraphics[width=1.1\textwidth, height=5.5cm]{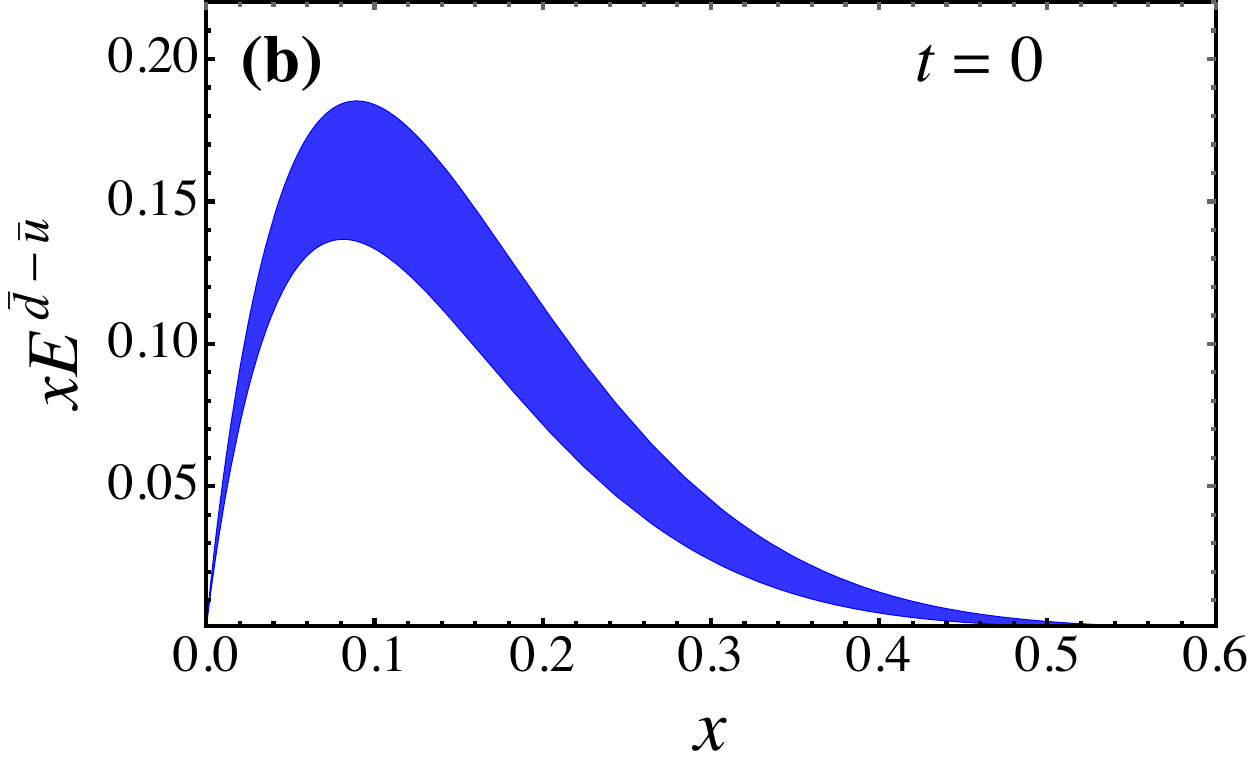} 
   \vspace{0pt}
\end{minipage}  
\\[-0.4cm]
\begin{minipage}[t]{.45\linewidth}
\hspace*{-0.5cm}\includegraphics[width=1.12\textwidth, height=5.5cm]{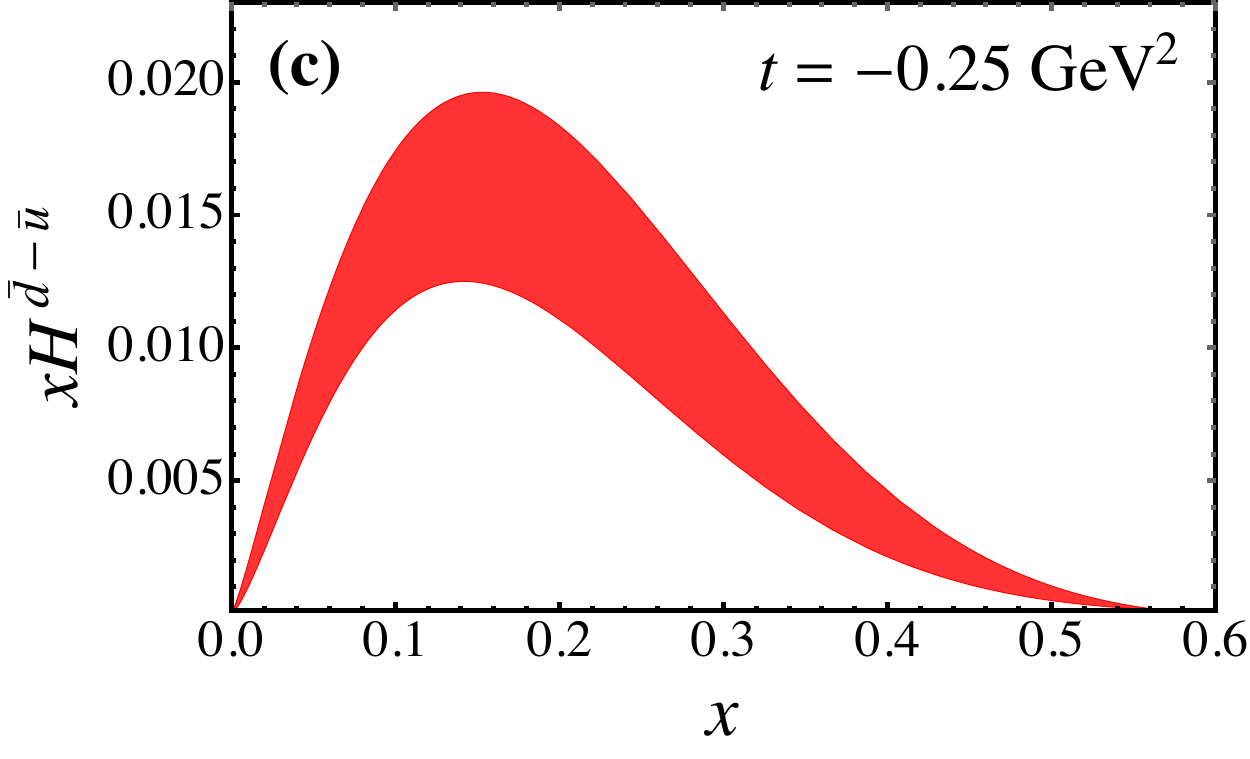}   
 \vspace{0pt}
\end{minipage}
\hfill
\begin{minipage}[t]{.45\linewidth}   
\hspace*{-0.85cm} \includegraphics[width=1.1\textwidth, height=5.5cm]{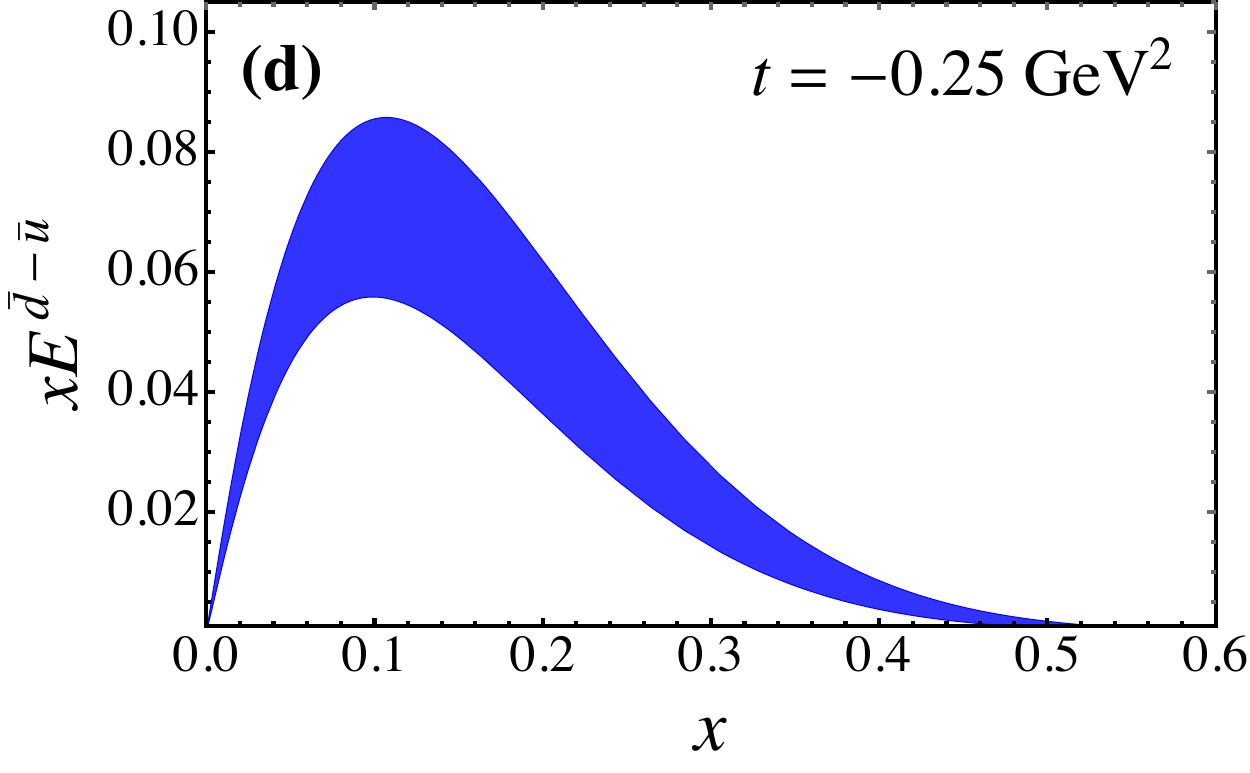}  
   \vspace{-12pt}
\end{minipage} 
\caption{Light antiquark asymmetries for the electric $xH^{\bar{u}-\bar{d}}$ (red bands) and magnetic $xE^{\bar{u}-\bar{d}}$ (blue bands) GPDs versus parton momentum fraction $x$ at four-momentum transfer squared of $t=0$ [{\bf (a)}, {\bf (b)}] and $t=-0.25$~GeV$^2$ [{\bf (c)}, {\bf (d)}], for cutoff parameter $\Lambda = 1.0(1)$~GeV. The asymmetries are shown at the input scale $Q_0=1$~GeV, except for the electric asymmetry at $t=0$, which is compared with the $x(\bar d-\bar u)$ PDF asymmetry from the JAM global QCD analysis~\cite{Cocuzza:2021cbi} (yellow band) evolved to the scale $Q=m_c$.}
\label{2du-dt0}
\end{figure}

To more clearly illustrate the shape and magnitude of the $\bar{d}-\bar{u}$ asymmetry, in Fig.~\ref{2du-dt0} we plot the $xH^{\bar{d}-\bar{u}}$ and $xE^{\bar{d}-\bar{u}}$ distributions at $t=0$, with the error bands corresponding to the 10\% uncertainty on the cutoff parameter $\Lambda$ that was set to 1~GeV.
The calculated electric asymmetry is compared with a recent parametrization of the $x(\bar d-\bar u)$ PDF from the JAM global QCD analysis of world data~\cite{Cocuzza:2021cbi} at a scale $Q=m_c=1.3$~GeV.
%
%In order to compare with the result in \cite{Towell:2001nh}, we use the APFEL Package \cite{Bertone:2013vaa} to evolve the result to $\mu^2=54$ GeV$^2$ for $x(\bar{d}(x)-\bar{u}(x))$. 
%
The numerical results are in good agreement with the phenomenological parametrization of $x(\bar d-\bar u)$, which is driven mostly by the Drell-Yan proton-proton and proton-deuteron scattering data~\cite{Towell:2001nh, SeaQuest:2021zxb}, and has a maximum of $\approx 0.3-0.4$ at $x \approx 0.05-0.10$.
%
% In our calculation, the value of $H^{\bar{d}}$ is always larger than $H^{\bar{u}}$ for any $x$. 
% As a result, $xH^{\bar{d}-\bar{u}}$ is positive for all $x$.
% The data from the experimental extraction are negative when $x$ is larger than about 0.3 which is different from our result.
% This difference at large $x$ needs further investigations.
Within our framework, for a cutoff parameter $\Lambda=1.0(1)$~GeV we find for the integrated values
    $\int_0^1 \dd{x} H^{\bar{d}-\bar{u}}(x,0) = 0.11(2)$ and 
    $\int_0^1 \dd{x} xH^{\bar{d}-\bar{u}}(x,0) = 0.009(2)$.
The magnetic GPD asymmetry $xE^{\bar{d}-\bar{u}}$ at $t=0$ has a similar shape, but is $\approx 4$ times larger than $xH^{\bar{d}-\bar{u}}$ at the peak. The fact that $xE^{\bar{d}-\bar{u}}$ exceeds $xH^{\bar{d}-\bar{u}}$ is also consistent with the prediction in the large-$N_c$ limit~\cite{Goeke:2001tz}.
After integrating over $x$, we find
    $\int_0^1 \dd{x} E^{\bar{d}-\bar{u}}(x,0) = 1.1(2)$ and
    $\int_0^1 \dd{x} xE^{\bar{d}-\bar{u}}(x,0) = 0.034(6)$.
A large magnitude for the magnetic asymmetry augurs well for future efforts to determine this asymmetry experimentally.

% \begin{figure}[] % Fig. 4
% \begin{center}
% \subfigure[]
% {
% 	\begin{minipage}[b]{0.45\linewidth}
% 	\centering 
% 	\includegraphics[scale=0.6]{figures/gFig6.pdf}
% 	\end{minipage}
% }
% \subfigure[]
% {
% 	\begin{minipage}[b]{0.45\linewidth}
% 	\centering  
% 	\includegraphics[scale=0.6]{figures/dbar-ubar.pdf}
% 	\end{minipage}
% }
% \subfigure[]
% {
% 	\begin{minipage}[b]{0.45\linewidth}
% 	\centering    
% 	\includegraphics[scale=0.6]{figures/HUD025.pdf} 
% 	\end{minipage}
% }
% \subfigure[] 
% {
% 	\begin{minipage}[b]{0.45\linewidth}
% 	\centering      
% 	\includegraphics[scale=0.6]{figures/EUD025.pdf} 
% 	\end{minipage}
% }
% \caption{Light antiquark asymmetries for the electric $xH^{\bar{u}-\bar{d}}$ and magnetic $xE^{\bar{u}-\bar{d}}$ GPDs versus parton momentum fraction $x$ at four-momentum transfer squared of $t=0$ [{\bf (a)}, {\bf (b)}] and $t=-0.25$~GeV$^2$ [{\bf (c)}, {\bf (d)}], for cutoff parameter $\Lambda = 1.0(1)$~GeV, at a scale $Q_0=1$~GeV. The electric asymmetry at $t=0$ is compared with the $x(\bar d-\bar u)$ PDF asymmetry from the JAM global QCD analysis~\cite{Cocuzza:2021cbi} evolved to the scale $Q_0=m_c$.}
% \label{2du-dt0}
% \end{center}
% \end{figure}

\begin{figure}[tp]
\begin{center}
\includegraphics[scale=0.45]{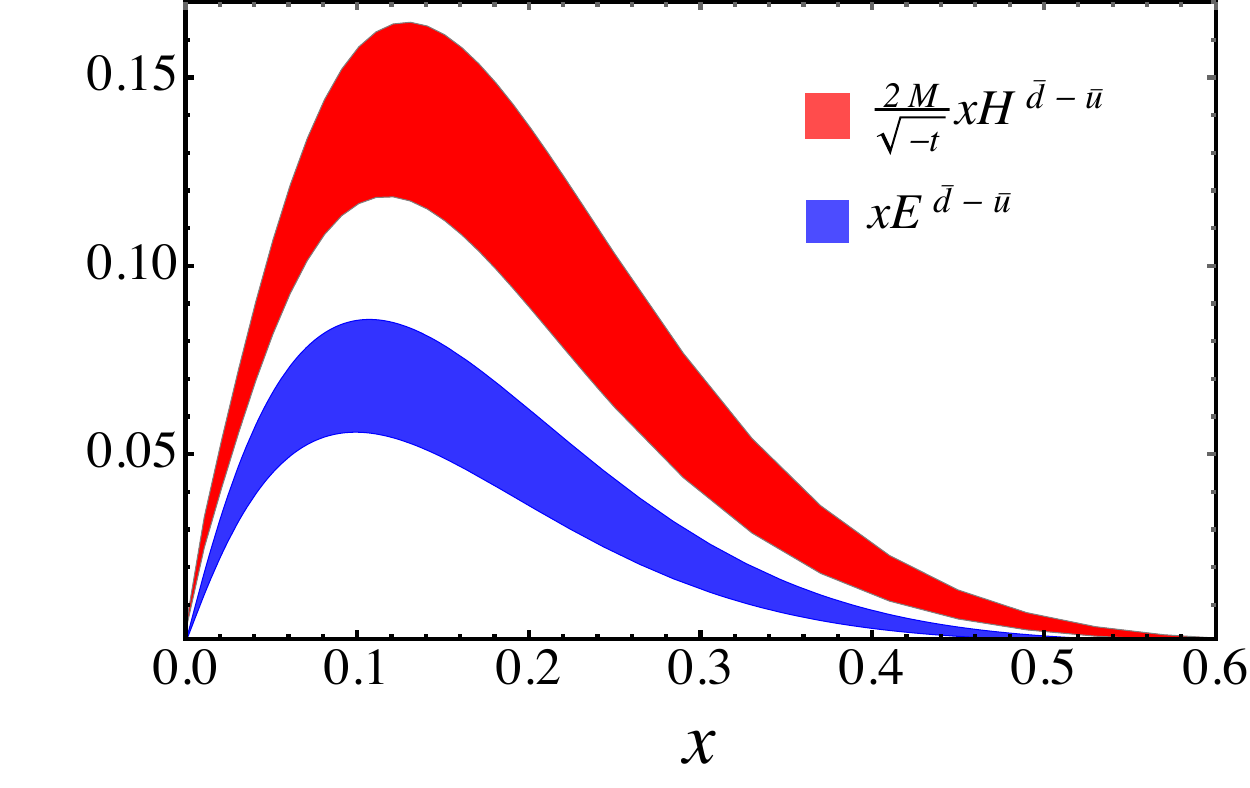}
\caption{Comparison of the GPDs $\frac{2M}{\sqrt{-t}}\, xH^{\bar d-\bar u}(x,0)$ (red band) and $xE^{\bar d-\bar u}(x,t)$ (blue band) at $-t = 0.25$~GeV$^2$.} 
\label{fig:ineq_ud}
\end{center}
\end{figure}

The $xH^{\bar{d}-\bar{u}}$ and $xE^{\bar{d}-\bar{u}}$ GPD asymmetries at finite $t$ are also shown in Fig.~\ref{2du-dt0}, for $-t = 0.25$~GeV$^2$.
As expected from the 3-dimensional plots in Fig.~\ref{3du-d}, the distributions are suppressed at larger $-t$ values, with the magnitudes of the functions about half of those at $t=0$.
This is consistent with the GPD inequality $H^q(x,t) \leq H^q(x,0)$~\cite{Pobylitsa:2002gw, Radyushkin:1998es}.
The peaks in both functions also shift to slightly larger $x$ values with increasing four-momentum transfer squared.  
We also compare the GPD $E^{\bar{d}-\bar{u}}(x,t=-0.25~\text{GeV}^2)$ with $H^{\bar{d}-\bar{u}}(x,0)$ in Fig.~\ref{fig:ineq_ud}, and find that our results satisfy the additional inequality $E^q(x,t) \leq \frac{2M}{\sqrt{-t}}\, H^q(x,0)$~\cite{Pobylitsa:2002gw}.

% ...........................................................................
\subsection{Strange quark GPDs}

\begin{figure}[] % Fig. 5
\begin{center}
    \begin{minipage}{0.45\linewidth}
        \centering
        \centerline{
        \includegraphics[width=1\textwidth]{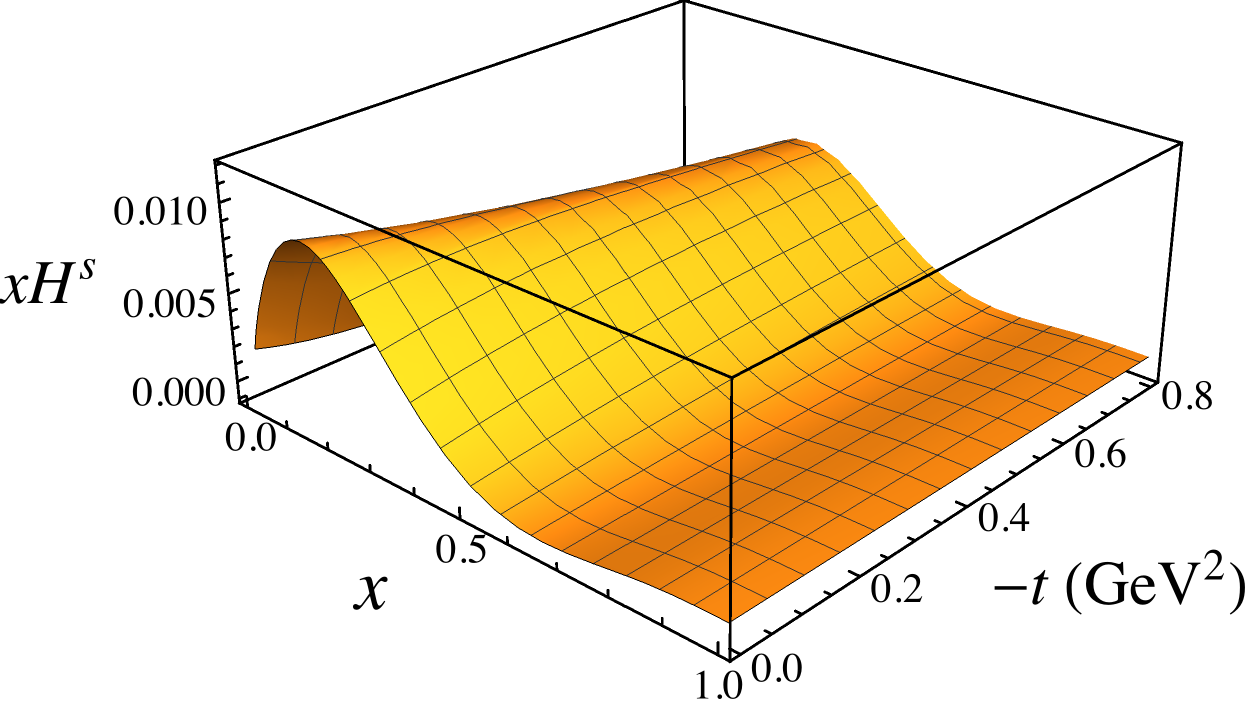}
        }
        \centerline{\small{\bf{(a)}}}
    \end{minipage}
    \begin{minipage}{0.45\linewidth}
        \centering
        \centerline{
        \includegraphics[width=1\textwidth]{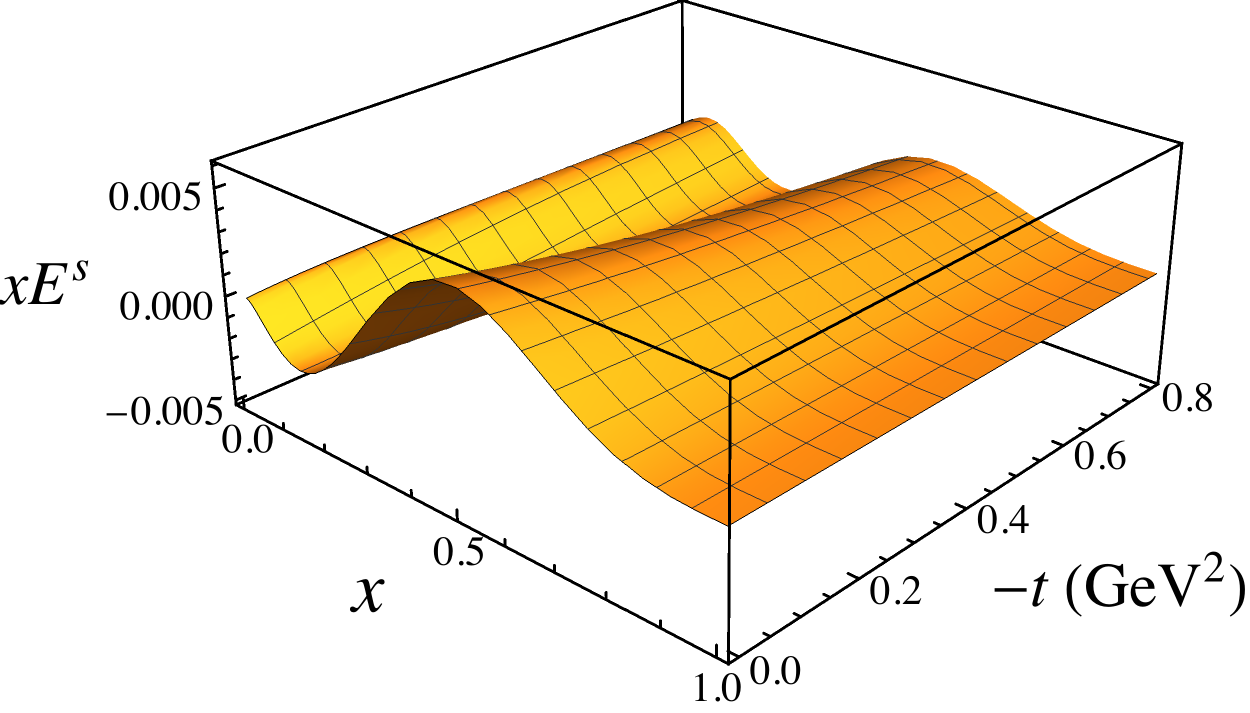}
        }
        \centerline{\small{\bf{(b)}}}
    \end{minipage}    
\\[0.4cm]
        \begin{minipage}{0.45\linewidth}
        \centering
        \centerline{
        \includegraphics[width=1\textwidth]{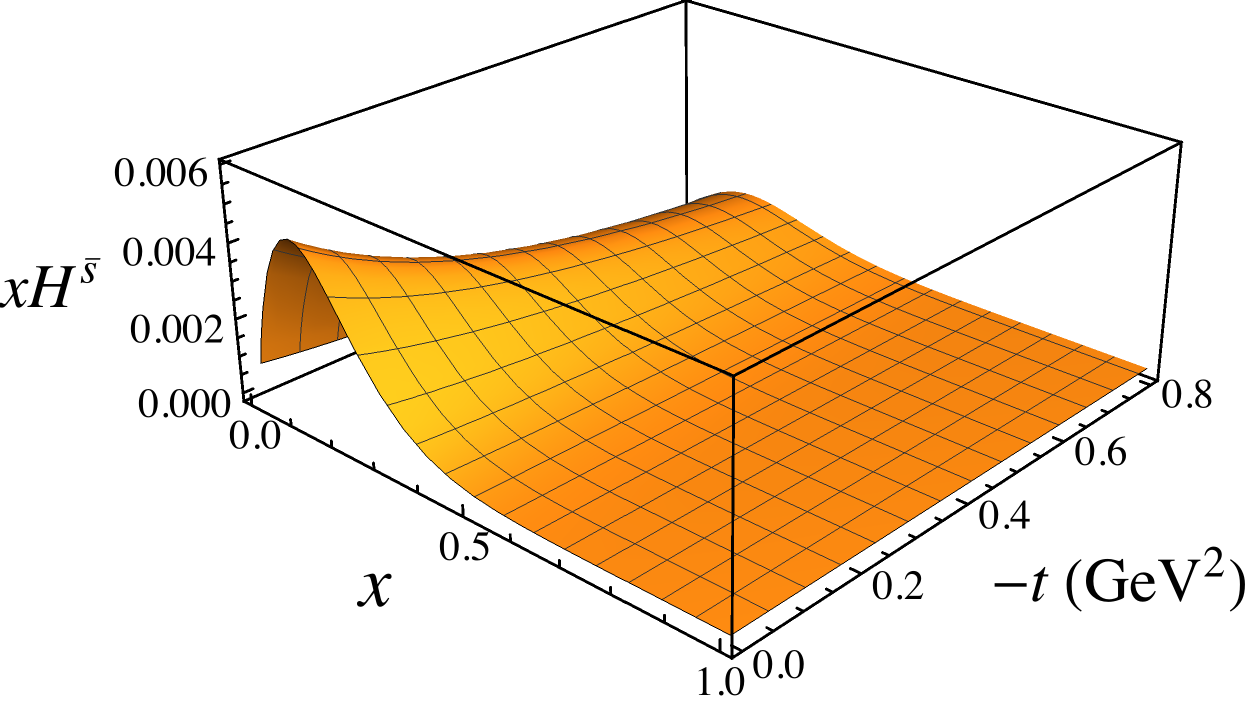}
        }
        \centerline{\small{\bf{(c)}}}
    \end{minipage}
        \begin{minipage}{0.45\linewidth}
        \centering
        \centerline{
        \includegraphics[width=1\textwidth]{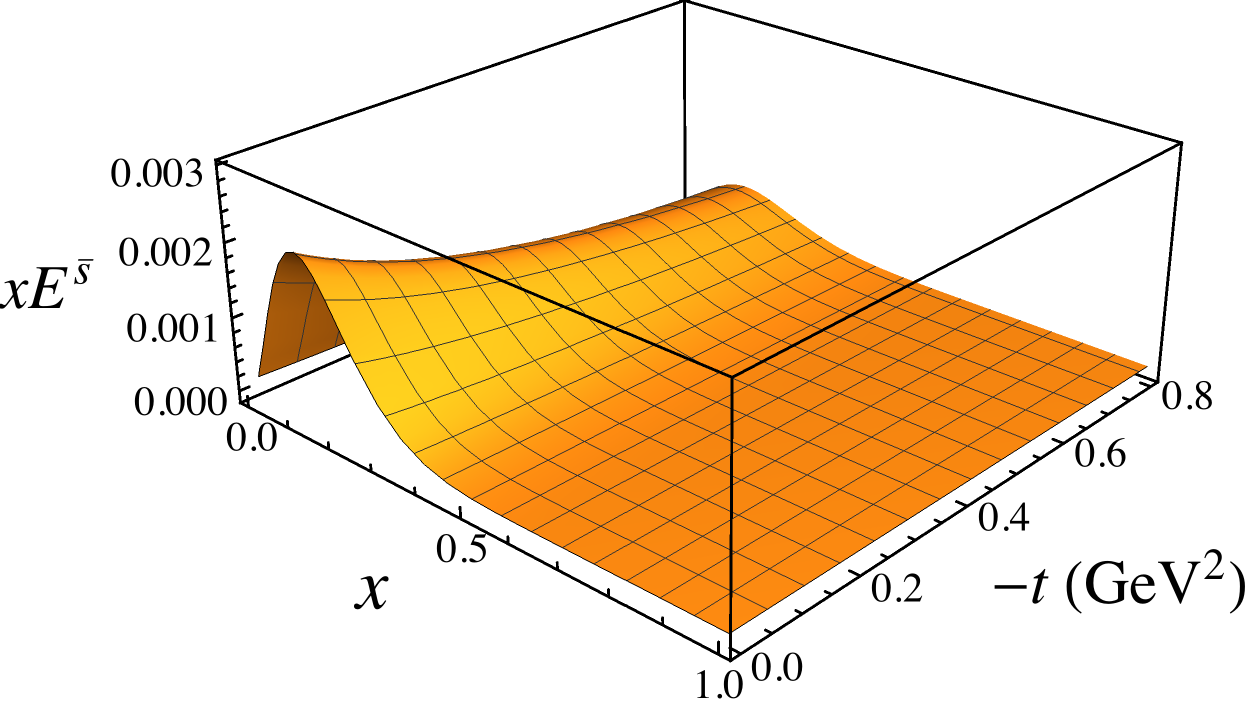}
        }
        \centerline{\small{\bf{(d)}}}
    \end{minipage}
\caption{Electric and magnetic GPDs for the strange and antistrange quarks: {\bf (a)}~$xH^s$, {\bf (b)}~$xE^s$, {\bf (c)}~$xH^{\bar{s}}$, and {\bf (d)}~$xE^{\bar{s}}$ versus the parton momentum fraction $x$ and four-momentum transfer squared $-t$, for $\Lambda=1$~GeV, at the scale $Q_0=1$~GeV.}
\label{3dssbar}
\end{center}
\end{figure}

The kaon loop contributions to the strange quark GPDs are shown in Fig.~\ref{3dssbar}.
Compared with the GPDs for the light antiquarks, the strange GPDs are smaller in magnitude, but display some interesting features.
As for the light antiquark GPDs, the signs of the electric GPDs $H^s$ and $H^{\bar{s}}$ are both positive.
%Therefore, for any flavor of quark $H^q(x,t)$ is always positive, which is related to the number density of the corresponding quark.
While the shapes of the $s$ and $\bar{s}$ distributions are expected to be almost identical perturbatively~\cite{Catani:2004nc}, the kaon loop contributions to these can be quite different due to their different origins.
Assuming the SU(3) symmetric relations for the GPDs in the hadronic intermediate states discussed in Sec.~\ref{ssec.gpdhad}, the $\bar s$ antiquark GPD arises from diagrams with a direct coupling to the kaon, as in Fig.~\ref{diagrams}(a), while contributions to the $s$ quark GPD come from couplings to the intermediate state hyperons, such as in~Fig.~\ref{diagrams}(b)~\cite{Signal:1987gz, Melnitchouk:1996fj}.

As evident from Fig.~\ref{3dssbar}, at small values of $x$ the strange $H^s$ GPD is larger than the antistrange $H^{\bar{s}}$, while for larger $x$ values, $x \gtrsim 0.5$, the antistrange contribution exceeds the strange.
However, the $x$ integrals of $H^s$ and $H^{\bar{s}}$ at zero momentum transfer can be shown to be identical with the inclusion of the $\delta$-function term, as is necessary for the requirement of zero net strangeness in the nucleon.
Since the $t$ dependence of $H^s$ is different from that of $H^{\bar{s}}$, at finite values of $t$ the lowest moments of the strange and antistrange GPDs need not be the same, which corresponds to nonzero values of the strange electric form factor at $-t > 0$.
The behaviors of the magnetic GPDs $E^s$ and $E^{\bar{s}}$ are, on the other hand, rather different.
While the sign of $E^{\bar{s}}$ is the same as that of $E^{\bar{d}}$ because of the positive orbital angular momentum of the meson, the strange GPD $E^s$ changes sign with $x$, from negative at small $x$ values to positive at $x \gtrsim 0.3$.

%We have shown the 3-dimensional plots for the sea quarks $\bar{u}$, $\bar{d}$, $s$ and $\bar{s}$. 
%All the distributions are sensitive to $x$ when it is small and they approach zero when $x$ is larger than about 0.5, except for $E^s$. 
%For $H^q$, they are all positive, while for $E^q$, which are more complicated and are related to the quark spin in baryons and orbit angular momentum in mesons, their signs are flavour dependent. 
%$E^{\bar{u}}$ are negative, while $E^{\bar{d}}$ and $E^{\bar{s}}$ are positive.
%We find that $E^s$ is negative at small $x$ and positive at large $x$.

\begin{figure}[t] % Fig. 6
\centering 
    \begin{minipage}{0.45\linewidth}
        \centering
        \centerline{
        \includegraphics[width=1\textwidth]{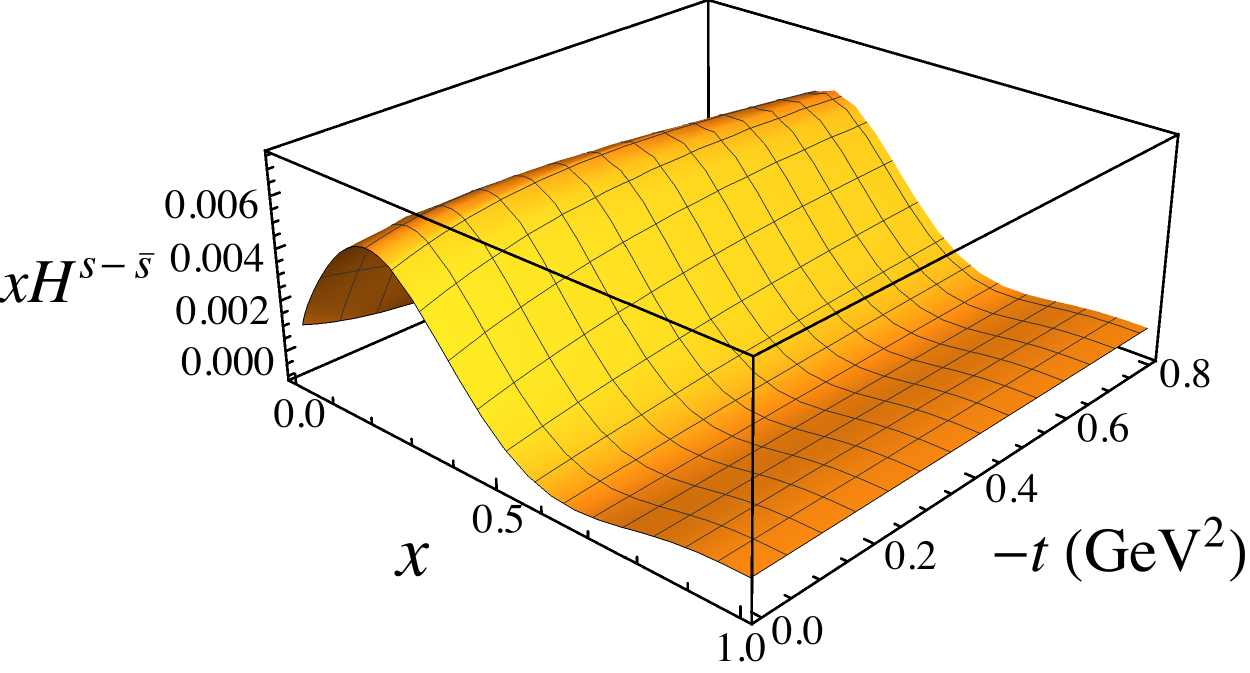}
        }
        \centerline{\small{\bf{(a)}}}
    \end{minipage}
    \begin{minipage}{0.45\linewidth}
        \centering
        \centerline{
        \includegraphics[width=1\textwidth]{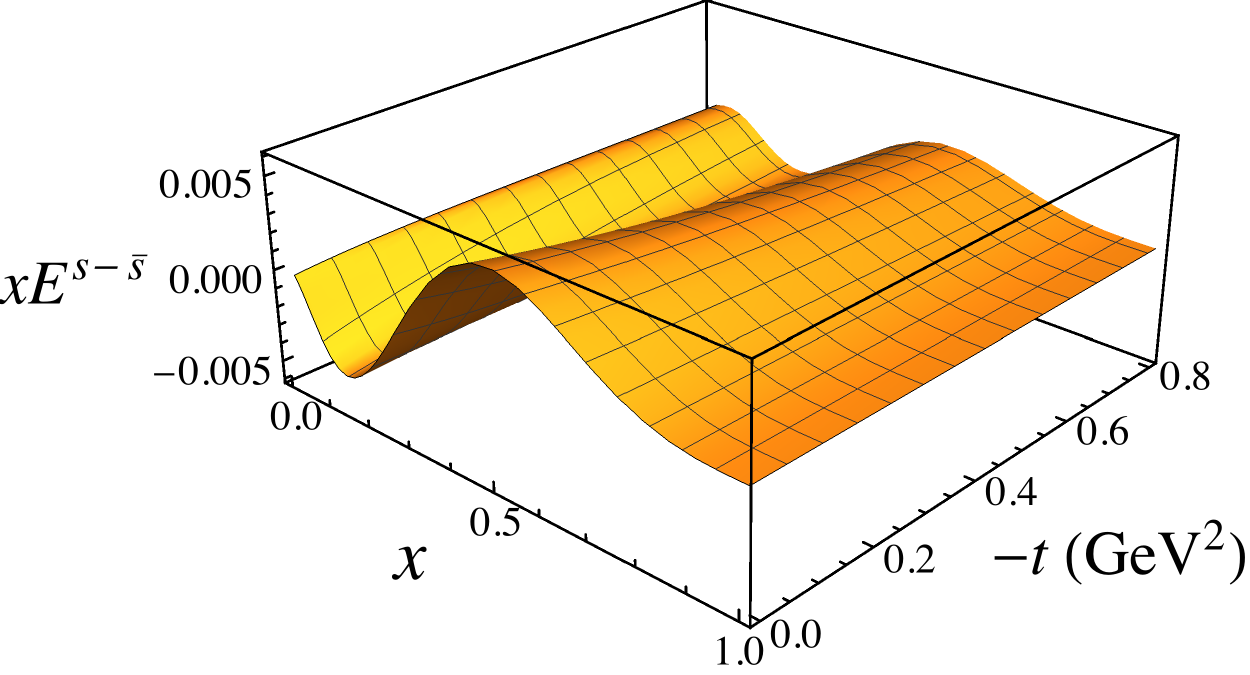}
        }
        \centerline{\small{\bf{(b)}}}
    \end{minipage}
\caption{The strange quark asymmetry for {\bf (a)} the electric $xH^{s-\bar{s}}$ and {\bf (b)} magnetic $xE^{s-\bar{s}}$ GPDs, versus momentum fraction $x$ and four-momentum transfer squared $-t$, for $\Lambda = 1$~GeV.}
\label{3ds-sbar}  
\end{figure}

In Fig.~\ref{3ds-sbar} we show the strange--antistrange asymmetries $xH^{s-\bar{s}}$ and $xE^{s-\bar{s}}$ versus $x$ and $-t$, for a fixed value of $\Lambda = 1$~GeV.
At nonzero values of $x$, the $xH^s$ GPD is generally larger than $xH^{\bar{s}}$, with a maximal asymmetry at $x \approx 0.2-0.3$. 
Unlike the individual $s$ and $\bar s$ contributions, for a given value of $x$ the asymmetry $xH^{s-\bar{s}}$ does not decrease monotonously with $-t$, and in fact increases at higher $-t$ in some cases.
For the magnetic asymmetry $xE^{s-\bar{s}}$, the change of sign with $x$ is driven by the behavior of the strange contribution, $xE^s$.
Generally, the $s-\bar{s}$ asymmetry is much smaller than the $\bar{d}-\bar{u}$ asymmetry in the nucleon for both the electric and magnetic GPDs.

\begin{figure}[htbp] % Fig. 7
\begin{minipage}[b]{.45\linewidth}
\hspace*{-0.55cm}\includegraphics[width=1.125\textwidth, height=5.5cm]{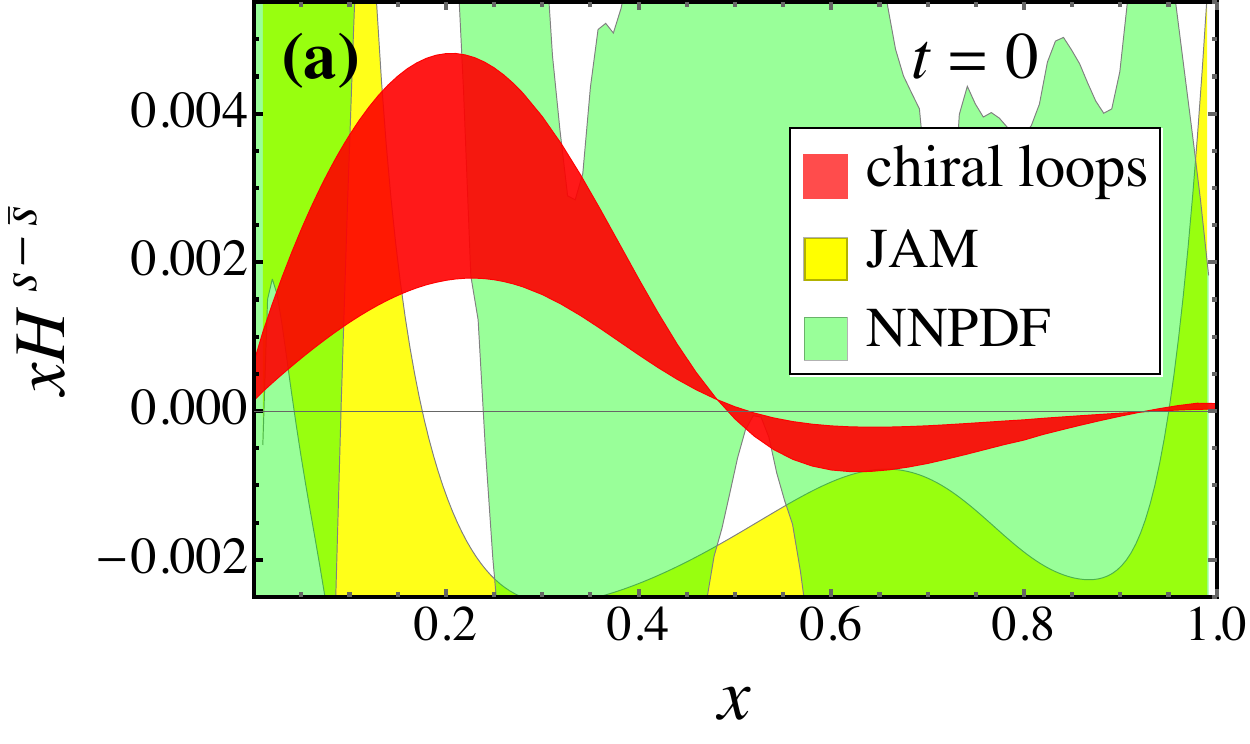} 
 \vspace{0pt}
\end{minipage}
\hfill
\begin{minipage}[b]{.45\linewidth}   
\hspace*{-1.1cm} \includegraphics[width=1.1\textwidth, height=5.5cm]{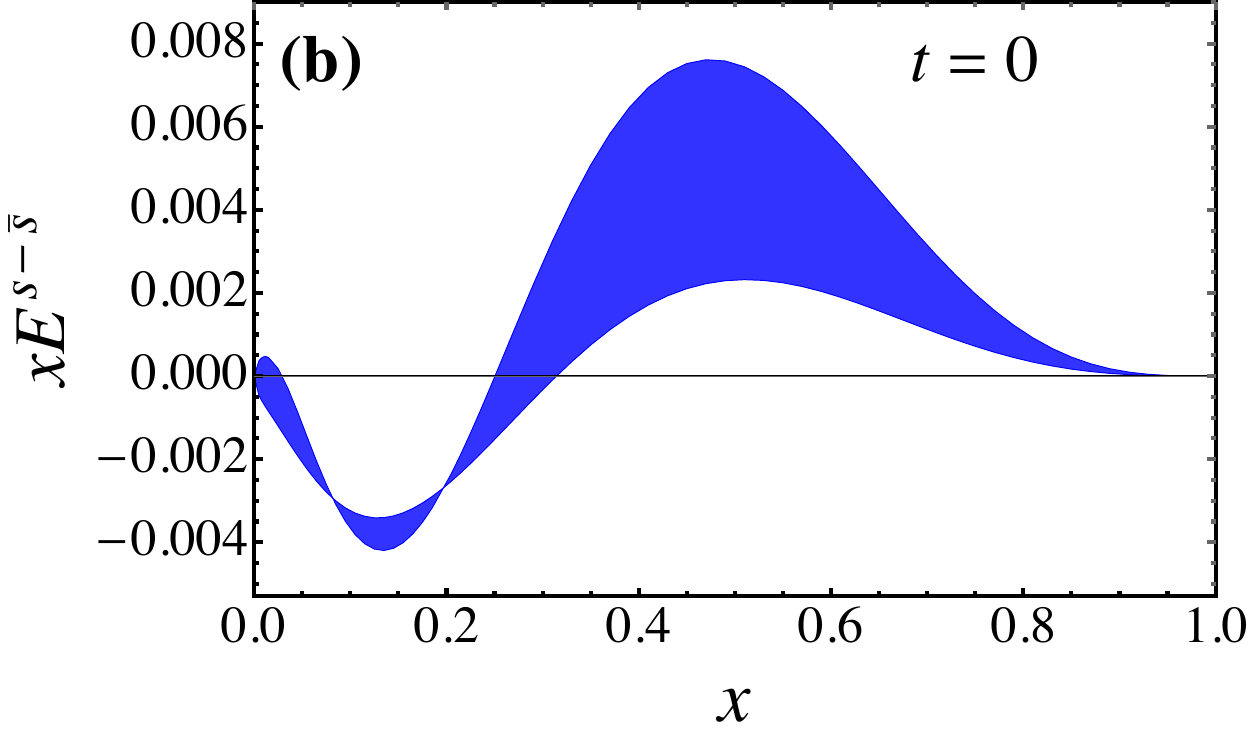} 
   \vspace{14pt}
\end{minipage}  
\\[-0.4cm]
\begin{minipage}[t]{.45\linewidth}
\hspace*{-0.3cm}\includegraphics[width=1.1\textwidth, height=5.5cm]{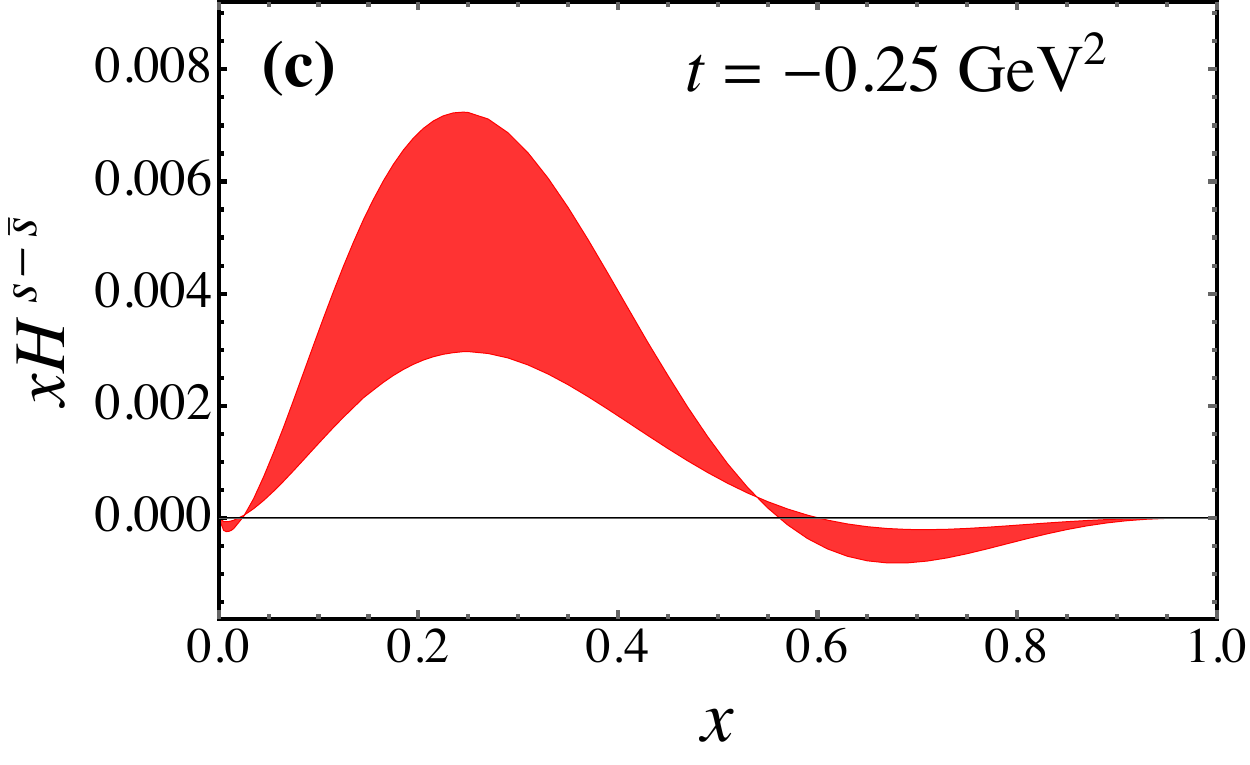}   
 \vspace{0pt}
\end{minipage}
\hfill
\begin{minipage}[t]{.45\linewidth}   
\hspace*{-1.15cm} \includegraphics[width=1.1\textwidth, height=5.5cm]{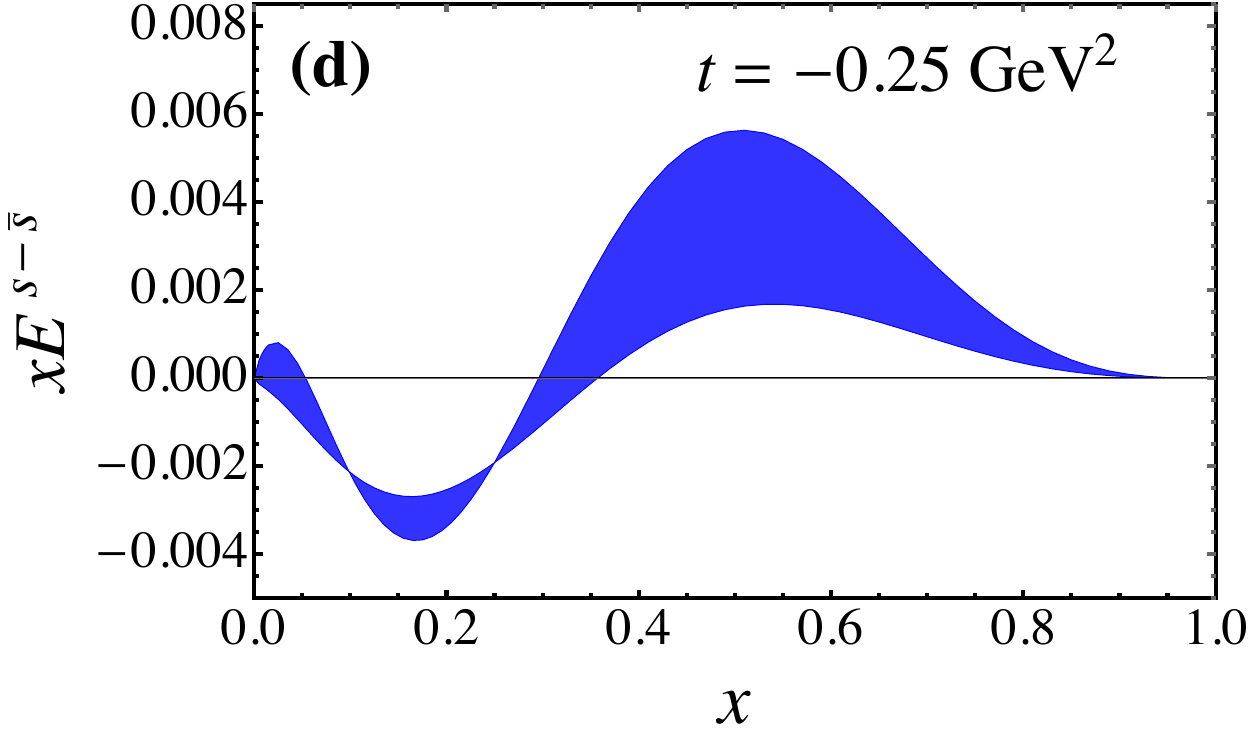}  
   \vspace{0pt}
\end{minipage} 
\caption{Strange quark asymmetry for the $xH^{s-\bar{s}}$ (red bands) and $xE^{s-\bar{s}}$ (blue bands) GPDs versus $x$ at squared momentum transfers $t=0$ [{\bf (a)}, {\bf (b)}] and $-t = 0.25$~GeV$^2$ [{\bf (c)}, {\bf (d)}], with the bands corresponding to cutoff mass $\Lambda=1.0(1)$~GeV. The asymmetries are shown at $Q_0=1$~GeV, except for the strange electric asymmetry at $t=0$, which is compared with PDF parametrizations of $x(s-\bar s)$ from JAM~\cite{Cocuzza:2021cbi} (yellow band) and NNPDF~\cite{Faura:2020oom} (green band) evolved to $Q=m_c$.}
\label{2ds-sbart0}
\end{figure}

% \begin{figure}[t] % Fig. 7
% \begin{center}
% \subfigure[]
% {
% 	\begin{minipage}[b]{0.45\linewidth}
% 	\centering 
% 	\includegraphics[scale=0.6]{figures/gFig8.pdf}
% 	\end{minipage}
% }
% \subfigure[]
% {
% 	\begin{minipage}[b]{0.45\linewidth}
% 	\centering  
% 	\includegraphics[scale=0.6]{figures/s-sbar.pdf}
% 	\end{minipage}
% }
% \subfigure[]
% {
% 	\begin{minipage}[b]{0.45\linewidth}
% 	\centering    
% 	\includegraphics[scale=0.6]{figures/HS025.pdf} 
% 	\end{minipage}
% }
% \subfigure[] 
% {
% 	\begin{minipage}[b]{0.45\linewidth}
% 	\centering      
% 	\includegraphics[scale=0.6]{figures/ES025.pdf} 
% 	\end{minipage}
% }
% \caption{Strange quark asymmetry for the $xH^{s-\bar{s}}$ and $xE^{s-\bar{s}}$ GPDs versus $x$ at squared momentum transfers $t=0$ [{\bf (a)}, {\bf (b)}] and $-t = 0.25$~GeV$^2$ [{\bf (c)}, {\bf (d)}], with the bands corresponding to cutoff mass $\Lambda=1.0(1)$~GeV.}
% \label{2ds-sbart0}
% \end{center}
% \end{figure}

In analogy with the $\bar{d}-\bar{u}$ asymmetry in Fig.~\ref{2du-dt0} above, in Fig.~\ref{2ds-sbart0} we show the $xH^{s-\bar{s}}$ and $xE^{s-\bar{s}}$ asymmetries at $t=0$ and $-t=0.25$~GeV$^2$ for varying cutoff parameters between $\Lambda=0.9$~GeV and 1.1~GeV.
The change in sign of $xH^{s-\bar{s}}$ is evident, with the asymmetry being positive at small $x$, before turning negative at $x \gtrsim 0.5$.
The calculated asymmetry is compared with recent PDF parametrizations of $x(s-\bar s)$ from the JAM~\cite{Cocuzza:2021cbi} and NNPDF~\cite{Faura:2020oom} global QCD analyses, which show very large uncertainties relative to the magnitude of the computed result.
For the lowest nonzero moment, we find 
    $\int_0^1 \dd{x} xH^{s-\bar{s}}(x,0) = 0.0009^{(5)}_{(4)}$
for $\Lambda=1.0(1)$~GeV, which is comparable with other recent estimates of the strange asymmetry~\cite{Bentz:2009yy, Wang:2016ndh, Salamu:2019dok}.
For the magnetic asymmetry $xE^{s-\bar{s}}$, the situation is reversed, with the asymmetry trending negative at small $x$ and becoming positive at larger $x$ values, $x \gtrsim 0.3$.
For comparison, the analogous integrated magnetic GPD asymmetry is
    $\int_0^1 \dd{x} xE^{s-\bar{s}}(x,0) = 0.0009^{(12)}_{(8)}$
for the $x$-weighted moment, while for the lowest moment, which corresponds to the strange quark contribution to the proton's magnetic moment, we find
    $\int_0^1 \dd{x} E^{s-\bar{s}}(x,0) = \mu_s = -0.033^{(11)}_{(13)}$.

At nonzero values of $t$, the strange asymmetry is not as strongly suppressed as the nonstrange $\bar d-\bar u$ asymmetry.
At $-t=0.25$~GeV$^2$, for instance, as also shown in Fig.~\ref{2ds-sbart0}, the magnetic GPD asymmetry $xE^{s-\bar{s}}(x,t)$ is only slightly smaller in magnitude than that at $t=0$, while for electric GPD asymmetry the peak value of the magnitude of $xH^{s-\bar{s}}(x,t)$ at $-t=0.25$~GeV$^2$ is even larger than that at $t=0$.
% Both of the integrated values of $xH^{s-\bar{s}}$ and $xE^{s-\bar{s}}$ are positive.  

\begin{figure}[] % Fig. 8
\graphicspath{{figures/}}
\includegraphics[width=8.0cm]{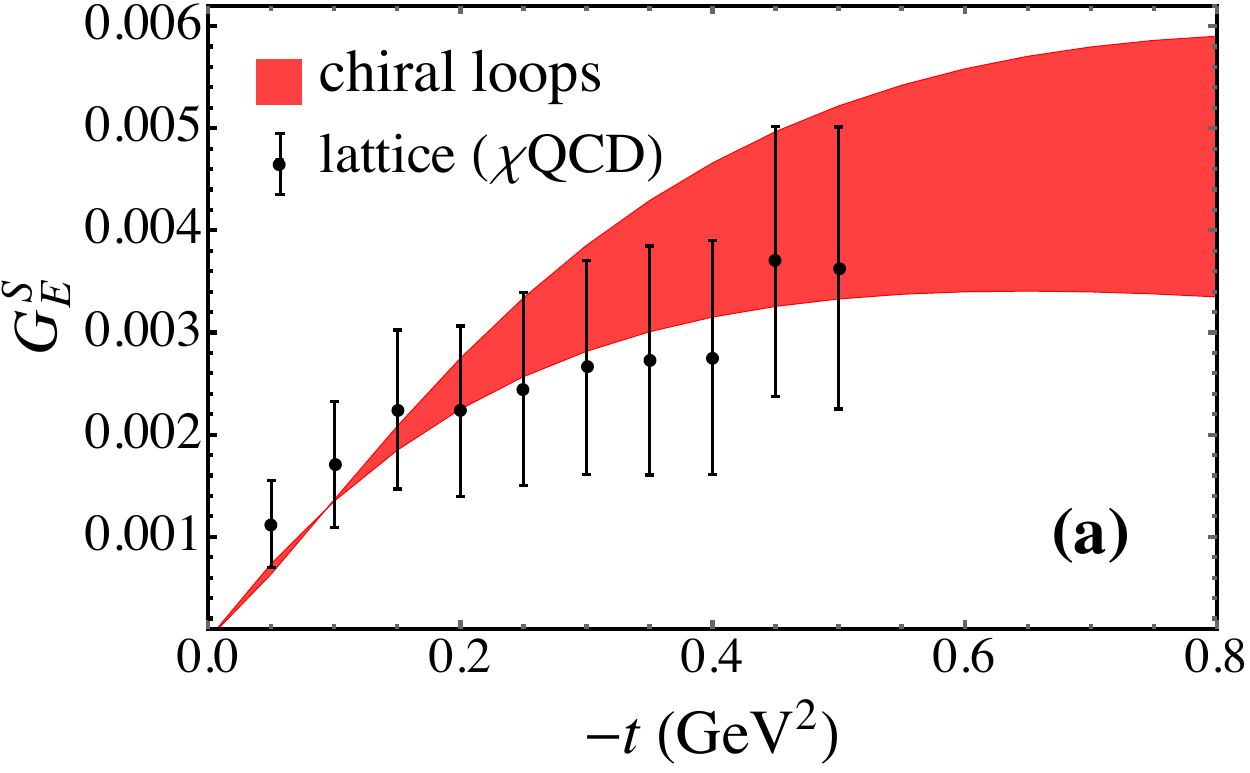} \hspace*{-0.1cm}
\includegraphics[width=8.0cm]{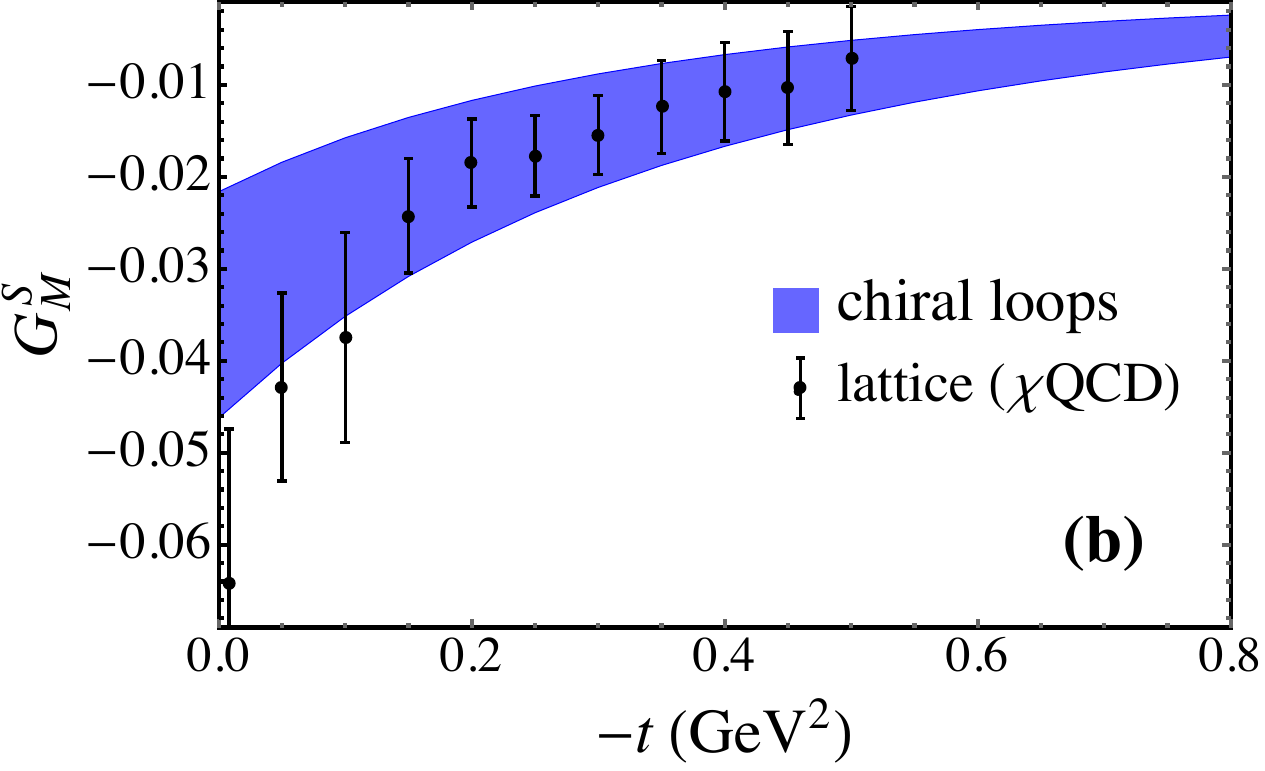} \\[0.3cm]
\caption{Strange quark contributions to the {\bf (a)} electric $G_E^s$ and {\bf (b)} magnetic $G_M^s$ form factors of the proton versus four-momentum transfer squared $-t$, with the uncertainty band corresponding to cutoff values $\Lambda=1.0(1)$~GeV, compared with the lattice simulation from Ref.~\cite{Sufian:2017osl}.}
\label{gems}
\end{figure}

A more direct representation of the $x$-integrated strange GPD asymmetries is given in Fig.~\ref{gems}, where the strange quark contributions to the proton's electric and magnetic form factors as in Eqs.~(\ref{eq.F12q})--(\ref{eq.GEMN}) are plotted versus $t$.
The uncertainty bands for the computed $G_{E,M}^s(t)$ form factors correspond to the results with $\Lambda=1.0(1)$~GeV, and the form factors are compared with recent lattice simulations at the physical pion mass~\cite{Sufian:2017osl}.
While the strange electric form factor in Fig.~\ref{gems}(a) at $t=0$ is normalized to zero, at finite momentum transfer $G_E^s(t)$ is positive and saturates at around $+0.004$ over the range $-t < 1$~GeV$^2$.
The strange charge radius can be evaluated from the slope of the electric form factor at $t=0$, 
\begin{eqnarray}
\langle (r^s_E)^2 \rangle &=& 6 \frac{\dd{G}^s_E(t)}{\dd{t}}\bigg|_{t=0}.
\end{eqnarray}
The value found in the present calculation, $\langle (r^s_E)^2 \rangle \approx -0.003$~fm$^2$, is very similar to that reported from the lattice simulation in Ref.~\cite{Leinweber:2006ug}.

The strange magnetic form factor $G_M^s(t)$ is shown in Fig.~\ref{gems}(b) as a function of $t$, also compared with the lattice calculation from Ref.~\cite{Sufian:2017osl}.
As for $G_E^s(t)$, the absolute value of the strange magnetic form factor increases with increasing values of the cutoff $\Lambda$, and decreases with $-t$, consistent with the lattice simulations from Ref.~\cite{Sufian:2017osl}.
The radius associated with the strange magnetic form factor is defined as
\begin{eqnarray}
\langle (r^s_M)^2 \rangle &=& 6 \frac{\dd{G}^s_M(t)}{\dd{t}}\bigg|_{t=0},
\end{eqnarray}
and is estimated to be $\langle (r^s_M)^2 \rangle = -0.023(7)$~fm$^2$ for $\Lambda=1.0(1)$~GeV. 
Our results are also consistent with the direct calculation of the strange form factors with a nonlocal chiral Lagrangian from Ref.~\cite{He:2018eyz}. \\

\begin{figure}[]
\graphicspath{{figures/}}
\includegraphics[width=8.0cm]{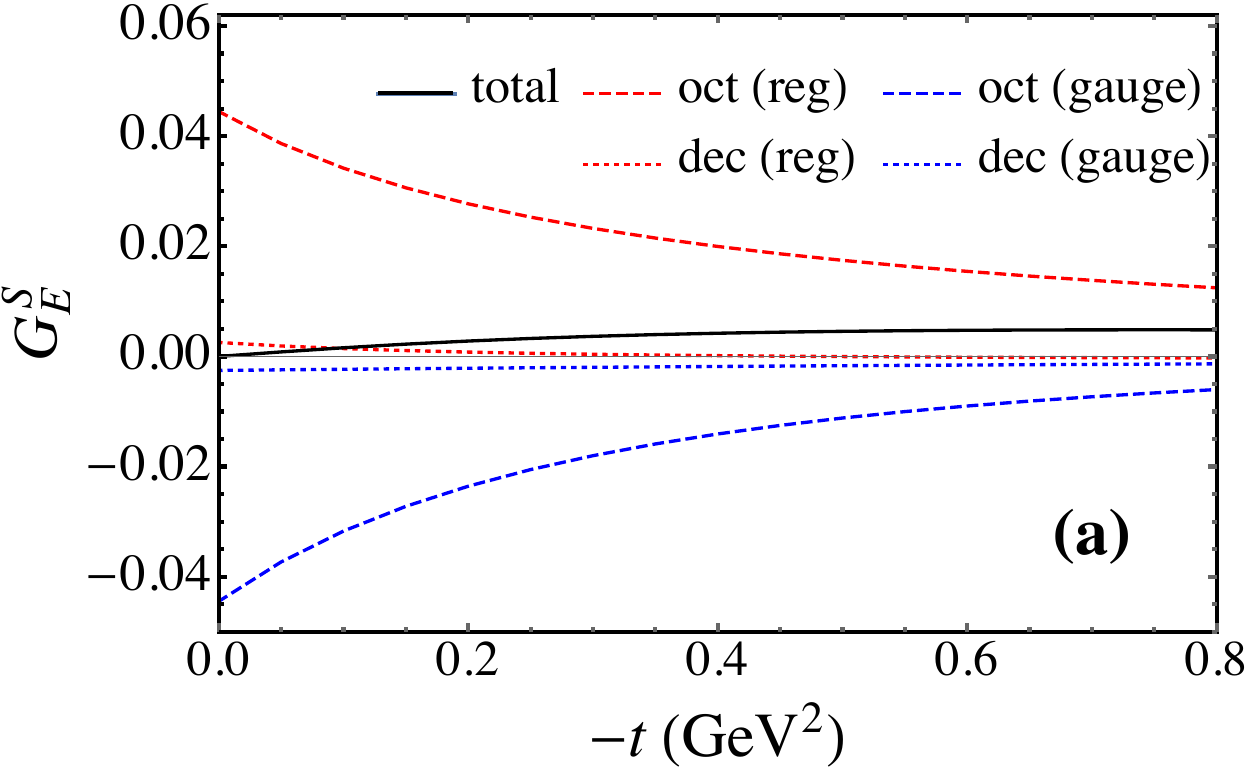} \hspace*{-0.1cm}
\includegraphics[width=8.0cm]{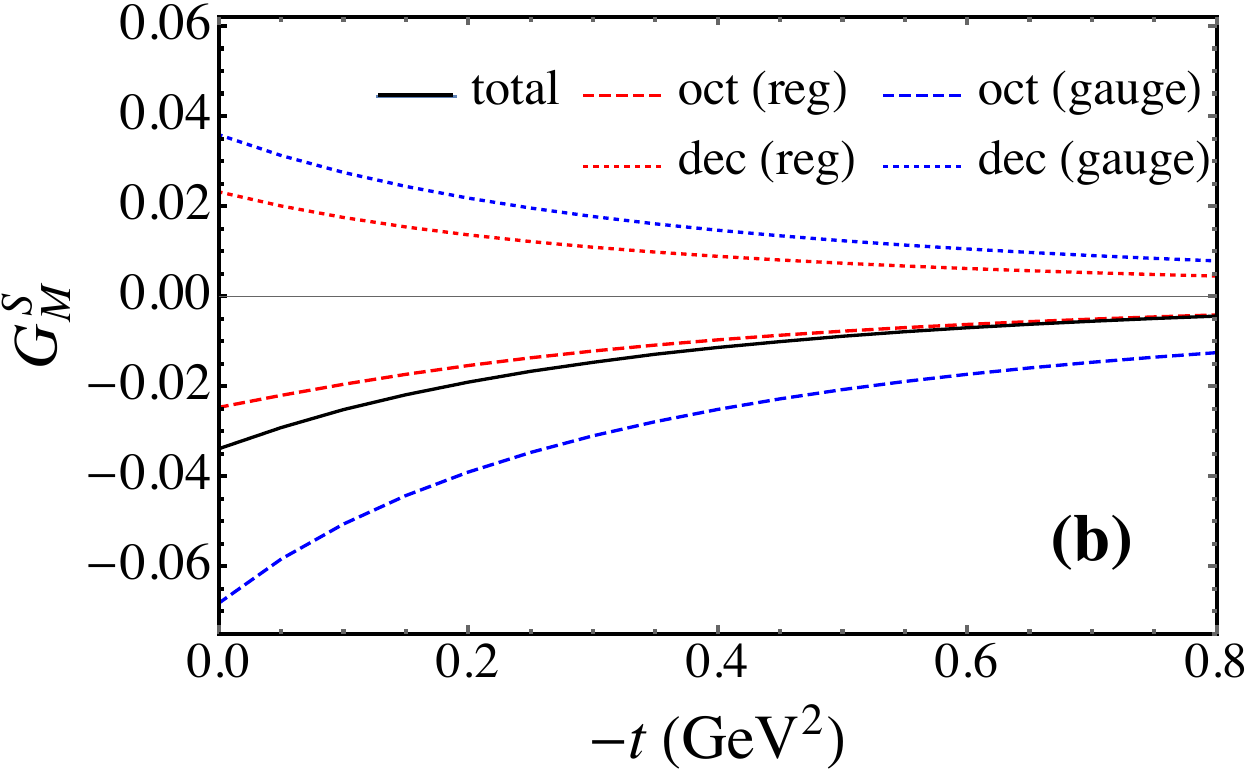} \\[0.3cm]
\caption{Contributions from different intermediate hadronic configurations to {\bf (a)} the strange electric $G_E^s$ and {\bf (b)} strange magnetic $G_M^s$ form factors, with $\Lambda=1.0$~GeV, including the octet (dashed lines) and decuplet (dotted lines) states and total (solid lines), for the regular (red lines) and additional gauge link (blue lines) diagrams.} 
\label{gems1GeV}
\end{figure}

In Fig.~\ref{gems1GeV}, we separate the contributions of different intermediate hadronic configurations to the strange electric and magnetic form factors for $\Lambda=1.0$~GeV.
Specifically, the contributions from intermediate state octet and decuplet baryons are shown, for the regular diagrams and the additional gauge link interaction diagrams in Fig.~\ref{diagrams}.
For the strange electric form factor, there is a cancellation between the contributions from the regular and gauge link diagrams when $t=0$, which is guaranteed in fact by the gauge invariance of the nonlocal Lagrangian.
The result is that the net strange charge in the proton is zero.
For the strange magnetic form factor, the gauge link contributions are actually larger in magnitude compared with the regular diagrams, and some cancellation is found between the negative octet and positive decuplet terms, resulting in an overall negative $G_M^s$.

%%%%%%%%%%%%%%%%%%%%%%%%%%%%%%%%%%%%%%%%%%%%%%%%%%%%%%%%%%%%%%%%%%%%%%%%%%%%%
\section{Summary}
\label{sec.summary}

This paper has presented a detailed account of unpolarized GPDs of sea quark and antiquarks in the proton arising from pseudoscalar meson loops whose interactions with octet and decuplet baryons are described within a nonlocal chiral effective theory with a finite range regularization.
We have restricted this initial study to the special case of zero skewness, $\xi = 0$, although the calculation can be straightforwardly extended to the  $\xi > 0$ case.
Within the convolution formulation, the dependence of the electric $H^q \equiv H^q(x,\xi\!=\!0,t)$ and magnetic $E^q \equiv E^q(x,\xi\!=\!0,t)$ sea quark GPDs on the parton momentum fraction $x$ and four-momentum transfer squared $t$ has been computed from the derived nonforward hadronic splitting functions and input GPDs of hadronic configurations constrained by SU(3) flavor symmetry.
%
%A~relativistic regulator is generated from the nonlocal Lagrangian where a gauge link is introduced to guarantee local gauge invariance, with additional diagrams from the expansion of the gauge link ensuring conservation of electric charge and strangeness.
%Flavor asymmetries for sea quarks at zero and finite momentum transfer, as well as strange form factors, are obtained from the calculated GPDs, and results compared with phenomenological extractions and lattice QCD.
%
%The splitting functions were derived including both intermediate octet and decuplet states at {\color{blue} the} one loop level.
% The parameters $c_1$ and $c_2$ were determined by the experimental nucleon magnetic moments.
%The mass parameter $\Lambda$ in the nonlocal vertex was chosen to be around 1~GeV, which gives a reasonable description of the nucleon electromagnetic and strange form factors up to {\color{blue} a rather} large momentum transfer. 
% input of valence quark distributions {\color{blue} taken from the parametrizations in Refs.~\cite{Martin:1998sq, Diehl:2004cx,  Leader:2006xc}} . 

For all light and strange quark flavors the electric, spin-nonflip GPDs $H^q$ are positive.
For the magnetic spin-flip GPDs, the $E^{\bar{d}}$ and $E^{\bar{s}}$ distributions are positive, while the $E^{\bar{u}}$ GPD is negative.
The strange magnetic GPD $E^s$, on the other hand, displays nontrivial $x$ dependence with changes of sign as a function of $x$.
The electric and magnetic sea quark flavor asymmetries for the light quarks consequently remain positive across all $x$ values, decreasing in magnitude with increasing momentum transfer squared $-t$. 
Interestingly, the magnetic asymmetry $E^{\bar{d}-\bar{u}}$ is some four times greater than the corresponding electric asymmetry $H^{\bar{d}-\bar{u}}$, which presents opportunities for phenomenological studies of this function with future experiments.
The shape of the electric asymmetry is constrained at $t=0$ by Drell-Yan and other measurements, and is quite comparable with the $\bar d-\bar u$ PDF asymmetry from global QCD analysis~\cite{Cocuzza:2021cbi}.

For the strange quark GPDs, both the electric and magnetic $s-\bar{s}$ asymmetries are significantly smaller than for the nonstrange case, and change sign with $x$.
The integral of $xH^{s-\bar{s}}$ favors positive values, and has a magnitude at finite $t$ that may be even larger than the value at $t=0$.
The results are also qualitatively consistent with current phenomenological determinations from global QCD analyses, although within rather large uncertainties.
% Without fine-tuning the parameters, our results are in reasonable agreement with previous calculations of PDFs, as well as the phenomenological extractions of PDFs. 
The electric and magnetic $s-\bar s$ GPD asymmetries, integrated over $x$, are also broadly consistent with the strange electromagnetic form factors as a function of $t$, obtained from recent lattice QCD simulations~\cite{Sufian:2017osl} as well as from direct calculations within nonlocal chiral effective theory~\cite{He:2018eyz}.

While the present analysis has been performed at zero skewness, $\xi=0$, in future it will be important to extend the GPD calculations to nonzero skewness.
Such calculations will naturally be rather more complicated, but should provide further insights into the three-dimensional structure of the nucleon.
The current analysis can also be easily extended to the case of spin-dependent GPDs of sea quarks in the proton, where we know from similar studies of helicity PDFs~\cite{Wang:2020hkn} that chiral loops play a somewhat different role for polarized and unpolarized distributions.
Experimentally, while facilities such as Jefferson Lab are expected to provide information on GPDs in the valence quark region at larger values of~$x$, distributions of sea quarks will be ideally suited for study at the future Electron-Ion Collider~\cite{AbdulKhalek:2021gbh}. \\

%\clearpage
%%%%%%%%%%%%%%%%%%%%%%%%%%%%%%%%%%%%%%%%%%%%%%%%%%%%%%%%%%%%%%%%%%%%%%%%%%%%%
\acknowledgments

We thank C.~Cocuzza for providing the results from the JAM PDF analysis from Ref.~\cite{Cocuzza:2021cbi}.
This work was supported by NSFC under Grant No.~11975241, the DOE Contract No.~DE-AC05-06OR23177, under which Jefferson Science Associates, LLC operates Jefferson Lab, DOE Contract No.~DE-FG02-03ER41260, and by the University of Adelaide (CSSM) and the Australian Research Council through the ARC Centre of Excellence for Dark Matter Particle Physics (CE200100008) and Discovery Project DP180100497.

\clearpage
%%%%%%%%%%%%%%%%%%%%%%%%%%%%%%%%%%%%%%%%%%%%%%%%%%%%%%%%%%%%%%%%%%%%%%%%%%%%
\appendix
\section{Splitting function integrals}
\label{sec.appendix}

In this appendix, we list the complete set of formulas for the numerators appearing in the integrals of the splitting function for the intermediate decuplet states.
Starting with the rainbow diagram in Fig.~\ref{diagrams}(m), % with coupling to the meson, 
the functions $F_{\phi T}^{\rm (rbw)}$ and $G_{\phi T}^{\rm (rbw)}$ in Eqs.~(\ref{eq.fg_phiTrbw}) are given by
\begin{subequations}
\label{eq.FG_phiTrbw}
\begin{eqnarray}
\hspace*{0cm}
F_{\phi T}^{{\rm (rbw)}}
&=& -\frac{y}{12 M_T^2} 
\bigg\{
  4\, k\cdot p\, k\cdot p'\, 
  \Big[ 4 k\cdot P - 2 M (2 M + 3 M_T) + (1+y)\, t 
  \Big]
\nonumber \\
&&
+\, 4 M (k\cdot p')^2 (\Delta_T-M)
+ 4 (k\cdot p)^2 \Big[ M (2M+M_T) + t \Big]
\nonumber\\
&&
+\, 2\, k\cdot p' 
  \Big[ 8 M^3 \MTbar 
    + (1+y)M\, (\Delta_T - M)\, t 
    + 3 M_T^2\,t 
    - 4 M^2 k^2
  \Big]
\nonumber\\
&&
-\, 2\, k\cdot p\,
  \Big[ 8 M^3 \MTbar
    + (1+11 y) M \MTbar\, t
    - 3 \MTbar^2 t 
    + 4 M^2 k^2
    - y t\, (5 M^2 + t)\, t
  \Big]
\nonumber \\
&&
+\, 4\, M \MTbar \big( 4 M^2-3 (1-y)\, t \big) 
    - 12 y M \MTbar^2\, \Delta_T\, t\,
    + 3 y \MTbar^2\, t^2
\nonumber \\
&&
+\, 4\, (1-2 y) M^2\, t\, k^2\,
    - y M \MTbar (4 M^2+5 t)\, t\, k^2 
\bigg\}\, ,
\\
\hspace*{0cm}
G_{\phi T}^{({\rm rbw})}
&=& \frac{yM}{3M_T^2} 
\bigg\{
4\, k\cdot p 
\Big[ k\cdot p'\, \big( y M + \MTbar \big)
    + k\cdot p\,  \big( 2\MTbar + M_T \big)
\Big]
  + 3 M \MTbar^2\, y \, t
\nonumber \\
&&
+\, 4\, k\cdot P
\Big[ 3 \MTbar^2 
    \big( \MTbar-2M \big)
  + \big( M-3\MTbar \big)\, k^2
\Big]
  + 2\, k\cdot p'\, M^2
\Big[ (3+y) \MTbar - 3 y M \Big]
\nonumber \\
&&
+\, 2\, k\cdot p \,
\Big[ y M (5 M^2+t) - \MTbar \big( (1+11 y) M^2 - t \big)
\Big]
  + \MTbar \big( 4(1+3y) M^2 - 3t \big)\, k^2
\nonumber \\
&&
-\, M \big( 8 y M^2 - t \big)\, k^2 
  - y M^2 \MTbar \big( 4M^2 - 12\MTbar (2M-\MTbar) + 5t \big) 
\bigg\}.
\end{eqnarray}
\end{subequations}
where $\MTbar \equiv M_T+M$ and $\Delta_T \equiv M_T-M$.

For the electric part of the photon coupling to the decuplet baryon in Fig.~\ref{diagrams}(n), the functions $F_{T\phi}^{\rm (rbw)}$ and $G_{T\phi}^{\rm (rbw)}$ in Eqs.~(\ref{eq.fg_Tphirbw}) are given by
\begin{subequations}
\label{eq.FG_Tphirbw}
\begin{eqnarray}
F_{T\phi}^{({\rm rbw})}
&=& \frac{1}{36 M_T^4}
\bigg\{
    16\, k\cdot p \, k\cdot p'\, (k\cdot \Delta)^2\,
    + 16\, k \cdot P
    \Big[ 
        \big( (k\cdot p')^2 + (k\cdot p)^2 \big) 
        (3M_T-2M) \MTbar
\nonumber\\
&& \qquad
        -\,k\cdot p\, k\cdot p'\,
        \big( 2 M (M_T - 2 M) + 3 (2+y) M_T^2 - 2 y\, t^2 \big)
    \Big]
\nonumber \\
&&  +\, 16\, (k\cdot\Delta)^2\, M^2 \MTbar^2\,
    +\, 48\, (k\cdot P)^2\, y M_T^2\, k^2
\nonumber \\
&&  +\, 4\, k\cdot p \, k\cdot p' 
    \Big[ 
        4 (4 M - 3 M_T ) M_T \MTbar^2
      - y^2 (8 M \MTbar + 3 M_T^2)\, t
      + 6 y \MTbar \Delta_T t
\nonumber \\
&& \qquad
    +\, 2 \MTbar^2\, t
    + 6 y (3 M - M_T) M_T^2 \MTbar
    + 2 (4 M + 9 M_T) M_T k^2
    - 2 (1+y) t\, k^2
    + y^2 t^2
    \Big]
\nonumber \\
&& 
    +\, 4 \big( (k\cdot p')^2 + (k\cdot p)^2 \big) 
\nonumber \\
&& \qquad \times
    \Big[ M_T \MTbar^2\, (9 M_T-8 M) 
    - 2 y (2 M-3 M_T) \MTbar\, t
    - 3 y M_T^2 \MTbar \Delta_T
    - 4 M M_T k^2 
    \Big]
\nonumber \\
&&
    +\, 4\, k \cdot P\, 
    \Big[2 y (2 M^2 + 3 M_T \MTbar)\, t\, k^2
    - 12 M_T \MTbar\, t\, k^2
    - 12 (3 M-M_T) M_T^2 \MTbar k^2 
\nonumber \\
&& \qquad
    -\, 3 y^2 (3 M^2 + \MTbar^2) \MTbar \Delta_T\, t^2
    + 2 y (2 M^2-7 M M_T+6 M_T^2) \MTbar^2\, t
\nonumber \\
&& \qquad
    -\, y^2 (2 M-3 M_T) \MTbar\, t^2
    + 12 y M M_T^2 \MTbar^2 \Delta_T
    - 12 M_T^2 k^4 
    + 4 y^2 M^2\, t\, k^2
\nonumber \\
&& \qquad
    +\, 3 y M_T (M+\MTbar) (y t - 4 M M_T) k^2
    \Big]
\nonumber \\
&&  +\, 48\, y M^3  M_T^2\, \MTbar k^2 
    - 2 y^2 M (8 M^2 \MTbar-3 M_T^3)\, t\, k^2
    + (2M - 3M_T)^2 \MTbar^2\, (y^2 t - 2 k^2)\, t
\nonumber \\
&&  +\, 2\, y  
    \big( 4 M^3 \MTbar - 3 M_T^2 (5M^2 + 3M_T^2 - 2M M_T)
    \big)\, t\, k^2
    + 24 (M-2M_T) M M_T^2 \MTbar^2\, k^2
\nonumber \\
&&  -\, 2\, y^2 M
      \big( 8 M^2 (M-2M_T) + 3 (2M+M_T) M_T^2 \big) \MTbar^2\, t
\nonumber \\
&&  +\, 2 (1-y) (2M+3M_T)^2\, t\, k^4
    + 24\, M M_T^2 (3M+2M_T)\, k^4
\bigg\}\, ,
\\
G_{T\phi}^{({\rm rbw})}
&=& \frac{M}{9 M_T^4\, t}
\bigg\{
    - 32\, k\cdot p\, k\cdot p'\, (k\cdot \Delta)^2\, \MTbar
\nonumber \\
&& 
+\, 8\, k \cdot P 
\Big[ (k\cdot\Delta)^2 (2M+3M_T) k^2
    - 3 \big((k\cdot p')^2 + (k\cdot p)^2\big)(2M+M_T) \MTbar \Delta_T
\nonumber \\
&& \qquad
-\, 2\, k\cdot p\, k\cdot p'\, 
    \big(\MTbar (3 M_T \MTbar - 6 M^2 - t)
        + y (3 M + 2M_T)\, t
    \big)
\Big] 
\nonumber \\
&& 
-\, 2\, (k\cdot\Delta)^2 
\Big[ 
    \big( 8 M^2 \MTbar - 3 (2M+M_T) M_T^2 \big) k^2
\nonumber \\
&& \qquad
+\, \left( 8 M^3 + \left(3 (2 M+M_T) M_T - 16 M^2\right) M_T
    \right) \MTbar^2
\Big]
\nonumber \\
&&
+\, 4\, k\cdot p\, k\cdot p'\, t
\Big[
    3 y \big( 4 M^2 - (M+2M_T) M_T \big) \MTbar
    - M_T \MTbar^2
    - y \MTbar\, t
\nonumber \\
&& \qquad
    +\, y^2 M (8M \MTbar + 3M_T^2)
    + y (4M+3M_T)\, k^2
    + (8M+9M_T)\, k^2
    - y^2 M t 
\Big]
\nonumber \\
&& 
+\, 2 \big( (k\cdot p')^2 + (k\cdot p)^2 \big)\, t
\Big[
    y \big(10 M^2-3 (3M+M_T) M_T \big) \MTbar
\nonumber \\
&& \qquad   
    +\, (2M-3M_T) \MTbar^2
    + (1+y)\, (2M+3M_T)\, k^2
\Big]
\nonumber \\
&& 
- 2\, k\cdot P\, t
\Big[ - 6 y^2 M (4M^2 + 2M M_T + M_T^2) \MTbar \Delta_T
\nonumber \\
&& \qquad   
+\, y (2M-3M_T) 
    \big( 8 M^2 - 4 M M_T - 3 M_T^2 \big) \MTbar^2
    - y (2 M y + \MTbar) (2M-M_T) \MTbar\, t
\nonumber \\
&& \qquad   
    +\, (4 k^2 - y t)(M_T+2\MTbar) k^2
    + 2 y^2 M \big(4M^2 + 3(M+\MTbar) M_T\big) k^2
\nonumber \\
&& \qquad   
    +\, y \big(24 M^3+32 M^2 M_T-9 M_T^3\big) k^2
    + 2 \big(8 M^3+2 M^2 M_T-21 M M_T^2-15 M_T^3\big) k^2
\Big]
\nonumber \\
&&
+\, t
\Big[2 y M \MTbar \big( 4M^3 - 9M_T^2 \MTbar \big) k^2
    + 2 y^2 M^2 (8 M^2 \MTbar - 3M_T^3) k^2
\nonumber \\
&& \qquad   
    -\, y (4M^2-9M_T^2) \MTbar\, t\, k^2
    + (2M-3M_T)^2 \MTbar^2\, (2 \MTbar k^2 - y^2 M t)
\nonumber \\
&& \qquad      
    +\, 2 y^2 M^2 
    \Big( 8 M^3 + \big( 3 M_T (M+\MTbar) - 16 M^2 \big) M_T
    \Big) \MTbar^2
\nonumber \\
&& \qquad   
    +\, 2 \big( 4 M^3 - 15 M M_T^2 - 9 M_T^3 \big) k^4
    + 2 y M (M_T+2\MTbar)^2 k^4
\Big]
\bigg\}.
\end{eqnarray}
\end{subequations}

\clearpage
For the magnetic photon-decuplet baryon diagram in Fig.~\ref{diagrams}(o), the functions $F_{T\phi}^{\rm (rbw\, mag)}$ and $G_{T\phi}^{\rm (rbw\, mag)}$ in Eqs.~(\ref{eq.fg_Tphirbwmag}) are given by
\begin{subequations}
\label{eq.FG_Tphirbwmag}
\begin{eqnarray}
F_{T\phi}^{\rm (rbw\, mag)}
&=&\frac{1}{36 M_T^5}
\bigg\{
     8\, k\cdot p\, k\cdot p'\,
\Big[ M_T (k\cdot\Delta)^2
    + y\, k\cdot P\, \big( 2(1+y)M + 5M_T \big)\, t 
\Big]
\nonumber \\
&& 
+\,  8\, k\cdot P\, (k\cdot\Delta)^2 
\Big[ (2M + 3M_T)\, k^2 
    - M (2M + M_T) \MTbar \big)
\Big] 
\nonumber \\
&&
+\,  2\, (k\cdot\Delta)^2
\Big[ 4 M^3 \MTbar^2
    + 3 (M-2 M_T) M_T^2 \MTbar^2
    - 2 M^2 M_T \MTbar^2
\nonumber \\
&& \qquad
-\,  M_T \big( 4 M^2 - M M_T - 12 M_T^2\big) k^2
    -(2M+3M_T) (y\, t + 2 k^2) k^2
\Big]
\nonumber \\
&&     
-\, 2\, \big( (k\cdot p')^2 + (k\cdot p)^2 \big)\, t
\Big[ y \big(2 M^2 + 5 M M_T - 9 M_T^2\big) \MTbar
\nonumber \\
&& \qquad
    + (2M-3M_T) \MTbar^2
    - (2M+3M_T) k^2
    + 2 y^2 M^2 (2M-M_T)
\Big]
\nonumber \\
&&
-\,  4\, k\cdot p \, k\cdot p'\, t
\Big[ y \big(4 M^2 + 2 M M_T - 5 M_T^2\big) \MTbar
    - y (1+y) \MTbar\, t  
    - (2 M-M_T) \MTbar^2
\nonumber \\
&& \qquad
    +\, 2 y^2 M (3M-2M_T) \MTbar
    + \big( 2M + 5M_T + y(2M+3M_T) + 2y^2 M \big) k^2
\Big]
\nonumber \\
&&
-\,  2\, k\cdot P\, t
\Big[ 12 M_T^2 \MTbar\, k^2
    + y^2 \big( 4 M^2-2 M M_T-3 M_T^2 \big) \MTbar\, t
    + y (2M-3M_T) \MTbar^2\, t
\nonumber \\
&& \qquad
    -\, y (M-2M_T) \big( 4 M^2 + 6 M M_T - 3 M_T^2 \big) \MTbar^2
    -\, 2 y^2 M^2  \big( 8 M^3 - 9 M M_T^2 - M_T^3 \big)
\nonumber \\
&& \qquad
    -\, y \big( (2M+3M_T)(2 k^2 - t) 
                - 8 y M^3
                + (15 M M_T^2 + 12 M_T^3 - 16 M^2 \MTbar)\, 
          \big) k^2
\Big]
\nonumber \\
&&
-\, t\, k^2 
\Big[ 4 M \MTbar\, \big( y M (2M-5M_T) + M_T (2M-3M_T) \big)
    - 2y^2 M^3 (4M^2-M_T^2)
\nonumber \\
&& \qquad    
    - y\, (4M^2 - 9M_T^2)\, \MTbar\, t 
    + 4M (yM-M_T) (2M+3M_T)\, k^2
\Big]
\nonumber \\
&&
+\  y^2 M \big(4 M^2 - 8 M M_T + 3 M_T^2\big)\, \MTbar^2\, (t-2M^2)\, t
\bigg\}\, ,
\\
G_{T\phi}^{\rm (rbw\, mag)}
&=& \frac{M}{18 M_T^5\, t}
\bigg\{
	8\, k\cdot p\, k\cdot p'\, 
\Big[         
	\big( (k\cdot p)^2 + (k\cdot p')^2 \big) 
	\big( 4 \MTbar \Delta_T + (1+y)t - 2 (M^2+k^2) \big)
\nonumber \\
&& \qquad\qquad
    +\, 4\, (k\cdot\Delta)^2\, k\cdot P
    + 2\, k\cdot p\, k\cdot p'\, 
        \big( 3 M M_T -  4 \MTbar \Delta_T - (1-y) t + 2 k^2 \big)
\Big]
\nonumber \\
&&
	-\, 8\, \big( (k\cdot p')^4 + (k\cdot p)^4\big)\, M (2M-M_T)
\nonumber \\
&&
    +\,	8\, k\cdot P\,
\Big[ 4 (k\cdot P)^2 \big( 4 M^2 (2M^2+k^2) - M M_T^2\, (9M+M_T) \big)
\nonumber \\
&& \qquad\qquad
    +\,	k\cdot p \, k\cdot p'\, t\,
    \big( 2\, (5M_T-M) \MTbar
	    + 2 y\, (5 M^2 + 4 M M_T - 2 M_T^2)
\nonumber \\
&& \qquad\qquad\qquad\qquad\qquad
        +\, 4 y^2 M^2 - y\, t + 2(1+y)\, k^2 
    \big)  
\nonumber \\
&& \qquad\qquad
    - \big( (k\cdot p')^2 + (k\cdot p)^2 \big)\, t\,
      \Big( 2(1+y)M^2 - M_T \big( (1+y)M + 3M_T \big) \Big)
\Big]
\nonumber \\
&&
-\, 4\, (k\cdot\Delta)^2 M^2
\Big[ \big( 4 M^2 - 8 M M_T + 3 M_T^2\big) \MTbar^2
    + (4M^2-M_T^2) k^2
\Big]
\nonumber \\
&&
\nonumber \\
&& 
+\, 2 \big( (k\cdot p')^2 + (k\cdot p)^2 \big)\, t\,
\Big[ \big( 8 M \MTbar + 9 M_T^2 \big) k^2
\nonumber \\
&& \qquad
    +\, y M \big( 12 M^2 - 6 M M_T - 13 M_T^2 \big) \MTbar
	- y (2M-3M_T) \MTbar\, t
\nonumber \\
&& \qquad
    +\, 4 M (M-2M_T) \MTbar^2
	+ 2 y M (4M+3M_T) k^2
	+ 4 y^2 M^3 (2M-M_T)
\Big]
\nonumber \\
&&
+\,   4\, k\cdot p \, k\cdot p'\, t\,
\Big[ 12\, y M \MTbar\, k^2
     + 4 y^2 M^2 (3M-2M_T) \MTbar
\nonumber \\
&& \qquad    
+\,  y M \big( 16 M^2 - 4 M M_T - 5 M_T^2 \big) \MTbar
     - 2 \big( M^2 - 5 M M_T + 3 M_T^2 \big) \MTbar^2
\nonumber \\
&& \qquad     
-\,  \big( 2 y^2 M + y (4M-M_T) + \MTbar \big) \MTbar\, t
\nonumber \\
&& \qquad  
+\, \big( 12 M^2 + 22 M M_T + 9 M_T^2 \big) k^2
    + 4 y^2 M^2 k^2
	- t\, k^2 
    + 2 k^4 
\Big]
\nonumber \\
&&
-\, 2\, k\cdot P \, t\,
\Big[ 2 y\, (4M-3M_T) \MTbar^2 \Delta _T\, t
	- 2 y^2 M \big( 4 M^2 - 2 M M_T - 3 M_T^2 \big) \MTbar\, t
\nonumber \\
&& \qquad
    +\, 4\, y M \big( 4 M^3 - (5 M^2-3 M_T \Delta_T) M_T \big) \MTbar^2
    + 4 y^2 M^3 \big( 8 M^3 - M_T^2 (9M+M_T) \big)
\nonumber \\
&& \qquad
	+\, 16\, y^2 M^4 k^2
	+ 4 \big( 6 M^3 - 4 M M_T^2 - 9 M_T^3 \big) \MTbar\, k^2
    - 8 M \MTbar\, t\, k^2
\nonumber \\
&& \qquad
    +\, 4\, y M \big(5M+6M_T\big) \big(2M^2-M_T^2\big) k^2
    - y \big( 4 M^2 + 10 M M_T + 3 M_T^2 \big)\, t\, k^2
\nonumber \\
&& \qquad
    +\, 12\, \big( 2 M^2 + 4 M M_T + 3 M_T^2 \big)\, k^4 
	+ 4 y M (2M+3M_T)\, k^4
\Big]
\nonumber \\
&&
+\, t\,
\Big[ 2y^2 M^2 (2 M^2-t) \big( 4 M^2 - 8 M M_T + 3 M_T^2 \big) \MTbar^2
	- \big( 2M-3M_T \big)^2 \MTbar^2\, t\, k^2
\nonumber \\
&& \qquad
    +\, 2 \big(2M-3M_T\big) \big(2 M^3 - M^2 M_T - 3 M_T^3 \big) \MTbar^2\, k^2
	+ 4 y^2 M^4 \big(4M^2-M_T^2\big)\, k^2
\nonumber \\
&& \qquad
    -\, 2 y M \big(4 M \MTbar-9 M_T^2\big) \MTbar\, t\, k^2
	+ 8 y M^3 \big(2 M^2+M_T (M-5M_T)\big) \MTbar\, k^2
\nonumber \\
&& \qquad
    +\, 8 y M^3 \big(2M+3M_T\big)\, k^4
    + \big(2 k^2-t) \big(2M+3M_T\big)^2\, k^4
\nonumber \\
&& \qquad
    +\, 4 \big(4 M^4 + 6 M^3 M_T - 12 M M_T^3 - 9 M_T^4\big)\, k^4
\Big]
\bigg\}\, .
\end{eqnarray}
\end{subequations}

For the magnetic octet-decuplet transition diagrams in Figs.~\ref{diagrams}(p) and \ref{diagrams}(q), the numerator functions $F_{T\phi}^{\rm (rbw\, mag\, 1,2)}$ and $G_{T\phi}^{\rm (rbw\, mag\, 1,2)}$ in Eqs.~(\ref{eq.fg_BTphi}) are expressed as  
\begin{subequations}
\label{eq.FG_BTphi}
\begin{eqnarray}
F^{\rm (rbw\, mag\, 1)}_{BT}
&=&\frac{1}{12 M_B M_T^2}
\bigg\{
	8\, k\cdot p\, k\cdot\Delta
\Big[
    k\cdot p'\, \big( 3 \MTbar + \Delta_{TB} \big)
	- k\cdot p\, \MBbar
\Big]
+\, 8\, k\cdot p\, k\cdot p' 
\nonumber\\
&& \quad \times
\Big[ (M_B+\MTbar) k^2
	+ (M^2 + M_B \Delta_T - 3 M_T^2 + t) \MTbar
    + y \big((1-y) M + \Delta_{TB}\big)\, t
%    + y \big(\MTbar - M_B - y M\big)\, t
\Big]
\nonumber\\
&&
-\, 4\, (k\cdot p')^2 
\Big[ (2M-3M_T) (2M_T-\Delta_B) \MTbar 
    + 2 (2M+3M_T) k^2
\Big] 
\nonumber\\
&&
+\, 4\, (k\cdot p)^2\,
\Big[ M_T \MBbar \MTbar	+ 2 (2M_T-\Delta_B)\, k^2
\Big]
- 2\, k\cdot p\, t\,
\Big[ y \MBbar\, (y\, t + 2 M_T \MTbar)
\nonumber\\
&& \qquad
+\, 2 y^2 M (M+3M_T) \MBbar
    + 2 y (2M+M_T)k^2
    + 2 (\Delta_{TB} - y^2 M) k^2 
\Big]
\nonumber\\
&&
+\, 4\, k\cdot p'\, t
\Big[ y \MTbar \big(2M \Delta_B	+ (M\!-\!3M_B) M_T\big)
%\nonumber\\
%&& \qquad
    - (2M\!+\!3M_T) k^2
    + y^2 M^2 (2M\!-\!M_T)
\Big]
\nonumber\\
&& 
+\, 4\, k\cdot \Delta
\Big[ (2M+3M_T)\, k^4
	- \big( 2 M \Delta _T \MTbar 
	      + 3 M_T \Delta _T \MTbar
    	  + 2 M M_T \Delta_{TB}
     \big)\, k^2 
\Big]
\nonumber\\
&&
+\, \big( 2 y^2 M M_T (4M-3M_T) - (2 k^2 - y^2 t) (2M-3M_T) 
    \big)\, \MBbar \MTbar\, t\,
\nonumber\\
&&
+\, 2 y M 
    \big( 2M_T \MBbar - 4 \Delta_T \MTbar
        + y (M_B (2M+3M_T) + 4 M M_T)
    \big)\, t\, k^2
\nonumber\\
&&
+\, (4M+6M_T)\, t\, k^4
\bigg\}\, ,
%
% F^{\rm (rbw\, mag\, 1)}_{BT}
% &=&\frac{1}{12 M_T^2 M_B}
% \Bigg\{
%     8 k\cdot p \, k\cdot\Delta \Big[k\cdot p' \left(3 M+4 M_T-M_B\right)-k\cdot p \MBbar\Big]
% \nonumber\\
% &&
% +\, 8 k\cdot p k\cdot p' \Big[k^2     \left(\MTbar+M_B\right)+\MTbar \left(M_B \Delta _T+M^2-3 M_T^2+t\right)+t y \left(\MTbar-M_B-M y\right)\Big]   \nonumber\\
% &-&4\left(k\cdot p'\right)^2 \Big[\left(2 M-3 M_T\right) \MTbar \left(2 M_T-\Delta _B\right)+2 k^2 \left(3 M_T+2 M\right)\Big] \nonumber\\
% &+&4(k\cdot p)^2 \Big[M_T \MBbar \MTbar+2 k^2 \left(2 M__B\right)\Big]+4 k\cdot p' \Big[-t y \MTbar \left(-2 M \Delta _B-\left(M-3 M_B\right) M_T\right)
% \nonumber\\
% &+&k^2 \left(-2 M \Delta _T \MTbar-3 M_T \Delta _T \overline{M_T}-2 M M_T \Delta_{TB}\right)+k^4 \left(3 M_T+2 M\right)-k^2 t \left(3 M_T+2 M\right)
% \nonumber\\
% &+&M^2 t y^2 \left(2 M-M_T\right)\Big] -2k\cdot p \Big[t^2 y^2 \MBbar+2 M t y^2 \left(3 M_T+M\right) \MBbar+2 t y M_T \MBbar \MTbar
% \nonumber\\
% &+&2 k^2 \left(-2 M \Delta _T \overline{M_T}-3 M_T \Delta _T \overline{M_T}-2 M M_T \Delta _{TB}\right)+2 k^4 \left(3 M_T+2 M\right)+2 k^2 t y \left(M_T+2 M\right)
% \nonumber\\
% &-&2 k^2 M t y^2+2 k^2 t \Delta _{TB}\Big]
% +t \Big[4 k^2 M y \left(M_T \left(\MBbar-2 M_T\right)+2 M^2\right)-2 k^2 \left(2 M-3 M_T\right) \MBbar \MTbar
% \nonumber\\
% &+&t y^2 \left(2 M-3 M_T\right) \MBbar \MTbar+2 M y^2 M_T \left(4 M-3 M_T\right) \MBbar \MTbar
% \nonumber\\
% &+&2 k^2 M y^2 \left(M_B \left(3 M_T+2 M\right)
% +4 M M_T\right)+2 k^4 \left(3 M_T+2 M\right)\Big]
% \Bigg\},
\\
G^{\rm (rbw\, mag\, 1)}_{BT}
&=& 
-\frac{M}{6 M_B M_T^2\, t }
\bigg\{
    16\, k\cdot p\, k\cdot p'\, (k\cdot \Delta)^2 
\nonumber\\
&&
    -\, 8\, (k\cdot p)^2\ k\cdot p'
\Big[ M (4M+5M_T) + 2 M_B (M+3M_T) + (3-y) t - 2 k^2 
\Big]
\nonumber\\
&&
    +\, 8\, (k\cdot p')^2\, k\cdot p\,
\Big[ M (5M+\MTBbar)+ 3 M_B M_T + (1+y) t - k^2 
\Big]
\nonumber\\
&& 
    +\, 8\, (k\cdot p)^3
\Big[ (M+3M_T) \MBbar - k^2
\Big]
    -\, 8\, (k\cdot p')^3
\Big[ M (2M-M_T)
\Big]
\nonumber\\
&&
    -\, 4\, (k\cdot\Delta)^2
\Big[ (4M-3M_T) M_T \MBbar \MTbar
    + \big( M_B (2M+3M_T) + 4 M M_T \big)\, k^2 
\Big]
\nonumber\\
&&    
    -\, 4\, k\cdot p \, k\cdot p'\, t\, 
\Big[ (M_B+5\Delta_T) \MTbar 
    + 4 y^2 M^2 - y\, t + (1+y) k^2 
\nonumber\\
&& \qquad\qquad
    +\, y\, \big( M_B (M-3M_T) - M (M+8M_T) \big)
\Big]
\nonumber\\
&&
    -\, 4\, (k\cdot p')^2\, t\,
\Big[ (2M-3M_T) \MTbar + y M (2M-M_T)
\Big]
\nonumber\\
&& 
    +\, 4\, (k\cdot p)^2\ t\,
\Big[ (3-y) k^2 - (1-3y) \MBbar \MTbar
\Big]
\nonumber\\
&& 
    -\, 4\, k\cdot P\, y\, t\, k^2 
\Big[ M_B (2M+3M_T) + 4 M M_T
\Big] 
-\, 6\, k \cdot \Delta\, M_T\Delta_{TB}\, t\, k^2
\nonumber\\
&& 
-\, 2\, k\cdot p'\, t\, 
\Big[ y (2M-3M_T) \MTbar\, t
    + 2 M \big( 2 \Delta_T + \Delta_{TB} \big)\, k^2
    - 4 y^2 M^3 (2M-M_T)
\nonumber\\
&& \qquad\qquad
+\, y \MTbar 
    \big(2 M (3M+5M_B) M_T - 3 M_T^2 \MBbar - 4 M^2 \Delta_B \big)
\Big]
\nonumber\\
&&
-\, 2\, k\cdot p\ t\,
\Big[ 4 y^2 M^2 (M+3M_T) \MBbar
    + y \MBbar \MTbar \big(M_T (2M-3M_T) + t\big)\,
\nonumber\\
&& \qquad\qquad
+\, 2 y^2 M \MBbar\, t
    + 2 \MTbar (2M-\MTBbar)\, k^2
    + y \big( 4M (\MTbar - y M) + t \big)\, k^2 
\Big]
\nonumber\\
&&
+\, 2M \big( 2 y^2 M M_T (4M-3M_T) + y^2 (2M-3M_T)\, t + 4 M_T\, k^2 
       \big)\, \MBbar \MTbar\, t
\nonumber\\
&&
+\, 4 y M^2 
    \big( y M_B (2M+3M_T) + 4 y M M_T -  \MTBbar M_T - 2 M M_B
    \big)\, t\, k^2
\nonumber\\
&&
+\,  y \big( 2 M (\MBbar+\MTbar) - 3 M_T \Delta_{TB} \big)\, t^2\, k^2
    + 8 M M_T\, t\, k^4
\bigg\},    
\\
F^{\rm (rbw\, mag\, 2)}_{BT}
&=& \frac{1}{12 M_B M_T^2}
\bigg\{
    - 8\, k\cdot p'\, k\cdot\Delta\,
\Big[ k\cdot p\, \big(3\MTbar+\Delta_{TB}\big) - k\cdot p'\, \MBbar 
\Big]
+ 8\, k\cdot p\, k\cdot p'
\nonumber\\
&& \quad \times
\Big[ (M_B+\MTbar)\, k^2
    + \big( M^2 + M_B \Delta_T - 3 M_T^2 + t \big) \MTbar
    + y \big( (1-y)M + \Delta_{TB} \big)\ t
\Big]
\nonumber\\
&&
-\, 4\, (k\cdot p)^2 
\Big[ (2M-3M_T) (2M_T-\Delta_B) \MTbar 
    + 2 (2M+3M_T)\, k^2
\Big]
\nonumber\\
&&
+\, 4\, (k\cdot p')^2 
\Big[ M_T \MBbar \MTbar
    + 2 (2M_T-\Delta_B)\, k^2 
\Big]
- 2\, k\cdot p'\, t 
\Big[ y \MBbar\, (y t + 2 M_T \MTbar)
\nonumber\\
&& \qquad
	+ 2 y^2 M (M+3M_T) \MBbar
    + 2 y (2M+M_T)\, k^2
    + 2 (\Delta_{TB}-y^2 M)\, k^2 
\Big]   
\nonumber\\
&&
+\, 4\, k\cdot p\, t 
\Big[ y \MTbar \big( 2M \Delta_B + (M\!-\!3M_B) M_T \big)
    - (2M\!+\!3M_T)\, k^2
    + y^2 M^2 (2M\!-\!M_T)
\Big]
\nonumber\\
&&     
-\, 4\, k\cdot \Delta
\Big[ (2M+3M_T)\, k^4
	- \big( 2 M \Delta _T \MTbar 
	      + 3 M_T \Delta _T \MTbar
    	  + 2 M M_T \Delta_{TB}
     \big)\, k^2 
\Big]
\nonumber\\
&&
    +\, \big( 2 y^2 M M_T (4M-3M_T) - (2k^2 - y^2 t)(2M-3M_T) 
        \big) \MBbar \MTbar\, t
\nonumber\\
&&
    +\, 2 y M 
        \big( 2 M_T \MBbar - 4 \Delta_T \MTbar 
            + y (M_B (2M+3M_T) + 4 M M_T)
        \big)\, t\, k^2
\nonumber\\
&&     
    +\, (4M+6M_T)\, t\, k^4 
\bigg\}\, ,
\\
G^{\rm (rbw\, mag\, 2)}_{BT}
&=&
-\frac{M}{6 M_B M_T^2\, t}
\bigg\{
    16\, k\cdot p\, k\cdot p' (k\cdot \Delta)^2
\nonumber\\
&&
-\, 8\, (k\cdot p')^2\, k\cdot p
\Big[ M (4M+5M_T) + 2M_B (M+3M_T) + (3-y)t - 2 k^2  
\Big]
\nonumber\\
&&
+\, 8\, (k\cdot p)^2\, k\cdot p' 
\Big[ M (5M+\MTBbar) + 3 M_B M_T + (1+y)t - k^2
\Big]
\nonumber\\
&&
+\, 8\, (k\cdot p')^3 
\Big[ (M+3M_T) \MBbar - k^2
\Big]
  - 8\, (k\cdot p)^3
\Big[ M (2M-M_T)
\Big]
\nonumber\\
&&    
-\, 4\, (k\cdot\Delta)^2
\Big[ (4M-3M_T) M_T \MBbar \MTbar
    + \big(M_B (2M+3M_T) + 4 M M_T\big) k^2
\Big]
\nonumber\\
&&
-\, 4\, k\cdot p \, k\cdot p' t 
\Big[ (M_B+5 \Delta_T) \MTbar 
    + 4 y^2 M^2 
    - y\, t
    + (1+y) k^2
\nonumber\\
&& \qquad\qquad  
    +\, y \big( M_B (M-3M_T) - M (M+8M_T) \big)
\Big]
\nonumber\\
&&
-\, 4\, (k\cdot p)^2\, t\,
\Big[ (2M-3M_T) \MTbar + y M (2M-M_T)
\Big]
\nonumber\\
&&
+\, 4\, (k\cdot p')^2\, t\,
\Big[ (3-y) k^2 - (1-3y) \MBbar \MTbar
\Big]
\nonumber\\
&&
-\, 4\, k\cdot P\, y\, t\, k^2 
\Big[ M_B (2M+3M_T) + 4 M M_T
\Big]
-\, 6\, k\cdot\Delta\, M_T\Delta_{TB}\, t\, k^2
\nonumber\\
&& 
-\, 2\, k\cdot p\, t\,
\Big[
    y (2M-3M_T) \MTbar\, t
    + 2M (2\Delta_T+\Delta_{TB})\, k^2
    - 4 y^2 M^3 (2M-M_T)
\nonumber\\
&& \qquad\qquad
    + y \MTbar 
    \big( 2M (3M+5M_B) M_T
    - 3 M_T^2 \MBbar
    - 4 M^2 \Delta_B
    \big)
\Big]
\nonumber\\
&&
-\, 2 k\cdot p'\, t\, 
\Big[ 4 y^2 M^2 (M+3M_T) \MBbar
    + y \MBbar \MTbar (M_T (2M-3M_T) + t)
\nonumber\\
&& \qquad\qquad
+\, 2 y^2 M \MBbar\, t
    + 2 \MTbar \big( 2 M - \overline{M}_{TB} \big)\, k^2
    + y \big( 4 M (\MTbar-yM) + t \big)\, k^2 
\Big]
\nonumber\\
&&
+\, 2M \big( 2 y^2 M M_T (4M-3M_T) + y^2 (2M-3M_T)\, t + 4 M_T\, k^2 \big)\,
    \MBbar \MTbar\, t
\nonumber\\
&&
+\, 4 y M^2 
    \big( y M_B (2M+3M_T) + 4 y M M_T - \MTBbar M_T - 2 M M_B    
    \big)\, t\, k^2
\nonumber\\
&&
+\, y\, \big( 2M (\MBbar+\MTbar) - 3M_T \Delta_{TB} \big)\, t^2\, k^2
    + 8 M M_T\, t\, k^4
\bigg\}\, ,
\end{eqnarray}
\end{subequations}
where we define $\Delta_{TB}=M_T-M_B$ and $\overline{M}_{TB}=M_T+M_B$.

\clearpage
Finally, for the KR diagrams with decuplet baryon intermediate states in Fig.~\ref{diagrams}(r) and~\ref{diagrams}(s), the numerator functions
$F^{\rm (KR\, 1,2)}_{T\phi}$ and $G^{\rm (KR\, 1,2)}_{T\phi}$ in Eqs.~(\ref{eq.fg_KRTphi}) are expressed as
\begin{subequations}
\label{eq.FG_KRTphi}
\begin{eqnarray}
F^{\rm (KR\, 1)}_{T\phi}
&=&\frac{1}{12 M_T^2} 
\bigg\{
- 4\, k\cdot p'\,
    \Big[ 2 y\, k\cdot P + 3 k\cdot p'- k\cdot p \Big]
\nonumber\\
&&
+\, 2\, k\cdot p'\,
    \Big[ 4 k^2 + 5 y M M_T + 6 (1+y) M^2
        - 3 M_T \Delta_T + y (1-y)\, t
    \Big]
\nonumber\\
&&    
-\, 2\, k\cdot p\ 
    \Big[ (1-y) M \big(2 M - M_T\big) - 3 M_T^2 \Big]
- 8 M \MTbar k^2 
\nonumber\\
&&      
-\, y (1-y) M (2 M - M_T)\, t
    + y \big( 3 M_T^2\, t - 8 M^3 \MTbar \big) 
\bigg\}\, ,
\\
G^{\rm (KR\, 1)}_{T\phi}
&=&\frac{M}{3 M_T^2\, t} 
\bigg\{
    2\, k \cdot p'\, k\cdot\Delta\, \big( 4 M+5 M_T \big)
\nonumber\\
&&    
-\, k \cdot \Delta\,
    \Big[ 2 M \big( 2 k^2 - 3 M_T^2 \big)
        + 6 M_T \big( k^2 - M_T^2 \big)
        + 4 M^2 \MTbar
    \Big]
\nonumber\\
&&    
-\, k\cdot p'\, t\, 
    \Big[ 2 y \big(2 - y\big) M + 3 y M_T + 2\MTbar
    \Big]
\nonumber\\
&& 
-\, t\,
    \Big[ 3 y M M_T^2  
        - \big(2 M + 3 M_T \big) k^2 
        - y \big( (1+y) M^2 M_T + 2 (2-y) M^3\big)
    \Big]
\bigg\}\, ,
\\
F^{\rm (KR\, 2)}_{T\phi}
&=&\frac{1}{12 M_T^2} 
\bigg\{
- 4\, k\cdot p \Big[ 2 y\, k\cdot P + 3 k\cdot p - k\cdot p' \Big]
\nonumber\\
&&
+\, 2\, k\cdot p\
    \Big[ 4 k^2 + 5 y M M_T + 6 (1+y) M^2 - 3 M_T\Delta_T + y(1-y)\, t 
    \Big]
\nonumber\\
&&    
-\, 2\, k\cdot p'\,
    \Big[ (1-y) M \big(2 M - M_T\big) - 3 M_T^2 \Big] 
- 8 M \MTbar k^2 
\nonumber\\
&&      
-\, y (1-y) M (2 M - M_T)\, t
+ y \big(3 M_T^2\, t - 8 M^3 \MTbar \big) 
\bigg\}\, ,
\\
G^{\rm (KR\, 2)}_{T\phi}
&=&\frac{M}{3 M_T^2\, t} 
\bigg\{
-2\, k\cdot p\, k\cdot\Delta\, \big( 4 M + 5 M_T \big)
\nonumber\\
&&
+\, k\cdot\Delta
\Big[ 2 M \big(2 k^2 - 3 M_T^2\big)
    + 6 M_T \big(k^2-M_T^2\big)
    + 4 M^2 \MTbar
\Big]
\nonumber\\
&&
-\, k\cdot p\, t\,
\Big[ 2 y \big(2- y\big) M + 3 y M_T + 2 \MTbar
\Big]
\nonumber\\
&& 
-\, t\,
\Big[ 3 y M M_T^2 
    - \big(2 M + 3 M_T \big) k^2
    - y \big( (1+y) M^2 M_T - 2 (y-2) M^3 \big)
\Big]
\bigg\}\, .~~
\end{eqnarray}
\end{subequations}

\clearpage
For the additional gauge link generated KR diagrams in Fig.~\ref{diagrams}(t) and \ref{diagrams}(u), the functions $\delta F^{\rm (KR\, 1,2)}_{T\phi}$ and $\delta G^{\rm (KR\, 1,2)}_{T\phi}$ in Eqs.~(\ref{eq.fg_delKRTphi}) are given by
\begin{subequations}
\label{eq.FG_delKRTphi}
\begin{eqnarray}
\delta F^{({\rm KR\, 1})}_{T\phi}
&=& \frac{y}{12 M_T^2}
\bigg\{
    - 16\, k\cdot p \, k\cdot p' \, k\cdot P
    +  4\, k\cdot p \, k\cdot p' 
    \Big[2M \big( 2 M + 3 M_T \big) - \big( 1+y \big) t \Big]
\nonumber\\
&& 
-\,   4 \left(k\cdot p'\right)^2
        \Big[ M \big( M_T + 2 M \big) + t \Big]
    - 4 (k\cdot p)^2 M \big(M_T-2 M\big)
\nonumber\\
&&
+\,   4\, k\cdot P\, 
        \Big[ 4 M^2 k^2 - 3 M_T^2\, t 
        \Big]
    + 4\, k\cdot\Delta\, M^2 \big( 4 M \MTbar - t \big)
\nonumber\\
&&
+\,   2\, k\cdot p'\,  t \Big[ 6 y M^2 -\, (5-11 y) M M_T - y t \Big]
    + 2\, k\cdot p\, M t \Big[ 2 y M -\,  (1+y) M_T \Big]
\nonumber\\
&&
+\,  t \Big[ 12 y M M_T \big( M_T \MTbar - k^2 \big)
        + 2 M^2 \big( y t + 2 (2-y) k^2 \big)
        + M M_T \big( 12 k^2 - y t \big) 
       \Big] 
\nonumber\\
&&     
    - 8 M^3\MTbar \big( 2 k^2 + y t \big) 
    - 3 y M_T^2\, t^2
\bigg\}\, ,
\\
\delta G^{\rm (KR\, 1)}_{T\phi}
&=& \frac{yM}{3M_T^2} 
\bigg\{ 
    4\, k\cdot p' 
    \Big[ k\cdot p'\, \big( 2 M + 3 M_T \big)
        + k\cdot p\,  \big( y M + \MTbar \big)
    \Big]
\nonumber\\
&&
+\, 4\, k\cdot P\, 
    \Big[ 3 M_T^2 \MTbar 
        - \big(2 M + 3 M_T\big) k^2 
    \Big]
\nonumber\\
&&
+\, 2\, k\cdot p'\,
    \Big[ (y M + \MTbar) t  - (4+11y) M^2 M_T - 2 (2+3y) M^3
    \Big]
\nonumber\\
&&
+\, 2\, k\cdot p\, y M^2 \big( M_T - 2 M \big)
    + 2 y M^2 \Big[ \MTbar \big( 4 M^2 - 6 M_T^2 \big)
                   + M \big( 2 k^2 - t \big)
                  \Big]
\nonumber\\
&&
+\,   4 M^2 \MTbar k^2
    + M t \big( 3 y M_T^2 - 2 k^2 \big)
    + y M^2 M_T \big( 12 k^2 + t \big)
    - 3 M_T\, t k^2
\bigg\}\, , 
\\
\delta F^{\rm (KR\, 2)}_{T\phi}
&=&\frac{y}{12M_T^2}
\bigg\{
    - 16\, k \cdot p\, k\cdot p'\, k\cdot P
    +  4\, k \cdot p\, k\cdot p' 
        \Big[ 2 M \big(2 M + 3 M_T\big) - \big(1+y\big) t \Big]
\nonumber\\
&&
-\,   4\, (k\cdot p)^2 \Big[ M \big( M_T+2 M \big) + t \Big]
    - 4\, (k\cdot p')^2 M \big(M_T-2 M\big)
\nonumber\\
&&
+\,   4\, k \cdot P 
        \Big[ 4 M^2 k^2 - 3 M_T^2 t \Big]
    - 4\, k \cdot \Delta\, M^2
        \Big[ 4 M \MTbar - t \Big]
\nonumber\\
&&
+\,   2\, k\cdot p\, t 
        \Big[ 6 y M^2 - (5-11y) M M_T - y t \Big]
    + 2\, k\cdot p'\, M t
        \Big[ 2 y M - (1+y) M_T \Big]
\nonumber\\
&&
+\, t \Big[12 y M M_T \big( M_T\MTbar - k^2 \big)
         + 2 M^2 \big( y t - 2 (y-2) k^2 \big)
         + M M_T \big( 12 k^2 - y t \big)
      \Big] 
\nonumber\\
&& 
-\, 8 M^3\MTbar \big(2 k^2+y t \big) 
    - 3 y M_T^2\, t^2 
\bigg\}\, , 
\\
\delta G^{\rm (KR\, 2)}_{T\phi}
&=& \frac{yM}{3 M_T^2}
\bigg\{
    4\, k\cdot p\,
    \Big[ k \cdot p\,  \big( 2 M + 3 M_T \big)
        + k \cdot p'\, \big( y M + \MTbar \big)
    \Big]
\nonumber\\
&&
+\, 4\, k\cdot P 
    \Big[ 3 M_T^2 \MTbar - \big(2 M + 3 M_T\big) k^2 
    \Big]
\nonumber\\
&&
+\, 2\, k\cdot p\ 
    \Big[ \big(y M + \MTbar\big) t - (4+11 y) M^2 M_T - 2 (2+3 y) M^3
    \Big]
\nonumber\\
&&
+\, 2\, k\cdot p'\, 
     y M^2 \big( M_T - 2 M \big) 
    + 2 y M^2 \Big[ \MTbar \big( 4 M^2 - 6 M_T^2 \big)
                    + M \big(2 k^2 - t\big)
              \Big]
\nonumber\\
&&
+\, 4 M^2 \MTbar k^2
    + M t \big( 3 y M_T^2 - 2 k^2 \big)
    + y M^2 M_T \big( 12 k^2 + t \big) - 3 M_T t\, k^2
\bigg\}\, .
\end{eqnarray}
\end{subequations}

\clearpage
\bibliography{ref}

\end{document}